\documentclass{ws-procs9x6_mod}
\usepackage{graphicx}
\usepackage{epsfig}
\usepackage{wrapfig}

\def\x{\chi}

\def\nt{\tilde \x^0}
\newcommand{\mnt}[1]   {m_{\tilde\x^0_{#1}}}

\def\sigva{\langle\sigma_A v\rangle}

\newcommand{\lang}{\left\langle}
\newcommand{\rang}{\right\rangle}

\newcommand{\GeV}      {~\mathrm{GeV}}
\newcommand{\TeV}      {~\mathrm{TeV}}

\newcommand{\GS}       {\mathrm{GS}}
\newcommand{\PL}       {\mathrm{pl}}

\newcommand{\re}{{\rm Re}}

\newcommand{\gappeq}{\mathrel{\rlap {\raise.5ex\hbox{$>$}}
{\lower.5ex\hbox{$\sim$}}}}
\newcommand{\lappeq}{\mathrel{\rlap{\raise.5ex\hbox{$<$}}
{\lower.5ex\hbox{$\sim$}}}}

\def\[{\left [}
\def\]{\right ]}
\def\({\left (}
\def\){\right )}

\renewcommand{\theequation}{\arabic{section}.\arabic{equation}}
\renewcommand{\thepage}{\arabic{page}}
\def\beq{\begin{equation}}
\def\eeq{\end{equation}}

\catcode`\@=11 

\def\lsim{\mathrel{\mathpalette\@versim<}}
\def\gsim{\mathrel{\mathpalette\@versim>}}
\def\@versim#1#2{\vcenter{\offinterlineskip
    \ialign{$\m@th#1\hfil##\hfil$\crcr#2\crcr\sim\crcr } }}
\def\etal{{\em et. al.}}

\def\t1{{\tilde 1}}

\def\GeV{\,{\rm GeV}}
\def\TeV{\,{\rm TeV}}

\def\to{\rightarrow}

\newcommand{\bear}{\begin{eqnarray}}
\newcommand{\eear}{\end{eqnarray}}

\newcommand{\goes}{\rightarrow}
\newcommand{\lapproxeq}{\lower .7ex\hbox{$\;\stackrel{\textstyle
<}{\sim}\;$}}
\newcommand{\gapproxeq}{\lower .7ex\hbox{$\;\stackrel{\textstyle
>}{\sim}\;$}}
\newcommand{\stackdown}[2]{\lower 1.4ex\hbox{$\;\stackrel{\textstyle{#1}}
{\scriptstyle{#2}}\;$}}
\newcommand{\lsp}{\tilde{\chi}}
\newcommand{\mlsp}{m_{\lsp}}

\newcommand{\be}{\begin{equation}}
\newcommand{\ba}{\begin{eqnarray}}
\newcommand{\ee}{\end{equation}}
\newcommand{\ea}{\end{eqnarray}}
\newcommand{\bpi}{\begin{picture}}
\newcommand{\bce}{\begin{center}}
\newcommand{\epi}{\end{picture}}
\newcommand{\ece}{\end{center}}

\begin{document}

\title{LHC Physics and Cosmology}

\author{N.E.~MAVROMATOS}

\address{King's College London, Department of Physics,\\
Strand, London WC2R 2LS, U.K.\\
E-mail: Nikolaos.Mavromatos@kcl.ac.uk}

\maketitle

\abstracts{In these Lectures I review possible constraints on particle physics models, obtained by means of combining the results of collider measurements with astrophysical data.
I emphasize the theoretical-model
dependence of these results. I discuss supersymmetric
dark matter constraints at colliders (mainly LHC) in various theoretical contexts: the
standard Cosmological-Constant-Cold-Dark-Matter ($\Lambda$CDM) model,
(super)string-inspired ones and
non-equilibrium relaxation dark energy models.
I then investigate the capability of LHC measurements in asserting whether supersymmetric matter (if discovered) constitutes part, or all, of the astrophysical dark matter. I also discuss prospects for improving the constraints in future precision facilities, such as the International Linear Collider. }

\section{Introduction}

In the past decade we have witnessed spectacular
progress in precision measurements in astrophysics as a result
of significant improvements in terrestrial and extraterrestrial
instrumentation. From the point of view of interest to particle physics,
the most spectacular claims from astrophysics came in 1999
from the study of distant (redshifts $z \sim 1$) supernovae (SNe) of type Ia
by two independent groups~\cite{supernovae}.
These observations pointed towards
a current era acceleration of our Universe, something that
could be explained
either by a non-zero cosmological {\it constant} in a Friedman-Robertson-Walker-Einstein Universe, or in general by a non-zero {\it dark energy} component,
which could even be relaxing to zero (the data are consistent
with this possibility). In the past five years~\cite{recentsn}
many more distant ($z > 1$) supernovae have been discovered,
exhibiting  similar features as the previous measurements,
thereby supporting the geometric interpretation of the acceleration
of the Universe today, and arguing against the nuclear physics or
intergalactic dust effects.

\begin{figure}[t]
\centering
\epsfig{file=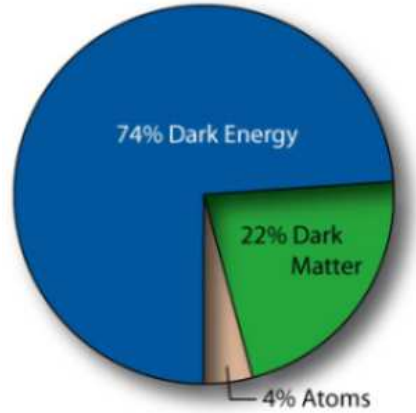, width=0.5\textwidth}
\caption{The energy content of our Universe as obtained
by fitting data of WMAP satellite. The chart is in
perfect agreement with earlier claims made by direct measurements
of a current era acceleration of the Universe from distant
supernovae type Ia (courtesy of
http://map.gsfc.nasa.gov/).}
\label{budget}
\end{figure}

Moreover, there is strong additional evidence from quite different in origin and thus independent astrophysical observations, those associated with the WMAP measurements of
the cosmic microwave background radiation (CMB)~\cite{wmap},
as well as baryon acoustic oscillations measurements~\cite{bao}.
After three years of running, WMAP
measured CMB anisotropies to an unprecedented accuracy
of billionth of a Kelvin degree, thereby correcting previous measurements
by the Cosmic Background Explorer (COBE) satellite~\cite{cobe}
by several orders of magnitude.
This new satellite experiment, therefore, opened up a new era
for astro-particle physics, given that such accuracies allow
for a determination (using best fit models of the Universe)
of cosmological parameters~\cite{spergel}, and
in particular cosmological densities,
which, as we shall discuss in this review,
is quite relevant for constraining models of particle physics
to a significant degree.
All these measurements point towards
the fact that (more than) $73$ \% of the Universe vacuum energy consists of
a dark (unknown) energy substance, in agreement with the (preliminary)
supernovae observations (see fig.~\ref{budget}).
This claim, if true, could revolutionize our
understanding of the basic physics governing fundamental interactions
in Nature. Indeed, only a few years ago, particle theorists were trying to
identify (alas in vain!) an exact symmetry of nature that could
set the cosmological constant (or more generally the vacuum energy) to zero.
Now, astrophysical observations point to the contrary.
The WMAP satellite experiment determined the most important
cosmological parameters that could be of relevance to
particle physicists, namely~\cite{spergel}: the
Hubble constant, and thus the
age of the Universe,
the thickness of the last scattering surface, the
dark energy and dark matter content of the Universe
(to an unprecedented accuracy) (c.f. figure \ref{budget}),
confirming the earlier claims from supernovae Ia data~\cite{supernovae},
and
provided evidence for early re-ionization ($z \sim 20$),
which, at least from the point of view of large scale structure
formation, excludes Warm Dark Matter particle theory models.
Its measurements have been independently confirmed recently by Baryon Oscillation measurements~\cite{bao}.

In this review I shall first describe
briefly the above-mentioned astrophysical measurements, and then use them
to constrain~\cite{lmnreview}  some particle physics
supersymmetric models, in particular (i) the minimally supersymmetric
model, constrained by its embedding
in a minimal supergravity model (mSUGRA)~\cite{msugra}, (ii) a class of  string-inspired cosmologies~\cite{heteropheno}, with non-trivial dilaton and/or moduli fields, and (iii) a class of stringy relaxation dark energy models~\cite{qcosmol}.
I shall give a critical discussion on the derived constraints,
and their strong theoretical-model dependence, and  discuss the capability of observing Supersymmetry
in colliders after the latest astrophysical data.
In this respect I will also discuss  the importance of the $g_{\mu}-2$ experiments
of the muon gyromagnetic ratio, in light of some very recent (2006) results~\cite{eidelman},
pointing towards a discrepancy between the measured $g_{\mu}-2$
value from the one calculated within the standard model.

\section{Astrophysical Measurements and Facts}

On Large scales our Universe looks isotropic and homogeneous.
A good formal description, which does not depend on the detailed underlying
microscopic model, is provided by the Robertson-Walker (RW) metric, according to which the
geometry of the Universe is described by means of the following
space-time invariant element~\cite{kolbturner}:
\begin{gather}\label{rw}
ds^2 = -dt^2 + a(t)^2~R_0^2 \left[\frac{dr^2}{1-k~r^2 }
+ r^2\left(d\theta^2 + {\rm sin}^2\theta d\varphi^2 \right)\right]
\end{gather}
where
$a(t) = \frac{R(t)}{R_0} =\frac{1}{1+z}$ is the scale factor,
$H \equiv \frac{\dot a}{a}$ is the Hubble Parameter,
$t$ is the Cosmological Observer time,
$R_0$ denotes the present-day scale factor,  $z=$ is the redshift,
 $k$ denotes the ~{\it Spatial Curvature}, which (by normalization) can take on the values:
 k=0 for a flat Universe (required by inflationary models),
 k=1 for a closed and  k=-1 for an open Universe.
In this section we shall outline
the main Cosmological Measurements and the pertinent quantities, of interest to us in these Lectures. For more details we refer the reader to the literature~\cite{kolbturner}.

\subsection{Model Independent (Geometric) Considerations}

An important quantity, which we shall make extensive use of in the following, when we use astrophysical data to
constrain theoretical models, is the so-called
{\it Luminosity Distance}, $d_L$, defined as:
$ d_L = \sqrt{\frac{{\mathcal L}}{4\pi  {\mathcal F}}}$,
 where ${\mathcal L}$ is the energy per unit time emitted by the source, at the source's rest frame, and
${\mathcal F}$ is the flux measured by detector, i.e. the energy per unit time per unit area measured by the detector. Taking into account
the time dilation induced by the expansion of the Universe:  $\delta t_{\rm detector} = (\delta t)_{\rm source} (1 + z) $,  as well as as the cosmic red-shift,  i.e. the fact that there is a reduced energy of photons at the detector as compared with that at emission from the source, we obtain:
$ {\mathcal F} = \frac{{\mathcal L}}{4\pi a(t_0)^2r_1^2(1 + z)^2}~.$

Another very commonly used quantity in Astrophysics is
the  {\it Angular Diameter}, which is defined as follows: A celestial object (cluster of galaxies etc.) has proper diameter $D$ at $r=r_1$ and emits light at $t=t_1$. The observed angular diameter by a detector at $t=t_0$ is: $ \delta = \frac{D}{a(t_1) r_1}~. $
 From this one defines the
 {\it Angular Diameter Distance}: $ d_A = \frac{D}{\delta} =a(t_1) r_1 = d_L (1 + z)^2~.$

The {\it Horizon Distance} (beyond which light cannot reach us) is defined as:
 $ ds^2 = 0 = dt^2 -a^2(t)\left(\frac{dr^2}{1-kr^2} + r^2 d\Omega ^2 \right)~.$ For  radial motion of light, pertinent to most observations,
 along null geodesics $ds^2 =0$, we have: $ \int_0^t \frac{dt'}{a(t')}=\int_0^{r_H}\frac{dr}{\sqrt{1 - kr}}$
 from which
 $  d_H =  a(t) \int_0^{r_H}\sqrt{g_{rr}} = a(t)\int_0^t \frac{dt'}{a(t')}$.
 In Standard Cosmology  $d_H \sim t_{\rm Age} < \infty $ due to the
 finite age of the Universe, i.e. there is an Horizon.

The above quantities are related among themselves~\cite{kolbturner}, as follows from the cosmic redshift phenomenon, the fact that photons follow null geodesics $ds^2=0$ {\it etc.}
These leads to relations among  $H_0, d_L $ and  the redshift $z$,
which are model independent, and they follow from pure geometrical considerations relying on the assumption of a RW homogeneous and isotropic cosmology. In the next subsection we discuss how a specific dynamical model of the Universe affects the cosmological measurements. In particular, as we shall show, model dependence is hidden inside the details of the dependence of the Hubble parameter on the various components of the Universe's  energy budget. This property is a consequence of the pertinent dynamical equations of motion of the gravitational field.

\subsection{Cosmological Measurements: Model Dependence}

Within the standard General-Relativistic framework, according to which
the dynamics of the gravitational field is described by the Einstein-Hilbert  action, the gravitational (Einstein) equations in a Universe with cosmological constant $\Lambda$ read:
$R_{\mu\nu} -\frac{1}{2}g_{\mu\nu}R +  g_{\mu\nu} \Lambda  = 8\pi G_N T_{\mu\nu}~,$
where $G_N$ is (the four-dimensional) Newton's constant, $T_{00} = \rho$ is the energy density of matter, and $T_{ii} =a^2 (t) p$ with $p=$ the pressure, and we assumed that the Universe and matter systems behave like ideal fluids in a co-moving cosmological frame, where all cosmological measurements are assumed to take place. From the RW metric (\ref{rw}),  we arrive at the
Friedman equation:
\begin{gather} \label{friedman}
3\left(\frac{\dot a}{a}\right)^2 + 3\frac{k}{a^2} -\Lambda =8\pi G_N \rho
\end{gather}
From this equation one obtains the expression for the Critical density (i.e. the total density required for flat $k=\Lambda=0$ Universe):
${\rho_c =\frac{3}{8\pi G_N}\left(\frac{\dot a}{a}\right)^2 }$.

From the dynamical equation (\ref{friedman}) one can obtain various relations between the Hubble parameter $H(z)$, the luminosity distance $d_L$, the deceleration parameter $q(z)$ and the energy densities $\rho$ at various epochs of the Universe. For instance, for matter dominating flat ($k=0$) Universes with $\Lambda > 0$ and
various (simple, $z$-independent) equations of state
 $p = w_i \rho~,$ ~($w_r = 1/3$ (radiation), $w_m =0$ (matter-dust), $~w_\Lambda =-1$ (cosmological constant (de Sitter)) we have
for the Hubble parameter:
 \begin{gather}\label{detailshubble}
{H(z) = H_0 \left( \sum_i \Omega_i (1 + z)^{3(1 + w_i)} \right)^{1/2}}
\end{gather}
with the notation: $\Omega_i \equiv \frac{\rho_i^0}{\rho_c}~,$ $i=r$(adiation), $m$(atter), $\Lambda, ...$

For the
deceleration parameter
we have at late eras, where radiation is negligible :
${q(z) \equiv -\frac{\ddot a a}{(\dot a)^2} =  \left(\frac{H_0}{H(z)}\right)^2 \left(\frac{1}{2}\Omega_m (1 + z)^3 - \Omega_\Lambda \right)}$, with $q_0 = \frac{1}{2}\Omega_m - \Omega_\Lambda $. Thus, it becomes evident that $\Lambda$ acts as ``repulsive'' gravity, tending to accelerate the Universe currently, and eventually dominates, leading to an eternally accelerating de Sitter type Universe, with a future cosmic horizon.
 At present in the data there is also evidence for past deceleration ($q(z) > 0$~, for some $z > z^{\star} >0$), which is to be expected if the dark energy is (almost) constant, due to matter dominance in earlier eras: $q(z) > 0 \Rightarrow (1 + z)^3 > 2\Omega_\Lambda/\Omega_m~ \Rightarrow z >  {z^\star = \left(\frac{2\Omega_\Lambda}{\Omega_m}\right)^{1/3} - 1}~. $

 Finally, for the luminosity distance we obtain the important relation:
 $d_L = (1 + z)\int_{t_1}^{t_0} \frac{a(t_0)dt}{a(t)} =
-(1 + z) \int_{a(t_1)}^{a(t_0)}\frac{d(\frac{a(t_0)}{a'})}{(\dot a/a)}~,$
from which
\begin{gather} \label{luminosityhublle} {d_L = (1 + z)\int_0^z \frac{dz}{H(z)}}
\end{gather}
We shall use this relation in the following, in order to constrain various theoretical cosmological models by means of astrophysical observations.

\subsection{Supernovae Ia Measurements of Cosmic Acceleration}

Type Ia Supernovae (SNe) behave as Excellent Standard Candles, and thus  can be used to measure directly the expansion rate of the Universe at high redshifts ($z \ge 1$) and compare it with the present rate, thereby providing direct
information on the Universe's acceleration.
SNe type Ia are  very bright objects, with absolute magnitude $M \sim 19.5$,
typically comparable to the brightness of the entire host galaxy!
This is why they can be detected at high redshifts $z \sim 1$, i.e.
$3000~Mpc, 1 pc \sim 3 \times 10^{16} m$.
Detailed studies of the luminosity profile~\cite{supernovae,recentsn}
of each SNe
suggests a strong relation between the width of the light curve
and the absolute luminosity of SNe. This implies an accurate determination of its absolute luminosity.
For each supernova one measures an effective (rest frame) magnitude in blue wavelength band, $m^{eff}_B$, which is then
compared with the theoretical expectation (depending on the underlying model for the Universe) to yield information on the various $\Omega^i$.
The larger the magnitude the dimmer the observed SNe.

\begin{figure}[htb]
\begin{center}
  \epsfig{file=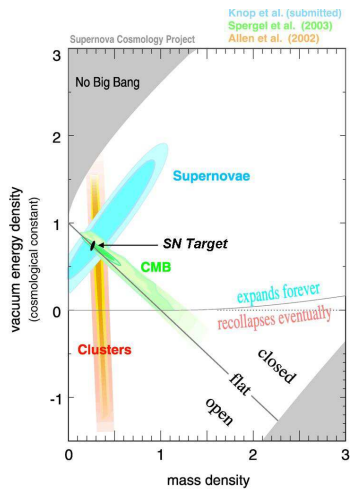, width=0.4\textwidth}
\end{center}
\caption{Supernovae (and other) measurements on the Universe's energy budget.}\label{fig:sneresults}
\end{figure}

To understand the pertinent measurements
recall the relation between the observed (on Earth) and
emitted wavelengths $\lambda_{\rm obs} = (1 + z) \lambda_{\rm emit}$, as a result of the cosmic redshift phenomenon.
In a magnitude-redshift graph, if nothing slowed down matter
blasted out of the Big Bang, we would expect a  straight line.
The data from High-redshift ($z \sim 1)$ SNe Ia,
showed that distant SNe lie  slightly above the straight line.
 Thus they are moving away  slower
than expected.
So at those  early days $(z \sim 1)$
the Universe was expanding
at a  slower rate than now.
{\it The Universe accelerates today! }
In such measurements, one needs the Hubble-Constant-Free Luminosity Distance:
${{\mathcal D}_L(z;\Omega_M, \Omega_\Lambda)
=\frac{H_0}{c}d_L, ~~d_L \equiv \sqrt{\frac{L}{4\pi F}}}$,
with $L$ the Intrinsic Luminosity of source, $F$ the
Measured Flux.
In Friedman models ${\mathcal D}_L$ is parametrically
known in terms of $\Omega_{M}, \Omega_\Lambda$.
An important quantity used in measurements is the Distance Modulus m - M,
where ${m = M + 25 + 5{\rm log}\left(\frac{d_L}{1~Mpc}\right)={\mathcal M} + 5{\rm log}{\mathcal D}_L }$,
 with $m$=Apparent Magnitude of the Source,
$M$ the Absolute Magnitude,  and ${\mathcal M} \equiv M -5{\rm log}H_0 + 25$
 the fit parameter.
 Comparison of theoretical expectations with data restricts $\Omega_M,\Omega_\Lambda$.
An important point to notice is that for  fixed redshifts z  the eqs.
${\mathcal D}_L(z;\Omega_M,\Omega_\Lambda) =$constant
yields {\it degeneracy curves} ${\mathcal C}$ in the $\Omega$-plane, of small curvature to which one associates a small slope, with the result that
even very accurate data can at best select
a narrow strip in $\Omega$-plane parallel to  ${\mathcal C}$.
The results (2004) are summarized in figure~\ref{fig:sneresults}
In the early works (1999) it was claimed that the best fit model, that of a FRW Universe with matter and cosmological constant for $z \le 3$ (where the SNe data are valid) yields the following values:
${0.8\Omega_M - 0.6\Omega_\Lambda \simeq -0.2 \pm 0.1~, \quad {\rm for}~\Omega_M \le 1.5 }$.
Assuming a flat model (k=0) the data imply:
${\Omega_M^{Flat} = 0.28^{+0.09}_{-0.08}~( 1\sigma~stat)^{+0.05}_{-0.04}~( {\rm identified~syst.})}$,
that is the {\it Universe accelerates today}
\begin{gather*}
{{q_0 =\frac{1}{2}\Omega_M - \Omega_\Lambda \simeq -0.6 < 0}}
\end{gather*}
Further support on these results comes, within the SNe measurement framework,  from the  recent ($> 2004$) discovery~\cite{recentsn},
by Hubble Space Telescope, ESSENCE  and SNLS Collaborations,
of more than  100 high-z ($ 2 > z \ge 1$) supernovae, pointing towards the fact that for the past 9 billion years the energy budget of the Universe is dominated  by an approximately constant dark energy component.

-

\subsection{CMB Anisotropy Measurements by WMAP1,3}

\begin{figure}
\centering
\includegraphics[angle=90,width=0.6\textwidth]{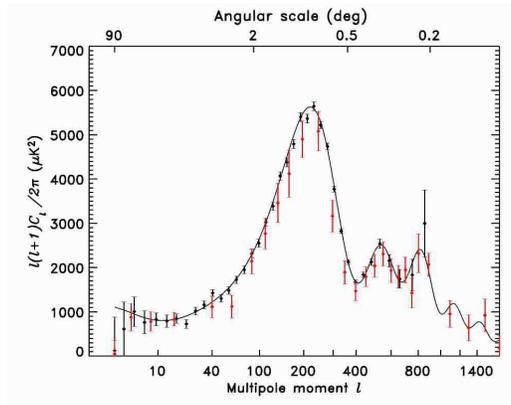}
\caption{Red points (larger errors) are previous measurements. Black points (smaller errors) are WMAP measurements
(G. Hinshaw, {\it et al.} arXiv:astro-ph/0302217).}
\label{wmapgaussian}
\end{figure}

After three years of running, WMAP provided
a much more detailed picture of the temperature fluctuations
than its COBE predecessor, which can be analyzed to
provide best fit models for cosmology, leading to
severe constraints on the energy content of
various model Universes, useful for particle physics,
and in particular supersymmetric searches.
Theoretically~\cite{kolbturner},
the temperature fluctuations in the CMB radiation
are attributed to: (i) our velocity w.r.t cosmic rest frame,
(ii) gravitational
potential fluctuations on the last scattering surface (Sachs-Wolf effect),
(iii) Radiation field fluctuations on the last scattering surface,
(iv) velocity of the last scattering surface, and
(v) damping of anisotropies if Universe
re-ionizes after decoupling.
A Gaussian model of fluctuations~\cite{kolbturner}, favored by inflation,
is in very good agreement with the recent WMAP data
(see figure \ref{wmapgaussian}). The perfect fit of the first few peaks
to the data allows a precise determination of the total density of the
Universe, which implies its spatial flatness.
The various peaks in the spectrum of fig.~\ref{wmapgaussian} contain interesting physical signatures:

(i) The angular scale of the first peak determines the curvature (but not the topology) of the Universe.

(ii) The second peak –--truly the ratio of the odd peaks to the even peaks-–- determines the reduced baryon density.

(iii) The third peak can be used to extract information about the dark matter (DM) density (this is a model-dependent result, though –--standard local Lorentz invariance assumed, see discussion in later sections on Lorentz-violating alternative to dark matter models).

The measurements of the WMAP~\cite{spergel} on the cosmological parameters
of interest to us here are given in \cite{spergel}, and reviewed in \cite{lmnreview}. The WMAP results constrain severely the
equation of state $p=w \rho $ ($p=$pressure),
pointing towards $w < -0.78$, if one fits the data with the assumption
$-1 \le w$  (we note for comparison that in the scenarios advocating the existence of a cosmological {\it constant}
one has $w=-1$).
Many quintessence models can easily satisfy
the criterion $-1 < w < -0.78$, especially the supersymmetric ones,
which we shall comment upon later in the article.
Thus, at present, the available data are not sufficient to distinguish
the cosmological constant model from quintessence (or more generally
from relaxation models of the vacuum energy).
The results lead
to the chart for the energy and
matter content of our Universe depicted in figure \ref{budget},
and are in perfect agreement
with the Supernovae Ia Data~\cite{supernovae}.
The data of the WMAP satellite
lead to a new determination of
$\Omega_{\rm total} =
1.02 \pm 0.02 $, where $\Omega _{\rm total} = \rho_{\rm total}/\rho_c$, due
to high precision measurements
of secondary (two more) acoustic peaks
as compared with previous CMB measurements
(c.f. figure \ref{wmapgaussian}). Essentially the value of $\Omega$
is determined by the position of the first acoustic peak in a
Gaussian model, whose reliability increases significantly
by the
discovery of secondary peaks and their excellent fit with the Gaussian
model~\cite{spergel}.

Finally we mention that the determination of the cosmological parameters by the WMAP
team~\cite{spergel}, after three years of running.
favors, by means of best fit procedure,
{\it spatially flat} inflationary
models of the Universe~\cite{wmapinfl}.
In general, WMAP gave values for important inflationary parameters, such as the
running spectral index, $n_s(k)$, of the primordial power spectrum of scalar density  fluctuations $\delta_{\vec k} $~\cite{runningspect}
$P(k) \equiv |\delta _{\vec k} |^2 \; . $
The running scalar spectral index $n_s (k)$ is
$ n_s (k) = \frac{d{\rm ln}P(k)}{d{\rm ln}k}~,$
where $k$ is the co-moving scale.
Basically inflation implies $n_s = 1$.
WMAP measurements yield
$n_s = 0.96$, thus favoring Gaussian primordial fluctuations, as predicted by inflation. For more details we refer the reader to the literature~\cite{spergel,lmnreview}.

\subsection{Baryon Acoustic Oscillations (BAO)}

 Further evidence for the energy budget of the Universe is obtained by  Detection of the baryon acoustic peak in the large-scale correlation function of SDSS luminous red galaxies~\cite{bao}. The underlying Physics of BAO can be understood as follows:  Because the universe has a significant fraction of baryons, cosmological theory predicts that the acoustic oscillations (CMB) in the plasma will also be  imprinted onto the late-time power spectrum of the non-relativistic matter:  from an initial point
perturbation common to the dark matter and the baryons, the dark matter
perturbation grows in place while the baryonic perturbation is carried outward
in an expanding spherical wave.
At recombination, this shell is roughly 150 Mpc in radius.  Afterwards,
the combined dark matter and baryon perturbation seeds the formation
of large-scale structure.  Because the central perturbation in the dark matter is dominant compared to the baryonic shell, the acoustic feature is
manifested as a small single spike in the correlation function at 150 Mpc separation~\cite{bao}.

 The acoustic signatures in the large-scale clustering of galaxies
yield three more opportunities to test the cosmological paradigm with the early-universe acoustic phenomenon:

\noindent{\bf (1)} They would provide smoking-gun evidence for the theory of gravitational clustering, notably the idea that large-scale
fluctuations grow by linear perturbation theory from $z\sim 1000$ to the present;

\noindent{\bf (2)} they would give another confirmation of
the existence of dark matter at $z\sim1000$, since a fully baryonic
model produces an effect much larger than observed;

\noindent{\bf (3)} they would provide
a characteristic and reasonably sharp length scale that can be measured
at a wide range of redshifts, thereby determining purely by geometry the
angular-diameter-distance-redshift relation and the evolution of the Hubble parameter.

In the current status of affairs of the BAO measurements it seems that there is an underlying-theoretical-model dependence of the interpretation of the results, as far as the predicted energy budget for the Universe is concerned. This stems from the fact that for small deviations from $\Omega_m=0.3$, $\Omega_\Lambda=0.7$, the
change in the Hubble parameter at $z=0.35$ is about half of that of
the angular diameter distance.  Eisenstein {\it et al.} in \cite{bao} modelled this  by treating the dilation scale
as the cubic root of the product of the radial dilation times the square
of the transverse dilation.  In other words, they defined
 \begin{equation}\label{eq:D}
D_V(z) = \left[ D_M(z)^2 {cz\over H(z)}\right]^{1/3}~, ~ H=H_0\left(\sum_i \Omega_i (1 + z)^{3(1 + w_i)}\right)^{1/2}
\end{equation}
 where $H(z)$ is the Hubble parameter and $D_M(z)$ is the co-moving angular
diameter distance.
As the typical redshift of the sample is $z=0.35$, we quote the result~\cite{bao}
for the dilation scale as $D_V(0.35)=1370 \pm 64 {\rm Mpc}$.
The BAO measurements from Large Galactic Surveys and their results for
the dark sector of the Universe are consistent with the WMAP data,
as far as the energy budget of the Universe is concerned,
but the reader should bear in mind that they based their parametrization on standard FRW cosmologies, so the consistency should be interpreted within that theory framework.

\subsection{Measuring H(z): an important constraint on models}

The previous results, based on SNe, CMB and BAO measurements, relied on the standard FRW Cosmological model for the Universe as the underlying theory.
However, in modern approaches to (quantum) gravity, such as brane and string theories, the underlying dynamics may no longer be described by the simple Einstein-Hilbert action. One may have extra fields, such as the dilaton or moduli fields in theories with extra dimensions, plus higher order curvature terms which could become important in the early Universe.
Moreover, there have been suggestions in the literature~\cite{riotto} that the claimed Dark Energy may not be there, but simply be the result of temperature fluctuations in a (flat) Universe filled with matter $\Omega_M = 1$ (``super-horizon model'').
\begin{figure}[t]
\begin{center}
\psfig{file=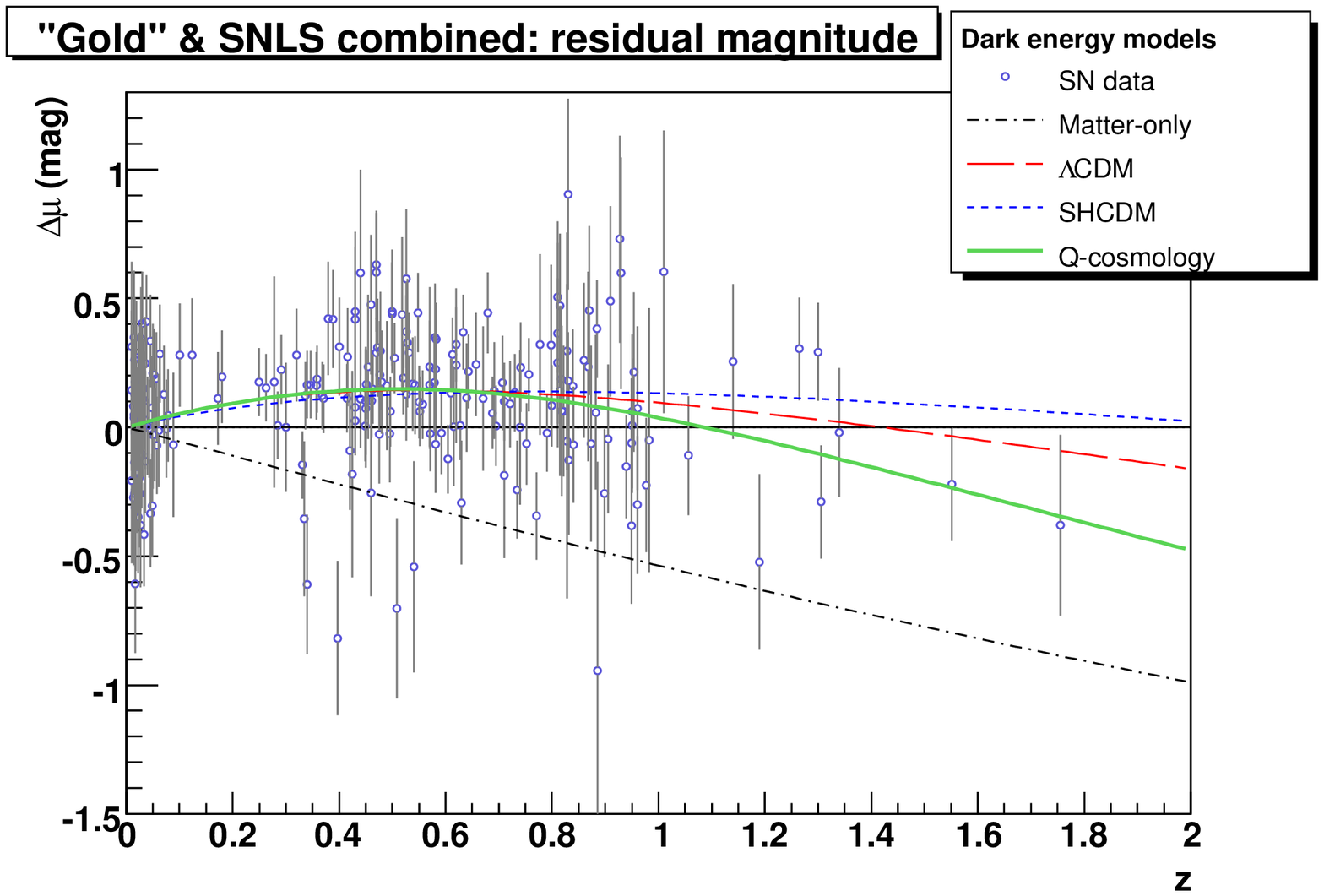,width=0.4\linewidth} \hfill
\psfig{file=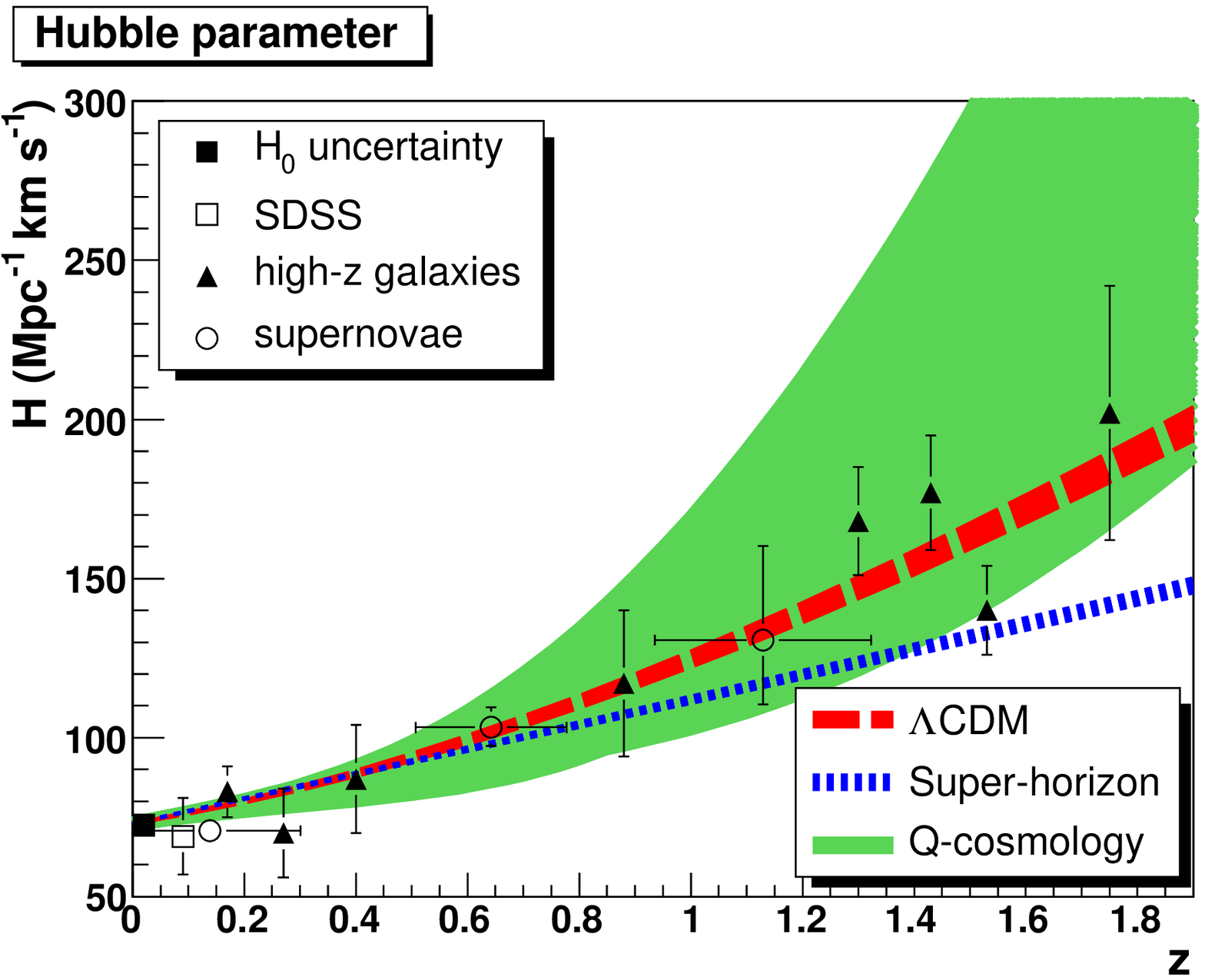,width=0.4\linewidth,clip=}
\end{center}
\caption{\underline{Left:}
 Residual magnitude versus redshift for supernovae from the `gold' and
the SNLS datasets for various cosmological models. \underline{Right: }  The Hubble-parameter vs. redshift relation for these models
and observational data. The bands represent 68\% confidence intervals derived
by the SN analysis for the standard $\Lambda$CDM, the
super-horizon (no DE) and the Q-cosmology models. The black rectangle shows the WMAP3 estimate for $H_0$, the squares show the measurements from SDSS
galaxies, the triangles result from high-$z$ red galaxies, and the circles correspond to a combined analysis
of supernovae data  (from [16]). }
\label{fig:HubbleCL}
\end{figure}
All such alternative theories should be tested against each one of the above-mentioned categories of measurements together  with an independent measurement of the behavior of the Hubble parameter vs. the redshift  $H(z)$, the latter coming from large galactic surveys. This
latter measurement provides
an {\it important constraint}  which could differentiate among the potential Dark Energy (DE)/Dark Matter (DM) models and their alternatives.
 This extra measurement has the potential of ruling out alternative models (to DM and DE) that otherwise fit the supernova data alone (in a $m_{\rm eff}$ vs z plot). This happens, for instance, with the super-horizon model of \cite{riotto}. I mention in passing that other non-equilibrium stringy cosmologies~\cite{qcosmol}, with relaxing to zero dark energy (quintessence-like due to the dilaton field) survive at present this constraint, as illustrated in figure~\ref{fig:HubbleCL}.
For more details I refer the reader to \cite{mitsou} and references therein.

\subsection{Cosmic Coincidence and Cosmological Constant Issues}

There may be several possible explanations regarding the Dark Energy part of the Universe's energy budget:

\noindent{\bf (i)}  The dark energy is an ``Honest''  Cosmological Constant  $\Lambda \sim 10^{-122}M_{\rm Pl}^4$,
strictly unchanging through space and time.
 This has been the working hypothesis of many of the best fits so far, but I stress it is {\it not} the only explanation consistent with the data.

\noindent{\bf (ii)} Quintessence: The Cosmological constant is mimicked
by a slowly-varying field, $\phi$,
whose time until it reaches its potential minimum
is (much) longer than the Age of Universe.
Simplest Quintessence models assume exponential potentials $V(\phi) \sim e^{\phi}$. In such a case the pertinent
equation of state reads: $w = \frac{\frac{({\dot \phi})^2}{2} - V(\phi)}
{\frac{({\dot \phi})^2}{2} + V(\phi)}$. For $\phi=-2{\rm ln}t$ one has a relaxing-to-zero vacuum energy
$\Lambda (t) \sim {\rm const} /t^2 $
(in Planck units), of the right order of magnitude today. Such a situation could be met~\cite{qcosmol} in some models of string theory, where the r\^ole of the quintessence field could be played by the dilaton~\cite{veneziano}, i.e. the scalar field of the string
gravitational multiplet.

\noindent{\bf (iii)} Einstein-Friedman model is incorrect, and one could have modifications in the gravitational law at galactic or supergalactic scales. Models of this kind have been proposed as alternatives to dark matter, for instance Modified Newtonian Dynamics (MOND) by Milgrom~\cite{milgrom}, and its field theory version by Bekenstein~\cite{TeVes}, known as Tensor-Vector-Scalar (TeVeS) theory, which however, is Lorentz Violating, as it involves a preferred frame.
Other modifications from Einstein theory, which however maintain Lorentz invariance of the four-dimensional world, could be brane models for the Universe, which are characterized by non-trivial, and in most cases time dependent, vacuum energy. It should be noted that such alternative models
may lead to completely different energy budget~\cite{skordis,liguori}. We shall discuss one such case of a non-critical string inspired (non-equilibrium, relaxation) cosmology (Q-cosmology) in a subsequent section, where we shall see that one may still fit the astrophysical data with exotic forms of ``dark matter'', not scaling
like dust with the redshift at late epochs, and different percentages of dark (dilaton quintessence) energy (c.f. also fig.~\ref{fig:HubbleCL}).

Given that from most of the standard best fits for the Universe it follows that
the energy budget of our Cosmos today is characterized by $73-74 \% $ vacuum energy, i.e. an energy density of order $\rho_{vac} \simeq (10^{-3}~eV)^4 = 10^{-8}~erg/cm^3 $,
and about $27-26 \%$ matter (mostly dark),
this implies {\it the Coincidence Problem}:
``The vacuum energy density  today
is  approximately equal (in order of magnitude) to the current
matter density.''  As the Universe expands, this relative balance is lost  in models with a cosmological constant, such as
the standard $\Lambda$CDM model,
since the matter density scales with the scale factor as $ \frac{\Omega_\Lambda}{\Omega_M} =\frac{\rho_\Lambda}{\rho_M} \propto a^{3} $. In this framework,
at early times we have that the Vacuum Energy is much more suppressed as compared with that of Matter and Radiation, while at
 late times it dominates. There is only one brief epoch for which the transition from domination of one
component to the other can be witnessed, and this epoch, according to the
$\Lambda$CDM model, happened to be the present one!
This calls for a microscopic Explanation, which is still lacking.

The smallness of the value of the Dark Energy today is another big mystery of particle physics. For several years the particle physics community thought that the vacuum energy was exactly zero, and in fact they were trying to devise microscopic explanations for such a vanishing by means of some symmetry.
One of the most appealing, but eventually failed in this respect, symmetry justifications
for the vanishing of the vacuum energy was that of {\it supersymmetry} (SUSY): if unbroken, supersymmetry implies strictly a vanishing vacuum energy, as a result of the cancelation among boson and fermion vacuum-energy contributions, due to opposite signs in the respective quantum loops.
However, this cannot be the correct explanation, given that SUSY, if it is to describe Nature, must be broken below some energy scale $M_{\rm susy}$, which  should be higher than a few TeV, as partners have not been observed as yet.
In broken SUSY theories, in four dimensional space times,
there are contributions to vacuum energy
$\rho_{\rm vac} \propto \sim \hbar M_{\rm susy}^4 \sim ({\rm few~TeV})^4$,
which is by far greater than the observed value today of the dark energy
$\Lambda \sim 10^{-122}$$~M_P^4$, $M_P \sim 10^{19}$ GeV.
Thus,
SUSY does not solve the {\it Cosmological Constant Problem}, which at present remains one of the greatest mysteries in Physics.

In my opinion, the smallness today of the value of the ``vacuum'' energy
density might point towards a relaxation problem.
Our world may have not yet reached equilibrium, from which it departed during an early-epoch cosmically catastrophic event, such as a Big Bang, or ---in the modern  version of string/brane theory ---a collision between two brane worlds.
This non equilibrium situation might be expressed today by a quintessence like exponential potential $\exp{(\phi)}$, where $\phi$ could be the dilaton field,
which in some models~\cite{qcosmol}
behave at late cosmic times as $\phi \sim -2{\rm ln}t$.
This would predict a vacuum energy today of order $1/t^2$, which has the right order of magnitude, if $t$ is of order of the Age of the Universe, i.e. $t \sim 10^{60}$ Planck times. Supersymmetry in such a picture may indeed be a symmetry of the vacuum, reached asymptotically, hence the asymptotic vanishing of the
dark energy. SUSY breaking may not be a spontaneous breaking but an {\it obstruction}, in the sense that only the excitation particle spectrum has mass differences between fermions and bosons. To achieve phenomenologically realistic situations, one may exploit~\cite{gravanis} the string/brane framework, by compactifying the extra dimensions into manifolds with non-trivial ``fluxes'' (these are not gauge fields associated with electromagnetic interactions, but pertain to extra-dimensional unbroken gauge symmetries characterizing the string models).
In such cases, fermions and bosons couple differently, due to their spin,
to these flux gauge fields (a sort of generalized ``Zeeman'' effects). Thus, they exhibit mass splittings proportional to the square of the ``magnetic field'', which could then be tuned to yield phenomenologically acceptable SUSY-splittings, while the relaxation dark energy has the cosmologically observed small value today.
In such a picture, SUSY is needed for stability of the vacuum, although
today, in view of the landscape scenarios for string theory, one might not even have supersymmetric vacua at all.
However, there may be another reason why SUSY could play an important physical r\^ole, that of dark matter. I now come to discuss this important issue, mainly from a particle physics perspective.

\section{Dark Matter (DM)}

In this section I will discuss issues pertaining to dark matter and supersymmetry. I will first make the case for Dark Matter, starting historically from discrepancies concerning rotational curves of galaxies.
Then I will move to describe possible candidates, and based on standard models for cosmology to exclude many of them, by means of WMAP data, arguing that supersymmetric dark matter remains compatible with such data.
I will again emphasize, however, the model dependence of such conclusions. Then I will proceed to discuss supersymmetric particle physics constraints in various frameworks by describing the underlying general framework for calculating thermal dark matter relics and compare them with WMAP data.
For a more complete discussion on direct searches for dark matter the reader is referred to \cite{zacek}, and references therein.

\subsection{The Case for DM}

Dark Matter (DM) is defined as a Non luminous massive matter, of unknown
composition, that does not emit or reflect enough electromagnetic radiation to be observed directly, but whose presence  can be inferred from gravitational effects on visible matter.
Observed phenomena consistent with the existence of dark matter are:

\noindent{\bf (i)} rotational speeds of galaxies and orbital velocities of galactic clusters,

\noindent{\bf (ii)} gravitational lensing of background objects by galaxy clusters such as the Bullet cluster of galaxies, and

\noindent{\bf (iii)} the temperature distribution of hot gas in galaxies and clusters of galaxies.

\noindent{\bf (iv)} As we have seen, DM also plays a central role in structure formation and galaxy evolution, and has measurable effects on the anisotropy of the cosmic microwave background, especially the third peak in the anisotropy spectrum (c.f. fig.~\ref{wmapgaussian}).
\begin{figure}[ht]
\centering
\includegraphics[width=6.5cm]{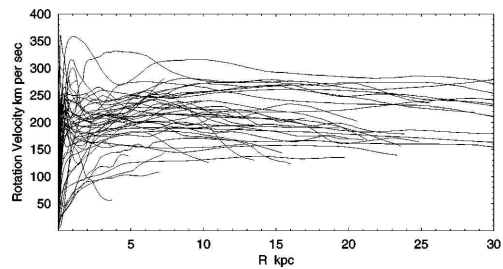}
\caption{Collage of Rotational Curves of nearby spiral galaxies obtained by combining Doppler data from CO molecular lines for the central regions, optical lines for the disks, and HI 21 cm line for the outer (gas) disks.   Graph from Y. Sophue and V. Rubin (Annual Review of Astronomy and Astrophysics, Volume 31 (c)2001, 127).}
 \label{fig:rc}
 \end{figure}

\begin{figure}[ht]
\begin{center}
\includegraphics[width=4.5cm]{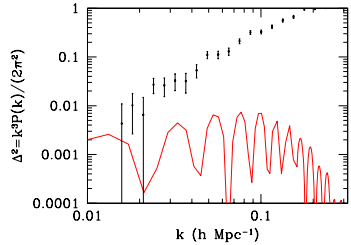}
\end{center}
\caption{Power spectrum of matter fluctuations (red curve, with wiggles) in a theory without dark matter as compared to observations of the galaxy
power spectrum. }
\label{baryongs}
\end{figure}

Historically, the first evidence for DM came~\cite{zwicky} from
discrepancies concerning the Rotational Curves (RC) of Galaxies.
If all matter were luminous then the rotational speed of the galactic disc would fall with the (radial) distance $r$ from the center as  $v(r) \sim r^{-1/2}$  but observations show  that $v(r) \sim $~const, as seen clearly in  figure~\ref{fig:rc}, where the rotation velocity
in units of km s$^{-1}$ is plotted vs galactocentric radius $R$ in kiloparsecs (kpc); 1 kpc $\approx$ 3000 light years.   It is seen that the RCs are flat to well beyond the edges of the optical disks ($\sim 10$ kpc).
Further Evidence for DM is provided by the Matter  oscillation spectrum in galaxies, depicted in figure~\ref{baryongs}.
The observed spectrum does not have the pronounced wiggles predicted by a baryon-only model, but it also has significantly
higher power than does the model. In fact,  $\Delta^2=k^3P(k)/(2\pi^2)$~, which is a dimensionless measure of the clumping, never rises above one in
a baryon-only model, so we could not see any large structures (clusters, galaxies, people, etc.) in the universe in such a model~\cite{Tegmark2003}.

However, at this stage we should mention the alternatives to Dark Matter models, the MOND~\cite{milgrom}, and its Lorentz-violating TeVeS field theory version~\cite{TeVes}, which could also reproduce the rotational curves of galaxies, by assuming modified Newtonian dynamics at  galactic scales for small gravitational accelerations, smaller than a universal value
$\gamma < \gamma_0\sim (200 {\rm km\,sec}^{-1})^2/(10\,{\rm kpc})$.
MOND theories have been claimed to fit most of the rotational curves of galaxies (fig.~\ref{fig:rc}),
with few notable exceptions, though, e.g. the bullet cluster.
It should be mentioned that TeVeS models, due to their preferred-cosmic-frame features, are characterized by
``Aether''-Lorentz violating  isotropic vector fields
$A_\mu = (f(t), 0,0,0)$, $A_\mu A^\mu = -1$,
whose cosmic instabilities are also claimed~\cite{liguori} to reproduce the enhanced growth of perturbations observed in galaxies (c.f. fig.~\ref{baryongs}). In these lectures I will not discuss such models.
It should be noted at this point that such issues, namely whether there are dark matter particles or not, could be resolved in principle by particle physics searches at colliders or direct dark matter searches, which I will now come to.

\subsection{Types of DM and Candidates}

From nucleosynthesis constraints we can estimate today the baryonic energy density contribution to be of order: $\Omega_{baryons} = 0.045 \pm .01$,
and this in fact is the dominant form of ordinary matter in the Universe.
Thus, barring alternatives, $90\%$ of the alleged matter content of the Universe seems to be dominated by DM of unknown composition at present.
There are several dark matter candidates, which can be classified into
two large categories depending on their origin and properties:

\noindent{\bf (I)} {\it Astrophysical}:
(i) MAssive Compact Halo ObectS (MACHOS): Dwarf stars  and Planets (Baryonic Dark Matter) and Black Holes, (ii) Non-luminous Gas Clouds.

\noindent{\bf (II)} {\it Particles} (Non-Baryonic Dark Matter):
Weakly Interacting Massive Particles (WIMP), which might be
the best candidates for DM: should not have
electromagnetic or strong interactions. May have weak
and gravitational interactions. WIMPS might include axions, neutrinos
stable supersymmetric
partners {\it etc}.
If these WIMPS are thermal relics from the Big Bang then we can
calculate their relic abundance today and compare with CMB and
other astrophysical data.
Non-thermal relics may also exist in some cosmological models but will not be
the subject of our discussion in these lectures.

There is an alternative classification of DM,
depending on the energetics of the constituting particles:

\noindent{\bf (i)} {\it Hot Dark Matter (HDM)}:  form of dark matter which consists of particles that travel with ultra-relativistic velocities: e.g. neutrinos.

\noindent{\bf (ii)} {\it Cold Dark Matter (CDM)}: form of dark matter consisting of slowly moving particles, hence cold,
e.g. WIMPS (stable supersymmetric particles (e.g. neutralinos {\it etc.}) or MACHOS.

\noindent{\bf (iii)} {\it Warm Dark Matter (WDM)}: form of dark matter with properties between those of HDM and CDM. Examples include sterile neutrinos, light gravitinos in supergravity theories {\it etc.}

Particle physics and/or astrophysics should provide candidates for DM and also
explain the relic densities of the right order as that predicted by the data.
Currently, the most favorite SUSY candidate for non baryonic CDM  are neutralinos~\cite{Hagelin}
${\tilde \chi}$.
These particles could be a WIMP if they are stable, which is the case in models where they are
the Lightest SUSY Particles (LSP) (with typical masses $m_{\tilde \chi} > 35~GeV$). Most of supersymmetric model constraints come from the requirement that a neutralino is the dominant astrophysical DM, whose
relic abundance can explain the missing Universe mass problem.
 I mention at this stage that direct searches for ${\tilde \chi}$ involve, among others, the recoil of nucleons
during their interaction with  ${\tilde \chi}$ in cryogenic materials.
In these lectures we shall concentrate mainly on colliders DM searches.
I refer the reader to ref.~\cite{zacek} for direct DM searches and other pertinent terrestrial and extraterrestrial experiments.

\subsection{WIMP DM: thermal properties and relic densities}

In all the searches we shall deal with in the present work, which are also the most commonly studied in the literature, one makes the standard assumption that the dark matter particle,
$\chi$, is a {\it thermal relic} of the Big Bang:
when the early Universe was dense and hot, with temperature $T\gg m_\chi$,
$\chi$  was in {\it thermal equilibrium};
annihilation of  $\chi$ and its antiparticle $\bar\chi$ into lighter particles, $\chi\bar\chi\to l\bar l$,
and the inverse process $l\bar l\to \chi\bar\chi$
proceeded with equal rates~\cite{kolbturner}.
As the Universe expanded and cooled down to a temperature $T<m_\chi$,
the number density of $\chi$ dropped exponentially, $n_\chi\sim e^{-m_\chi/T}$. Eventually, the temperature became too low for the annihilation to keep up with the expansion rate and the species $\chi$ `froze out'  with
the cosmological {\it abundance} (``relic'')  observed today.

\begin{figure}[t]
\centering \epsfig{file=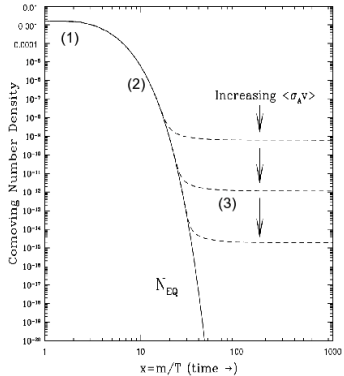, width=0.4\textwidth} \caption{The full
line is the equilibrium abundance; the dashed lines
 are the actual abundance after freeze-out. As the annihilation cross
section $\sigva$ is increased, the WIMP stays in equilibrium longer,
leading to a smaller relic density (from [12]).}
\label{fig:abun}
\end{figure}

The time evolution of the number density $n_\chi (t)$ is determined by the Boltzmann equation~\cite{kolbturner},
{ \begin{gather}\label{boltzmann}{
{\rm d}n_\chi/{\rm d}t + 3Hn_\chi =
  - \sigva \, [(n_\chi)^2-(n_\chi^{\rm eq})^2]\,},\end{gather}}
 where $H$ is the Hubble expansion rate,
$n_\chi^{\rm eq}$  the equilibrium number density and
$\sigva $ is the thermally averaged annihilation cross section
summed over all contributing channels.
It turns out that the relic abundance today is inversely
proportional to the thermally averaged annihilation cross section,  $\Omega_\chi h^2\sim 1/\sigva$. The situation is depicted in fig.~\ref{fig:abun}. When the properties and interactions of the WIMP are known, its thermal relic abundance can hence be computed from particle physics' principles and compared with cosmological data.

\subsection{Hot and Warm DM Excluded by WMAP}

The WMAP/CMB results on the cosmological parameters
discussed previously
disfavor strongly Hot Dark Matter (neutrinos), as a result of the
new determination of the upper bound on neutrino masses.
The contribution of neutrinos to the energy density of the
Universe depends upon the sum of the mass of the light neutrino
species~\cite{kolbturner,spergel}:
\begin{equation}
\Omega_\nu h^2 = \frac{\sum_{i} m_i }{94.0~{\rm eV}}
\label{neutrinodens}
\end{equation}
where the sum includes neutrino species
that are light enough to decouple while still relativistic.

The combined results from WMAP and other experiments~\cite{spergel}
on the cumulative likelihood of data as a function of the energy
density in neutrinos lead to $\Omega_\nu h^2 < 0.0067$ (at 95\% confidence limit).
Adding the Lyman $\alpha$ data, the limit weakens slightly~\cite{spergel}:
$\Omega_\nu h^2 < 0.0076 $
or equivalently (from (\ref{neutrinodens})):
$\sum_i m_{\nu_i} <0.69$ eV, where, we repeat again, the sum includes
light species of neutrinos.
This may then imply
an average upper limit on electron neutrino mass
$<m_\nu>_e < 0.23$ eV. These
upper bounds
strongly disfavors
Hot Dark Matter scenarios.

Caution should be exercised, however, when interpreting the above WMAP result.
There is the underlying-theoretical-model dependence of these results, which stems from the assumption of an Einstein-FRW Cosmology, characterized by
local Lorentz
invariance.
If Lorentz symmetry is violated, as, for instance, is the case of the TeVeS models alternative to DM, then neutrinos with (rest) masses
of up to 2 eV  could have
an abundance of $\Omega_\nu \sim 0.15$  in order to reproduce
the peaks in the observed CMB spectrum (fig.~\ref{wmapgaussian})~\cite{skordis} and thus being phenomenologically acceptable, at least from the CMB measurements viewpoint.

At this juncture we note that another important result of WMAP is the evidence for
early re-ionization of the Universe at redshifts $z \simeq 20$.
If one assumes that structure formation is responsible for
re-ionization, then such early re-ionization periods
are compatible only for high values of the masses $m_X$ of {\it
Warm Dark Matter }. Specifically, one can exclude models
with $m_X \le 10$ KeV
based on numerical simulations of structure formation for such models~\cite{warm}.
Such simulations imply that dominant structure
formation responsible for re-ionization, for Warm Dark Matter candidates
with $m_X \le 10$ KeV, occurs at much smaller $z$ than those observed
by WMAP. In view of this, one can therefore exclude popular
particle physics models
employing
light gravitinos ($m_X \simeq 0.5 $ KeV) as the Warm Dark Matter candidate.
It should be noted at this stage that such structure formation arguments
can only place a lower bound on the mass of the Warm Dark Matter candidate.
The reader should bear in mind that
Warm Dark Matter with masses $m_X \ge 100$ KeV becomes indistinguishable
from Cold Dark Matter, as far as structure formation is concerned.

\subsection{Cold DM in Supersymmetric Models: Neutralino}

After the exclusion of Hot and Warm Dark Matter, the only type of
Dark matter that remains consistent with the recent WMAP results~\cite{spergel}
is the {\it Cold Dark Matter }, which in general may consist of
axions, superheavy particles (with masses
$\sim 10^{14 \pm 5}$ GeV)~\cite{kryptons,chung}
and stable supersymmetric partners.
Indeed, one of the major and rather
unexpected predictions of Supersymmetry (SUSY), broken
at low energies $M_{SUSY} \thickapprox \mathcal{O}(1 \TeV) $,
while $R$-parity is conserved, is the existence of a stable, neutral
particle, the lightest neutralino  ($\lsp$),
referred to as the lightest supersymmetric particle (LSP)~\cite{Hagelin}.
Such particle is an ideal candidate
for the Cold Dark Matter in the Universe \cite{Hagelin}.
Such a prediction fits well with the fact that
SUSY is not only  indispensable in
constructing consistent string
theories, but
it also seems unavoidable at low energies ($\sim 1 \TeV$)
if the gauge hierarchy problem is to
be resolved.
Such a resolution provides a measure of the SUSY
breaking scale $M_{SUSY} \thickapprox \mathcal{O}(1 \TeV) $.

This type of Cold Dark Matter will be our focus from now on,
in association with the recent results from WMAP3 on relic
densities~\cite{wmap,spergel}. The WMAP3 results, combined with other
existing data, yield  for the
baryon and matter densities (including dark matter) at 2$\sigma$-level:
$\Omega_m h^2 = 0.1268^{+0.0072}_{-0.0095}~ {\rm (matter)}~,
100\Omega_b h^2 = 2.233^{+0.072}_{-0.091}~ {\rm (baryons)}~.$
One assumes that CDM is given by the difference of these two.
As mentioned already, in supersymmetric (SUSY) theories the
favorite candidate for CDM is
the lightest of the Neutralinos $\lsp$
(SUSY CDM), which is stable, as being the Lightest SUSY particle (LSP)
(There are cases where the stau or the sneutrino can be
the lightest supersymmetric particles. These cases are not favored~\cite{nickref}  and hence are not considered).
From the WMAP3 results~\cite{wmap}, then,  assuming $\Omega_{CDM}
\simeq \Omega_\chi$,  we
can infer stringent limits for the neutralino $\chi$ relic density:
\begin{equation}
0.0950 < \Omega_{\chi} h^2 < 0.1117~,
\label{relicdensity}
\end{equation}
It is important to notice that in this
inequality  only the upper limit is {\it rigorous}.
The lower Limit is {\it optional}, given
that there might (and probably do) exist other contributions
to the overall (dark) matter density.
It is imperative to notice that all the constraints we shall discuss
in this review are {\it highly model dependent}. The results on the minimal SUSY extensions of the standard model~\cite{lmnreview}, for instance, cannot apply to other models such as superstring-inspired ones, including non equilibrium cosmologies, which we shall also discuss here.
However, formally at least,
most of the analysis can be extrapolated to such models, with possibly different results, provided the SUSY dark matter in such models is thermal.
Before moving into such a discussion we consider it as instructive to
describe briefly various important properties of the Neutralino DM.

The Neutralino is a superposition of SUSY partner states. Its mass matrix in bino--wino--higgsinos basis
$\psi_j^0=(-i\lambda ',\,-i\lambda^3,\,\psi_{H_1}^0,\,\psi_{H_2}^0)$ is given by
\begin{equation}
  {\mathcal M}_N =
  \left( \begin{array}{cccc}
  M_1 & 0 & -m_Z s_W c_\beta  & m_Z s_W s_\beta \\
  0 & M_2 &  m_Z c_W c_\beta  & -m_Z c_W s_\beta  \\
  -m_Z s_W c_\beta & m_Z c_W c_\beta   & 0 & -\mu \\
   m_Z s_W s_\beta & - m_Z c_W s_\beta & -\mu & 0
  \end{array}\right)\nonumber
\label{eq:ntmassmat}
\end{equation}
where $M_{1}$, $M_{2}$: the U(1) and SU(2) gaugino masses,
$\mu$: higgsino mass parameter,
$s_W=\sin\theta_W$, $c_W=\cos\theta_W$, $s_\beta=\sin\beta$,
$c_\beta=\cos\beta$ and $\tan\beta = v_2/v_1$
($v_{1,2}$ v.e.v. of Higgs fields $H_{1,2}$).
The mass matrix is diagonalized by a unitary mixing matrix $N$,
$N^*{{\mathcal M}_N} N^\dagger =
  {\rm diag}(\mnt{1},\,\mnt{2},\,\mnt{3},\,\mnt{4})\,,$
where $\mnt{i}$, $i=1,...,4$, are the (non-negative) masses
of the physical neutralino states with $\mnt{1}<...<\mnt{4}$.
The lightest neutralino is then:
\begin{equation}{
  \nt_1= N_{11}\tilde{B}+ N_{12} \tilde{W} +N_{13}\tilde{H_1}+
  N_{14}\tilde{H_2} \,. \nonumber }
\end{equation}

To calculate relic densities it is assumed that the initial number
density of neutralinos $\chi$ particle in the Early Universe has been in thermal equilibrium:
interactions creating $\chi$ usually happen as frequently as the reverse
interactions which destroy them. Eventually the temperature of the expanding Universe drops to the order of the neutralino (rest) mass $T\simeq m_{\chi}$. In such a situation,
most particles no longer have
sufficient energy to create neutralinos. Now neutralinos
can only
{\it annihilate}, until their rate becomes smaller than the
Hubble expansion rate, $H\geq \Gamma_{\rm ann}$.
Then, Neutralinos are being separated apart from
each other too quickly to maintain equilibrium, and thus they reach their freeze-out temperature,  $T_{F}\simeq
m_{\chi}/20$, which characterizes  this type of cold dark matter.

\begin{figure}
  \centering
\epsfig{file=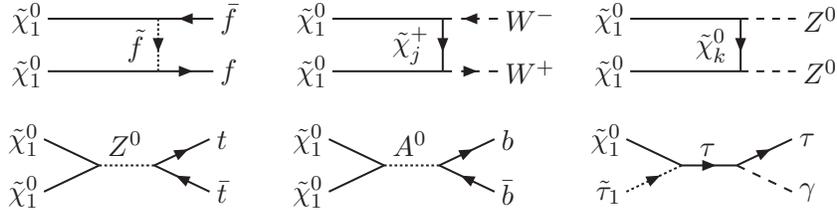,width=1\linewidth,clip=}
\caption{Basic Neutralino annihilations including stau co-annihilations in MSSM (from S. Kraml, Pramana 67, 597 (2006) [hep-ph/0607270]).}
\label{fig:co-an}\end{figure}

In most neutralino relic density calculations, the only interaction cross
sections that need to be calculated are annihilations of the type
$\chi \chi \rightarrow X$ where $\chi$ is the lightest neutralino and $X$ is any final state involving
only Standard Model particles.
However, there are scenarios in
which other particles  in the thermal bath have important effects
on the evolution of the neutralino relic density.   Such a
particle  annihilates with the neutralino into Standard Model particles and
is called a {\it co-annihilator} (c.f. figure \ref{fig:co-an}).
In order for a particle to be an {\it Effective co-annihilator},
it must have direct
interactions with the neutralino and must be {\it nearly
degenerate} in mass: Such degeneracy happens in the Minimal Supersymmetric Standard Model (MSSM), for
instance, with possible co-annihilators being the lightest
stau, the lightest stop, the
second-to-lightest neutralino or the lightest chargino.
When this
degeneracy occurs, the neutralino and all relevant co-annihilators form a
coupled system.

Without co-annihilations the evolution of a relic particle
number density, $n$, is governed, as mentioned previously, by a single-species
Boltzmann equation (\ref{boltzmann}).
It should be noted that the relic-particle
number density is modified by the Hubble expansion and
by direct and inverse annihilations of the relic particle.  The Relic particle is assumed stable, so relic
decay is neglected. Also commonly  assumed
is time-reversal (T) invariance, which relates annihilation and inverse annihilation processes.
In the presence of co-annihilators the Boltzmann equation
gets more complicated but it can be simplified using stability
properties of relic particle and co-annihilators (using
$n=\sum_{i=1}^{N} n_{i}$):
\begin{eqnarray}\label{bolcoan}
\frac{d n}{dt}=&-&3 H n -\sum_{i,j=1}^{N} \langle\sigma_{ij}
v_{ij}\rangle \left(n_{i} n_{j} - n_{i}^{\rm eq} n_{j}^{\rm
eq}\right)~.
\end{eqnarray}
To a very good approximation, one can use an effective single species
Boltzmann equation for this case if
$ \langle \sigma v\rangle = \sum_{i,j} \langle \sigma_{ij} v_{ij}
\rangle \frac{n_{i}^{\rm eq}}{n^{\rm eq}}\frac{n_{j}^{\rm
eq}}{n^{\rm eq}}.$

The Boltzmann equation (\ref{bolcoan}) can be solved numerically, but in most cases even analytically.
Details on how to solve the Boltzmann equation are given abundantly in the cosmology literature~\cite{kolbturner} and will not be repeated here.
We shall only outline the most important results that will be essential for our discussion in these lectures.
One should determine the freeze-out temperature $x_{F} =
m_{\chi}/T_{F}$: $
x_{F} = \ln \left(\frac{0.038\, g\, m_{\PL} \,m_{\chi} \,\langle
\sigma v \rangle }{\sqrt{g_{*} \,x_{F}}}\right)$,
with $m_{\PL}$ the Planck mass, $g$ the total number of
degrees of freedom of the $\chi$ particle (spin, color, etc.),
$g_{*}$ the total number of effective relativistic degrees of
freedom at freeze-out, and the thermally averaged cross section is
evaluated at the freeze-out temperature.  For most CDM candidates, $x_{F}\simeq 20$. The total
(co)annihilation depletion of neutralino number density
is calculated by integrating the thermally averaged cross section
from freeze-out to the present temperature:
\begin{eqnarray}
&&\Omega_{\chi} {\rm h}^2 = 40
\sqrt{\frac{\pi}{5}}\frac{h^2}{H_{0}^2}\frac{s_{0}}{m_{Pl}^{3}}
\frac{1}{\left(g_{*S}/g_{*}^{1/2}\right) \, J\left(x_{F}\right)}
= \frac{1.07 \times 10^9 \,
{\rm GeV}^{-1}}{g_{*}^{1/2} \, m_{\PL} \, J\left(x_{F}\right)}~,\nonumber \\
&& J\left(x_{F}\right)
= \int_{x_{F}}^{\infty} \langle\sigma v \rangle \, x^{-2} dx~.\label{standardbol}
\end{eqnarray}
where $s_0$ is the entropy density,
$g_{*S}$ denotes the number of effective relativistic d.o.f.
contributing to the (constant) entropy of the universe and $h$ is the reduced Hubble parameter: $H_0 = 100\, h\, {\rm \, km \, sec^{-1} \,
Mpc^{-1}}$. This is the expression one compares with the experimental determination of the DM abundance via, e.g., WMAP data.
It should be noted at this stage that the theoretical
assumptions leading to the above results may not hold in general for all DM models and candidates: the
missing non-baryonic
matter in the universe may only partially, or not at all, consist of relic neutralinos.
Also, as we shall discuss later on the article,
in some off-shell,
non-equilibrium relaxation stringy models of dark energy, the Boltzmann equation gets modified by off-shell, non-equilibrium terms as well as time-dependent dilaton-source terms. This leads to important modifications on the associated particle-physics models constraints.

\subsection{Model-Independent DM Searches in Colliders}

\begin{figure}[t]
\begin{center}
  \epsfig{file=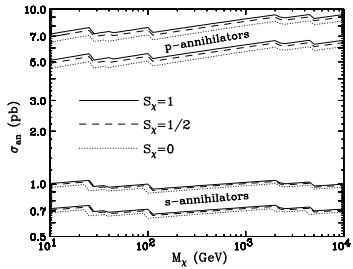,width=0.4\textwidth}
\end{center}
 \caption {Values of the quantity $\sigma_{an}$ allowed at $2\sigma$ level as a function of the DM mass.}
\label{fig:sandp}
\end{figure}

As we have discussed above, if dark matter comes from a thermal relic, its density
is determined, to a large extent,
by the dark matter annihilation cross section: $ \sigma \left( \chi \, \chi \rightarrow SM \, SM \right)$.  Indeed, as already mentioned, the present-day dark matter abundance is roughly inversely proportional to the thermally averaged annihilation cross section times velocity,
$\Omega_{\chi} h^2 \propto 1/\langle \sigma v \rangle$.
 This latter quantity can be conveniently expanded in powers of the relative dark matter particle velocity:
 \begin{equation}\label{velexpan}
 \sigma v = \sum_{J} \sigma_{an}^{\left(J\right)} v^{\left(2J\right)}~.
   \end{equation}
   Usually, only the lowest order non-negligible power of $v$ dominates.  For $J=0$, such dark matter particles are called $s$-annihilators, and for $J=1$, they are called $p$-annihilators; powers of $J$ larger than $1$ are rarely needed.:

Figure~\ref{fig:sandp} shows the constraint on the annihilation cross section as a function of dark matter mass that results from Eq.~(\ref{relicdensity})~\cite{birkendal}.
The lower (upper) band of fig.~\ref{fig:sandp} is for models where $s$-wave ($p$-wave) annihilation dominates.
It is important to notice~\cite{birkendal} that the total annihilation cross section $\sigma_{\rm an}$ is
  virtually insensitive to dark-matter mass.
  This latter effect
is due to the changing number of degrees of freedom at the time of freeze-out as the dark matter mass changes.
It also points to cross sections expected from weak-scale interactions (around $0.8\, pb$ for $s$-annihilators and $6\, pb$ for $p$-annihilators), hence
  implying the possibility that DM is connected to an explanation for the weak scale and thus WIMPs~\cite{birkendal}.  Such WIMPs exist not only in supersymmetric theories, of course, but in a plethora of other models such as theories involving extra dimensions and 'little Higgs' models.
The LHC and the ILC are specifically designed to probe the origin of the weak scale, so dark matter searches and future collider physics appear to be closely related.
\begin{figure}[t]
\begin{center}
  \epsfig{file=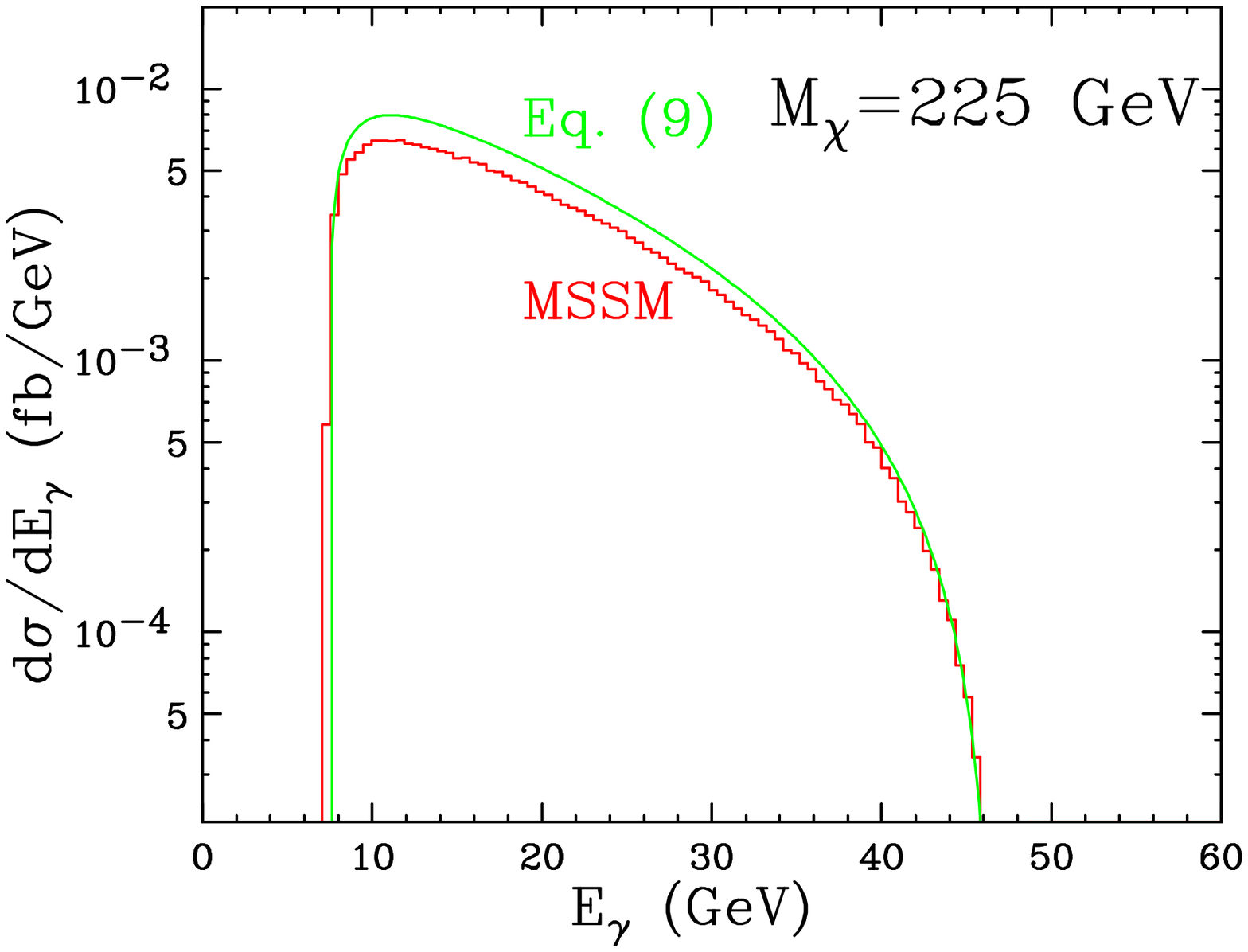,width=0.4\textwidth} \hspace{2cm}
\epsfig{file=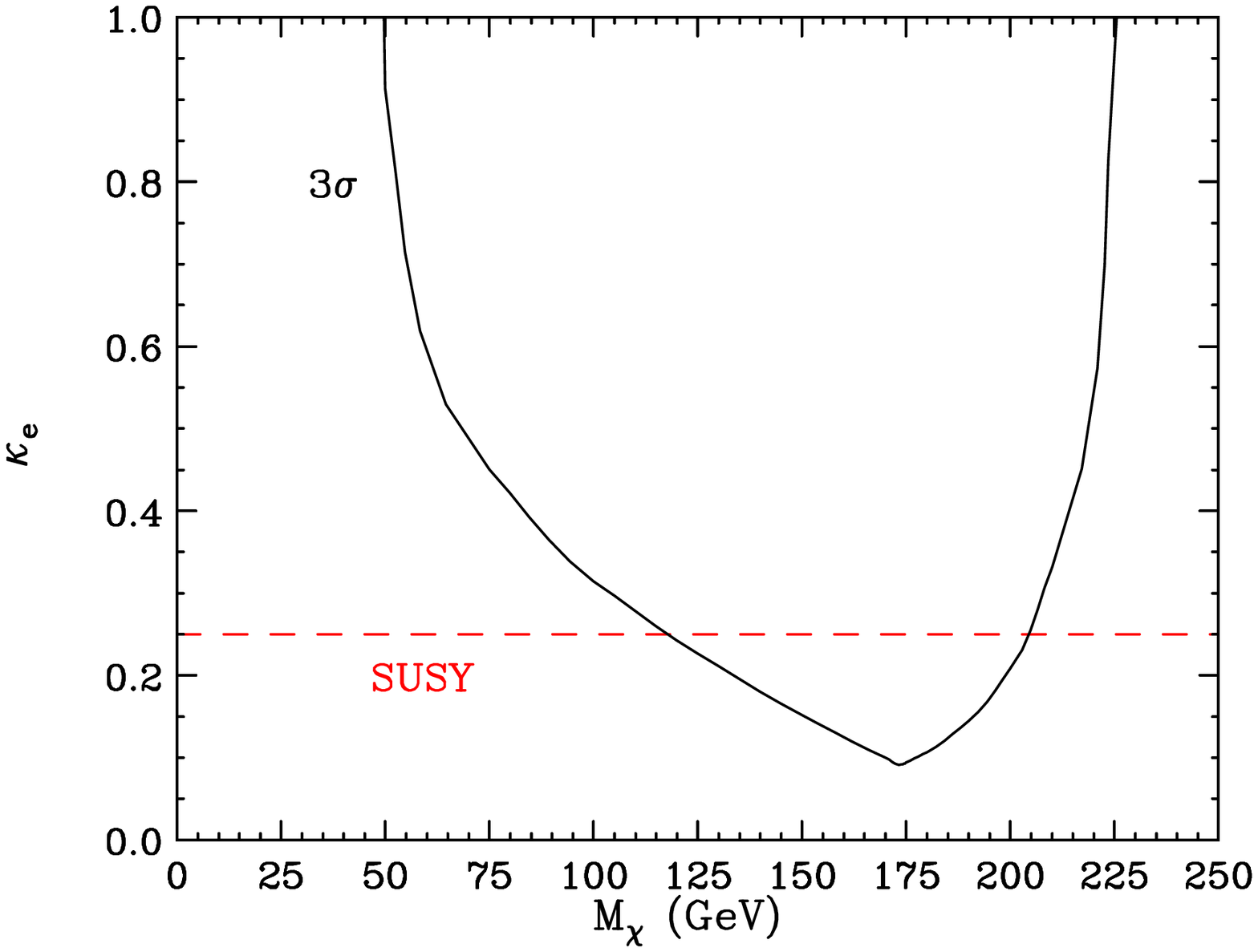,width=0.4\textwidth}
 \end{center}
\caption{ {\it Left panel:}  Comparison between the photon spectra from the process $e^+e^-\to2\chi_1^0+\gamma$ in the explicit supersymmetric models defined in A.~Birkedal, K.~Matchev and M.~Perelstein,
  Phys.\ Rev.\ D {\bf 70}, 077701 (2004)
 (red/dark-gray) and the predicted spectra for a
$p$-annihilator of the corresponding mass and $\kappa_e$
(green/light-gray).  {\it Right panel:}  The reach of a 500 GeV unpolarized electron-positron collider with
an integrated luminosity of 500 fb$^{-1}$ for the discovery of
$p$-annihilator WIMPs, as a function of the WIMP mass $M_\chi$ and the
$e^+e^-$ annihilation fraction $\kappa_e$. The 3 $\sigma$ (black) contour is shown, along with an indication of values one might expect from supersymmetric models (red dashed line, labelled 'SUSY'). Only statistical uncertainty is included.}
\label{carrots}
\end{figure}
The next question one could ask is whether the above cross section could be turned, within a WIMP working hypothesis framework, into a model-independent signature at colliders. This question was answered in the affirmative in \cite{birkendal}. One introduces the parameter
 $ \qquad \kappa_e \equiv \sigma (\chi \chi \to e^+e^-)/\sigma (\chi \chi \to SM|SM) $
 which relates dark matter annihilation processes to cross sections involving
 $e^+e^-$ in the final state.
Using crossing symmetries to relate $\sigma (\chi \chi \to e^+e^-)$ to
$\sigma (e^+e^- \to \chi \chi )$ and  co-linear factorization
one can relate $\sigma (e^+e^- \to \chi \chi )$ to
$\sigma (e^+e^- \to \chi \chi \gamma)$,  thus connecting
astrophysical data on $\sigma_{\rm an}$ to the process  $e^+e^- \to \chi \chi \gamma $. The resulting differential cross section reads~\cite{birkendal}
\begin{eqnarray}\label{crossect}
&& \frac{d\sigma}{dx d{\rm cos}\theta}(e^+e^- \to 2\chi + \gamma) \simeq \nonumber \\
&& \frac{\alpha ~\kappa_e \sigma_{\rm an}}{16\pi}
\frac{1 + (1 - x)^2}{x}\frac{1}{{\rm sin}^2\theta} 2^{J_0}(2S_\chi + 1)^2\left(1 - \frac{4M_\chi ^2}{(1 - x)s}\right)^{\frac{1}{2} + J_0}
\end{eqnarray}
with $\alpha$ the appropriate fine structure constant, $x=2E_\gamma/\sqrt{s}$, $\theta$ angle between photon and incoming electron,
$S_\chi $ spin of WIMP, $J_0$ is the dominant value of $J$ in the velocity expansion of (\ref{velexpan}) (as discussed above, commonly $J=0$ dominates, $s$-annhilator DM).
The accuracy of the method and its predictions are illustrated in fig.~\ref{carrots}, where the left panel illustrates the results obtained using the formula (\ref{crossect}), which are then compared with those of an exact calculation, based on a supersymmetric MSSM model, with WIMP masses 225 GeV, whilst  the right panel shows the expected reach in $\kappa_e$ for a $500$ GeV linear $e^+ e^-$ collider as a function of the WIMP mass. As we observe from such comparisons the results of the method and of the exact calculation are in pretty good agreement.

We note at this stage, however, that, although model independent, the above process is rarely the dominant collider signature of new physics within a given model.
It therefore makes sense to look for model dependent processes at colliders, which we now turn to.  In this last respect, it is important to realize~\cite{birkendal} that a calculation of slepton masses is essential for computing accurately relic abundances in theoretical models;  without a collider measurement of the slepton mass, there may be a significant uncertainty
in the relic abundance calculation.  This uncertainty results because the slepton mass should then be allowed to vary within
the whole experimentally allowed range.
We mention here that measuring slepton masses at LHC is challenging
due to
$W^+ W^-$ and $t \bar{t}$ production. However, as shown in \cite{birkendal},
it is possible through  the study of di-lepton mass distribution $m_{\ell \ell}$ in  the decay channel ${\tilde \chi_2 ^0} \to \ell^\pm\ell^\mp{\tilde \chi_1^0}$ and also at the International Linear Collider (ILC). The reader is referred to the literature~\cite{birkendal} for further details on these important issues. We are now ready to start our discussion on model-dependent DM signatures at LHC and future colliders.

\section{Model-Dependent WMAP SUSY Constraints }

We shall concentrate on DM signatures at colliders, using WMAP1,3 data.
To illustrate the underlying-theoretical-model dependence of the results we chose three representative theoretical models: (i) the mSUGRA (or constrained MSSM model)~\cite{msugra,lmnreview}, (ii) a heterotic string model with orbifold compactification~\cite{heteropheno}, and (iii) a non-critical (non-equilibrium) stringy cosmology (Q-cosmology) with running dilatons (implying a dilaton quintessence relaxation model for dark energy at late eras) and off-shell terms~\cite{qcosmol}.

\subsection{Constrained MSSM/mSUGRA Model}

MSSM has too many parameters to be constrained effectively by data.
To minimize the number of parameters one can ``embed'' this model by taking into account the gravity sector, which from a cosmological point of view is a physical necessity. Such an embedding in principle affects the dark energy sector of the cosmology, and in fact the minimal Supergravity model (mSUGRA)~\cite{msugra}, used to yield the Constrained MSSM (CMSSM)  predicts too large values of the cosmological constant at a quantum level, and hence it should not be viewed as the physical model. Nevertheless, as far as DM searches are concerned, such models give a pretty good idea of how astrophysical data can be used to constrain particle physics models, and this is the point of view we take in this work. mSUGRA is the best studied model so far as far as constraints on
supersymmetric models using astrophysical CMB data are concerned.
A relatively recent review on such approaches is given in \cite{lmnreview}, where we refer the reader for details and further material and references. In our presentation here we shall be very brief and concentrate only on the basic conclusions of such analyses.

\subsubsection{Basic Features: geometry of the parameter space}

Before embarking into a detailed analysis of the constraints
of the minimal supersymmetric standard model
embedded in a minimal
supergravity model (CMSSM)~\cite{msugra}, we consider it
useful to
outline the basic features of these models, which will be used
in this review.
The embedding of SUSY models into the minimal supergravity
(mSUGRA) model implies that there are five independent parameters:
Three of them, the scalar and gaugino masses $m_0, m_{1/2}$ as well as the trilinear soft coupling
$A_0=$, at the unification scale, set the size of the Supersymmetry breaking scale. In addition one can consider as input parameter ${\rm tan}\beta =\frac{<H_2>}{<H_1>}$, the ratio  of the v.e.v's of the Higgses $H_2$ and $H_1$ giving masses to up and down quarks respectively. The sign
( signature) of the Higgsino mixing parameter $\mu$ is also an input but not its size which is determined from the Higgs potential minimization condition~\cite{lmnreview}.
The parameter space of mSUGRA can be effectively described in terms of two branches:

\noindent{\bf (i)} \underline{An Ellipsoidal Branch (EB) of Radiative Symmetry Breaking}, which
exists for small to moderate values of tan$\beta \lsim 7$,
where the loop corrections are typically small. One finds that the
radiative symmetry breaking constraint demands that the allowed set of
soft parameters $m_0$ and a combination~\cite{lmnreview} $m_{12}' =f(m_{1/2}, A_0, {\rm tan}\beta )$ lie, for a given value of $\mu$,
on the surface of an {\it Ellipsoid}. This places
upper bounds on the sparticle masses for a given value of
$\Phi \equiv \mu^2/M_Z^2 + 1/4$.

\noindent{\bf (ii)} \underline{Hyperbolic Branch (HB) of Radiative Symmetry Breaking}. This branch is realized~\cite{chatto} for large values of tan$\beta \gsim 7$,
where the loop corrections to $\mu$ are significant.
In this branch,  $(m_0, m'_{1/2})$ lie now on the surface of a {\it hyperboloid}:
$\frac{{m_{1/2}'}^2}{\alpha^2(Q_0)} - \frac{{m_0}^2}{\beta^2(Q_0)}
\simeq \pm 1, \quad Q_0=0$ a fixed value of the running scale,
$\alpha, \beta $ constant functions of $\Phi, M_Z, A_0$.
For fixed $A_0$, the $m_0,m_{1/2}$ lie on a {\it hyperbola},
hence they can get large for fixed $\mu$ or $\Phi$.
What is interesting in the HB case is the fact that $m_0$ and/or  ${m}_{1/2}$ can become very large,  while much smaller values for $\mu$ can occur.

\noindent{\bf (iia)} A subset of HB is
the so-called {\it high zone}~\cite{chatto}. In this case electroweak symmetry breaking (EWSB) can occur in regions where $m_0$ and ${m}_{1/2}$ can be in the several TeV range, with much smaller values for the parameter $\mu$ which however is much larger than $M_Z$. This has important consequences for phenomenology, as we shall see. In this zone the lightest of the neutralinos, $\chi_1$, is almost a Higgsino having a mass of order $\mu$. This is called {\it{inversion phenomenon}} since the LSP is a Higgsino rather a Bino.
The inversion phenomenon has dramatic effects on the nature
of the particle spectrum and
SUSY phenomenology in this HB.
Indeed, as we discussed above, in mSUGRA one naturally has co-annihilation
with the sleptons when the neutralino mass extends to masses
beyond 150-200 GeV with processes of the type (c.f. fig.~\ref{fig:co-an}):
$\chi \tilde \ell_R^a \rightarrow \ell^a \gamma, \ell^a Z, \ell^a h$,
$\tilde \ell_R^a \tilde \ell_R^b \rightarrow \ell^a \ell^b$,
 and $\tilde \ell_R^a \tilde \ell_R^{b*} \rightarrow \ell^a\bar \ell^b,
\gamma \gamma, \gamma Z, ZZ, W^+W^-, hh$, where $\tilde{\it l}$ is
essentially a $\tilde \tau$.
Remarkably the relic density constraints can
be satisfied on the hyperbolic branch also by co-annihilation.
However, on the HB the co-annihilation is of
an entirely different nature as compared with the stau co-annihilations
discussed previously:
instead of a neutralino-stau co-annihilation, and stau - stau
in the HB one has co-annihilation processes
involving the second lightest neutralino and chargino states~\cite{gondolo},
$\chi_1^0-\chi_1^{\pm}$, followed by
$\chi_1^0-\chi_2^{0}$,$\chi_1^+-\chi_1^{-}$,$\chi_1^{\pm}-\chi_2^{0}~.$
Some of the dominant processes that contribute
to the above co-annihilation processes are~\cite{gondolo}:
$~\chi_1^0 \chi_1^{+}, \chi_2^0 \chi_1^{+}\rightarrow
u_i\bar d_i, \bar e_i\nu_i, AW^+,Z W^+,
W^+h $ and  $
\chi_1^{+} \chi_1^{-}, \chi_1^0 \chi_2^{0}\rightarrow
u_i\bar u_i, d_i \bar d_i,
W^+W^-~.$
Since the mass difference between the states
$\chi_1^+$ and $\chi_1^{0}$ is the smallest the $\chi_1^0 \chi_1^{+}$
co-annihilation dominates.
In such cases, the masses $m_0$ $m_{1/2}$
may be pushed beyond $10$ TeV, so that squarks and sleptons
can get masses up to several TeV, i.e.
beyond
detectability limits of immediate future accelerators such as LHC.

\noindent{\bf (iib)} Except the high zone where the inversion
phenomenon takes place the HB includes the so called {\it Focus Point}
(FP) region~\cite{focus},
which is defined as a region in which some renormalization group (RG)
trajectories
intersect (FP region would be only a point,
were it not for threshold effects which smear it out).
We stress that the FP is {\it not} a fixed point of the RG.
The FP region is a subset of the HB limited to relatively low values of ${m}_{1/2}$ and values of $\mu$ close to the electroweak scale, $M_Z$, while $m_0$ can be a few TeV but not as large as in the high zone due to the constraints imposed by the EWSB condition. The LSP neutralino in this region is a mixture of Bino and Higgsino and the Higgsino impurity  allows for rapid s-channel LSP annihilations, resulting to low neutralino relic densities at experimentally acceptable levels.
This region is
characterized by $m_0$ in the few TeV range, low values of $m_1{1/2}<< m_0$
and rather small values of $\mu$ close to
$M_Z$.
The LSP neutralino in this case is a mixture of Bino and Higgsino and its
Higgsino impurity is adequate to give rize to rapid s-channel LSP
annihilations so that the neutralino relic density is kept low at
experimentally acceptable values. Since $\mu$  is small the lightest
chargino may be lighter than 500 GeV and the FP region may be accessible to
future TeV scale colliders. Also due to the relative smallness of ${m}_{1/2}$
in this region gluino pair production may occur at a high rate making the
FP region accessible at LHC energies.

It should be pointed out that, although the HB may be  viewed as fine tuned,
nevertheless recent studies~\cite{baer}, based on a $\chi^2$ analysis, have
indicated that the WMAP data,
when combined with data on $b \to s\gamma$ and $g_\mu -2$,
seem to favor the Focus Point HB region and the large tan $\beta$
neutralino resonance annihilation of mSUGRA.

\subsubsection{Muon's anomaly and SUSY detection prospects}

Undoubtedly one of the most significant experimental results of the last
years is the measurement of the anomalous magnetic moment of the
muon~\cite{E821Expt}.
Deviation of its measured value from the Standard Model (SM) predictions is
evidence for new physics with Supersymmetry being the prominent candidate to
play that role.
Adopting Supersymmetry as the most natural extension of the SM, such
deviations may be explained and  impose at the same time severe constraints
on the predictions of the available SUSY models by putting upper bounds on
sparticle masses. Therefore knowledge of the value of $g_\mu - 2$  is of
paramount importance for Supersymmetry and in particular for the fate of
models including heavy sparticles in their mass spectrum, as for
instance those belonging to the Hyperbolic Branch.

Unfortunately the situation concerning the anomalous magnetic moment is
not clear as some theoretical uncertainties remain unsettled as yet.
Until last year, as far as I am aware of,
there were two theoretical estimates for the difference of the
experimentally measured~\cite{E821Expt} value of
$a_\mu = (g_\mu -2)/2$ from the theoretically
calculated one within the SM~\cite{narison},
\begin{itemize}

\item
{Estimate (I)} $\; \;a_\mu^{\rm exp} - a_\mu^{\rm SM} = 1.7(14.2) \times
10^{-10} \;  \quad [ 0.4(15.5) \times 10^{-10} ]  $

\item
{Estimate (II)} $a_\mu^{\rm exp} - a_\mu^{\rm SM} = 24.1(14.0) \times
10^{-10}  \quad [ 22.8(15.3) \times 10^{-10} ] $

\end{itemize}
In (I) the $\tau$-decay data are used in conjunction with Current Algebra
while in (II) the ${e^{-}}{e^{+}} \rightarrow$ Hadrons data are used in
order to extract the photon vacuum polarization which enters into the
calculation of $g_\mu - 2$.
Within square brackets are updated values of Ref.~\cite{narison}
{\footnote{Due to the rapid updates concerning $g_\mu - 2$ the values of
$a_\mu^{\rm exp} - a_\mu^{\rm SM}$ used in previous works quoted in this
article may differ from those appearing above.}}.
Estimate (I) is considered less reliable since it carries additional
systematic uncertainties and for this reason in many studies only the
Estimate (II) is adopted. Estimate (II) includes the contributions of
additional scalar mesons not taken into account in previous calculations.

In order to get an idea of how important the data on the muon anomaly might
be we quote Ref.\cite{chatto} where both estimates have been used.
If Estimate (II) is used at a 1.5$\sigma$ range
much of the HB  and all of the inversion region
can be eliminated. In that case the usually explored region of
SUSY in the EB is the only one that survives, which, as we shall discuss
below, can be severely constrained by means of the recent
WMAP data.
On the other hand, Estimate (I), essentially implies no difference
from the SM value,
and hence, if adopted, leaves
the HB, and hence its high zone (inversion
region), {\it intact}.  In such a case, SUSY may not be detectable
at colliders, at least in the context of the mSUGRA model, but may be
detectable in some direct dark matter searches, to which we shall turn to
later in the article.

For the above reasons, it is therefore {\it imperative}
to determine unambiguously the muon anomalous magnetic moment
$g_\mu -2$ by reducing the errors
in the leading order hadronic contribution, experimentally, and improving
the theoretical computations within the standard model.
In view of its importance for SUSY searches, it should also be necessary
to have further experiments in the future, that could provide
independent checks of the measured muon magnetic moment by the
E821 experiment~\cite{E821Expt}.
Quite recently (2006) a new measurement~\cite{eidelman} of the $g_\mu -2$
became available, which shows a clear discrepancy from the theoretically calculated Standard Model prediction by $3.4~\sigma$
\begin{equation}\label{gm2}
1.91 \times 10^{-9} < \Delta a_\mu  < 3.59 \times 10^{-9}
\end{equation}
thereby pointing towards the elimination
of the inversion region of the HB of mSUGRA,
according to the above discussion.

\subsubsection{WMAP mSUGRA Constraints in the EB}

\begin{figure}[htb]
\centering
\epsfig{file= 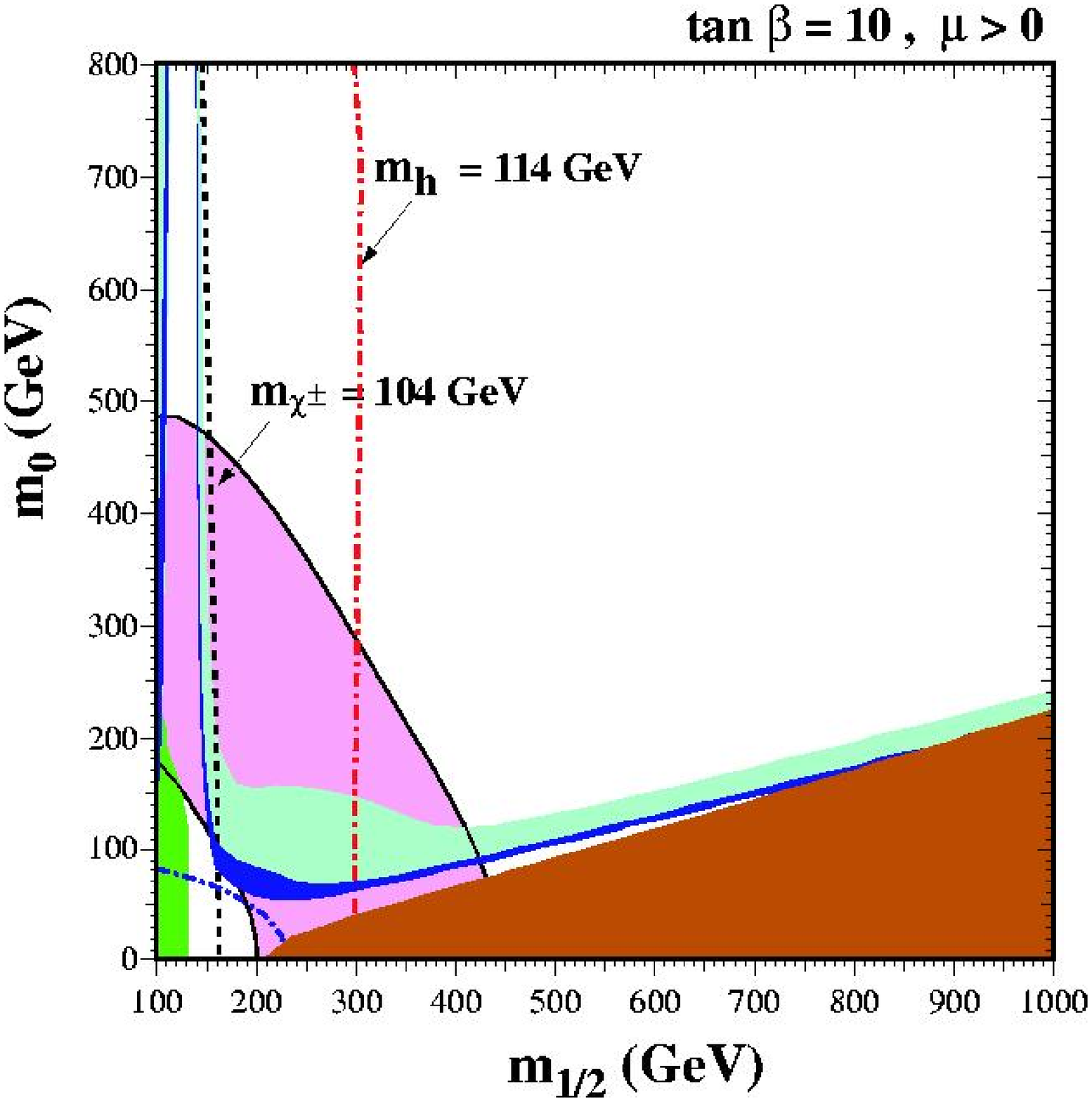,angle=0,width=0.35\linewidth}
\hfill \epsfig{file= 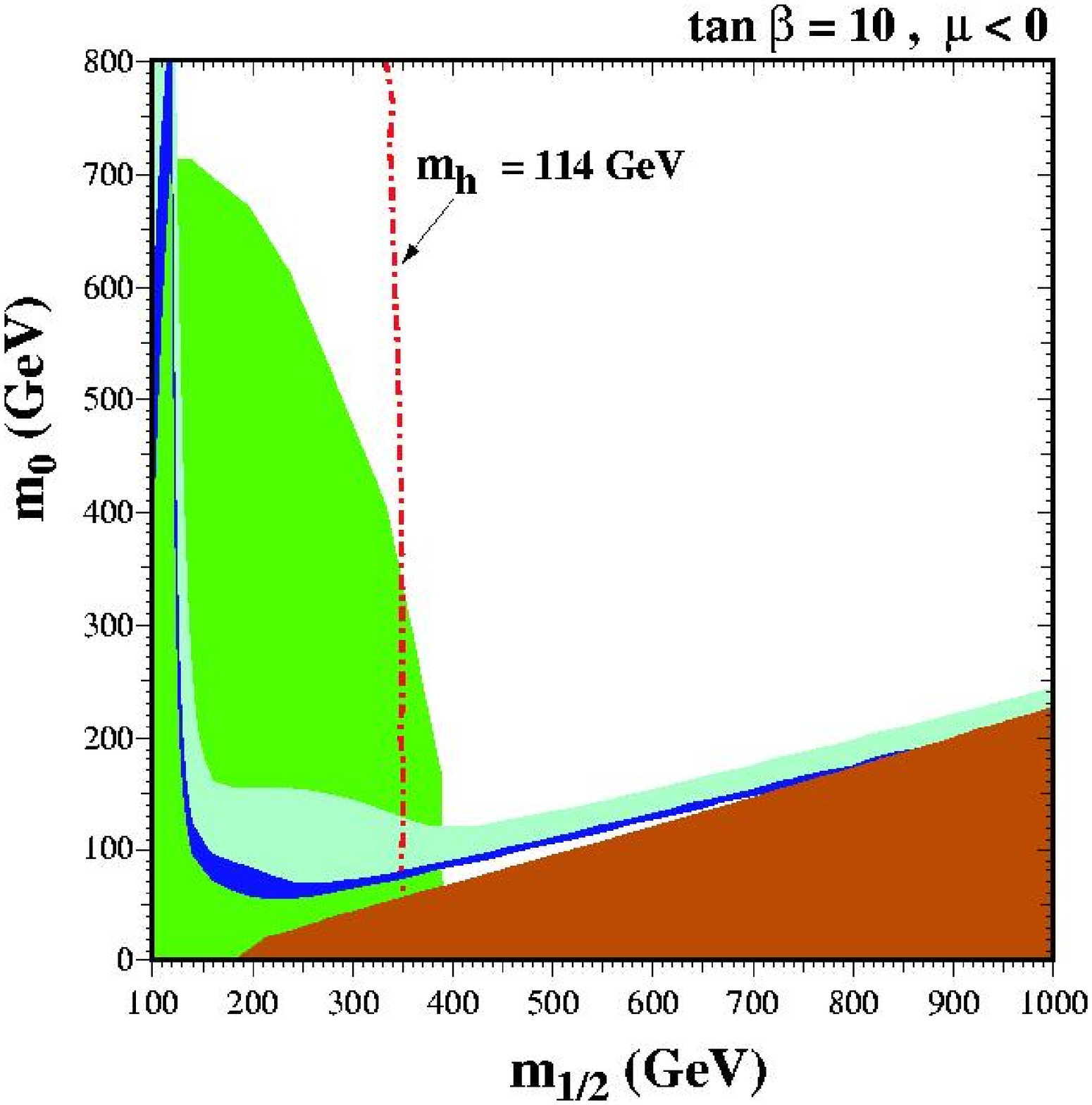,angle=0,width=0.35\linewidth}
\epsfig{file= 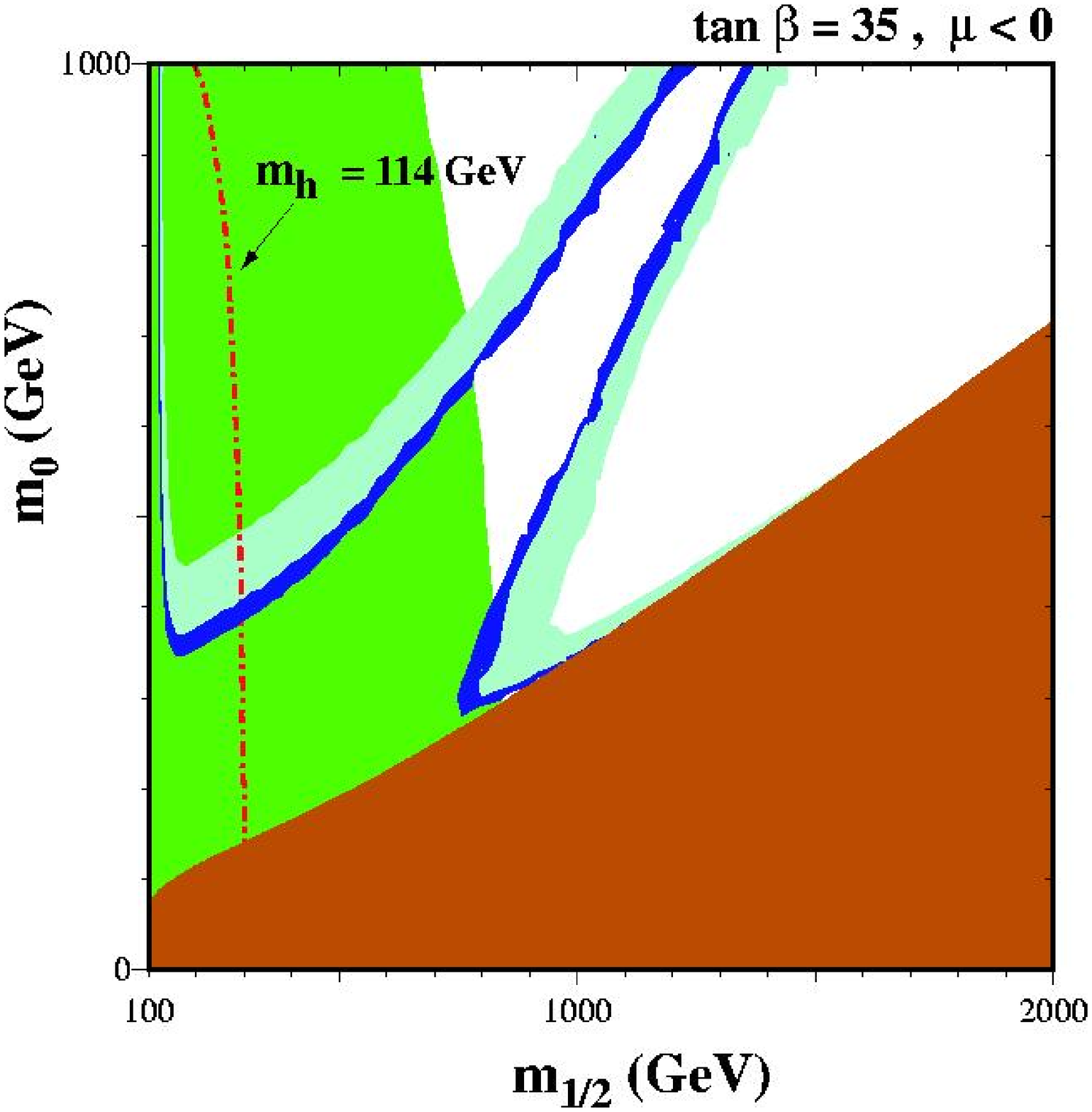,angle=0,width=0.35\linewidth}
\hfill \epsfig{file=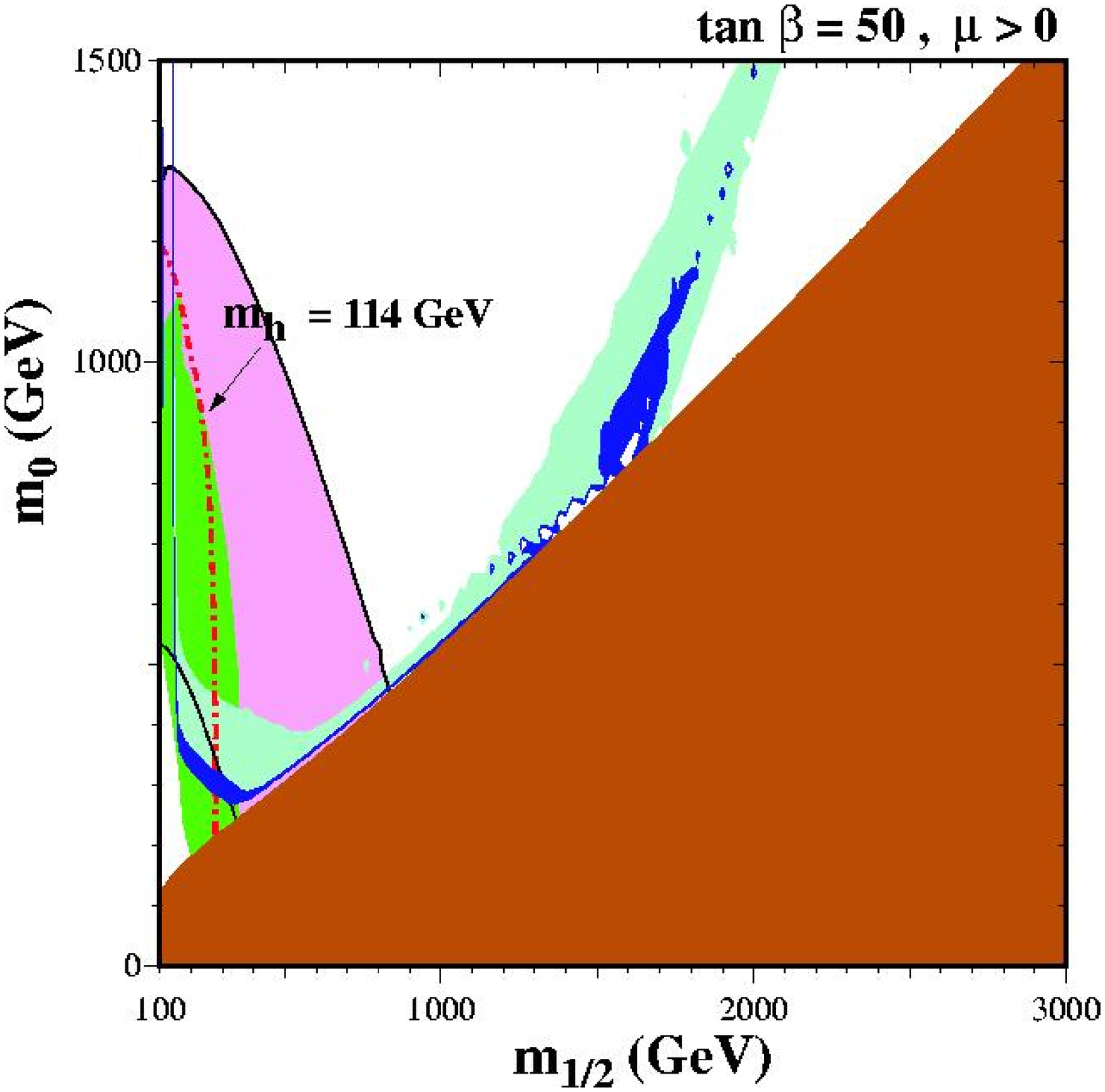,angle=0,width=0.35\linewidth}
\caption
{{\it mSUGRA/CMSSM constraints after WMAP from Ref. [38].
The Dark Blue shaded region is favored by WMAP1
( $0.094 \le \Omega_\chi h^2 \le 0.129$ ).
Turquoise shaded regions have $0.1 \le \Omega_\chi h^2 \le 0.3$.
Brick red shaded regions are excluded because LSP is charged.
Dark green regions are excluded by $b \to s\gamma $.
The Pink shaded region includes $2-\sigma$ effects of $g_\mu - 2$.
Finally, the dash-dotted line represents the
LEP constraint on ${\tilde e}$ mass.}}
\label{olivefig1}
\end{figure}

After the first year of running of WMAP,
there have been two independent groups working on this update of the
CMSSM in light of the WMAP data,
with similar results~\cite{olive,ln}
and below we summarize
the results of \cite{olive} in
fig.~\ref{olivefig1}
for some typical values of the
parameters ${\rm tan}\beta$ and signature of $\mu$. In such analyses one plots $m_0$ vs. $m_{1/2}$, taking into account the calculated relic abundance of neutralinos in the model and constraining it by means of the WMAP results (\ref{relicdensity}). Details are given in \cite{lmnreview} and references therein.

For the LSP, the lightest of the charginos, stops, staus and Higgses the upper
bounds on their masses of order of a few hundreds of GeV~\cite{lmnreview},
for various values
of the parameter $\; \tan \beta \;$, if the new WMAP determination~\cite{wmap,spergel} of the
Cold Dark Matter (\ref{relicdensity}) and the $2 \sigma$ bound
$\; 149 < 10^{-11}\;\alpha_{\mu}^{SUSY}<573\;$ of E821 is respected.
The lightest of the charginos has a mass
whose upper bound is $\; \approx 550 \; \mathrm{GeV} \;$,
and this is smaller than the upper bounds put on the masses of the
lightest of the other charged sparticles, namely the stau and stop.
Hence the prospects of discovering CMSSM
at a $e^{+} e^{-}$ collider with center of mass energy $\sqrt s = 800 \GeV$,
are  {\em not} guaranteed. Thus, a center of mass energy of at least
$\sqrt s \approx 1.1 \; \mathrm{TeV}\;$
is required to discover SUSY through chargino pair production.
Note that in the allowed regions
the next to the lightest neutralino, ${\tilde{\chi}^{\prime}}$, has a mass
very close to the lightest of the charginos and hence the process
$e^{+} e^{-} \goes {\tilde{\chi}} {\tilde{\chi}^{\prime}}$, with
${\tilde{\chi}^{\prime}}$ subsequently decaying to
$ {\tilde{\chi}} + {l^{+}} {l^{-}}$ or
$ {\tilde{\chi}}+\mathrm{2\,jets}$,
is kinematically allowed for such large $\tan \beta$, provided
the energy is increased to at least $\sqrt{s} = 860 \GeV$. It should
be noted however that this channel proceeds via the $t$-channel exchange
of a selectron and it is suppressed due to the heaviness of the exchanged
sfermion.
Therefore only if the center of mass energy is increased to
$\; \sqrt{s} = 1.1 \; \mathrm{TeV} \;$ supersymmetry can be
discovered in a $e^{+} e^{-}$
collider provided it is based on the Constrained scenario~\cite{ln}.

An important conclusion, therefore, which can be inferred by inspecting
the figures \ref{olivefig1} is
that the constraints implied by a possible discrepancy of $g_\mu - 2$
from the SM value, as seems to be supported by the 2006 data~\cite{eidelman} (\ref{gm2}),
( $\alpha_{\mu}^{SUSY} \gsim 15.0 \times 10^{-10}$ ),
when combined with the
WMAP restrictions on CDM (neutralino) relic densities (\ref{relicdensity}),
imply severe restrictions on the available parameter space of the EB
and lower significantly the upper bounds on the allowed neutralino masses
$\mlsp$.

\subsubsection{WMAP mSUGRA Constraints in the HB}

\begin{figure}[htb]
\centering
\epsfig{file=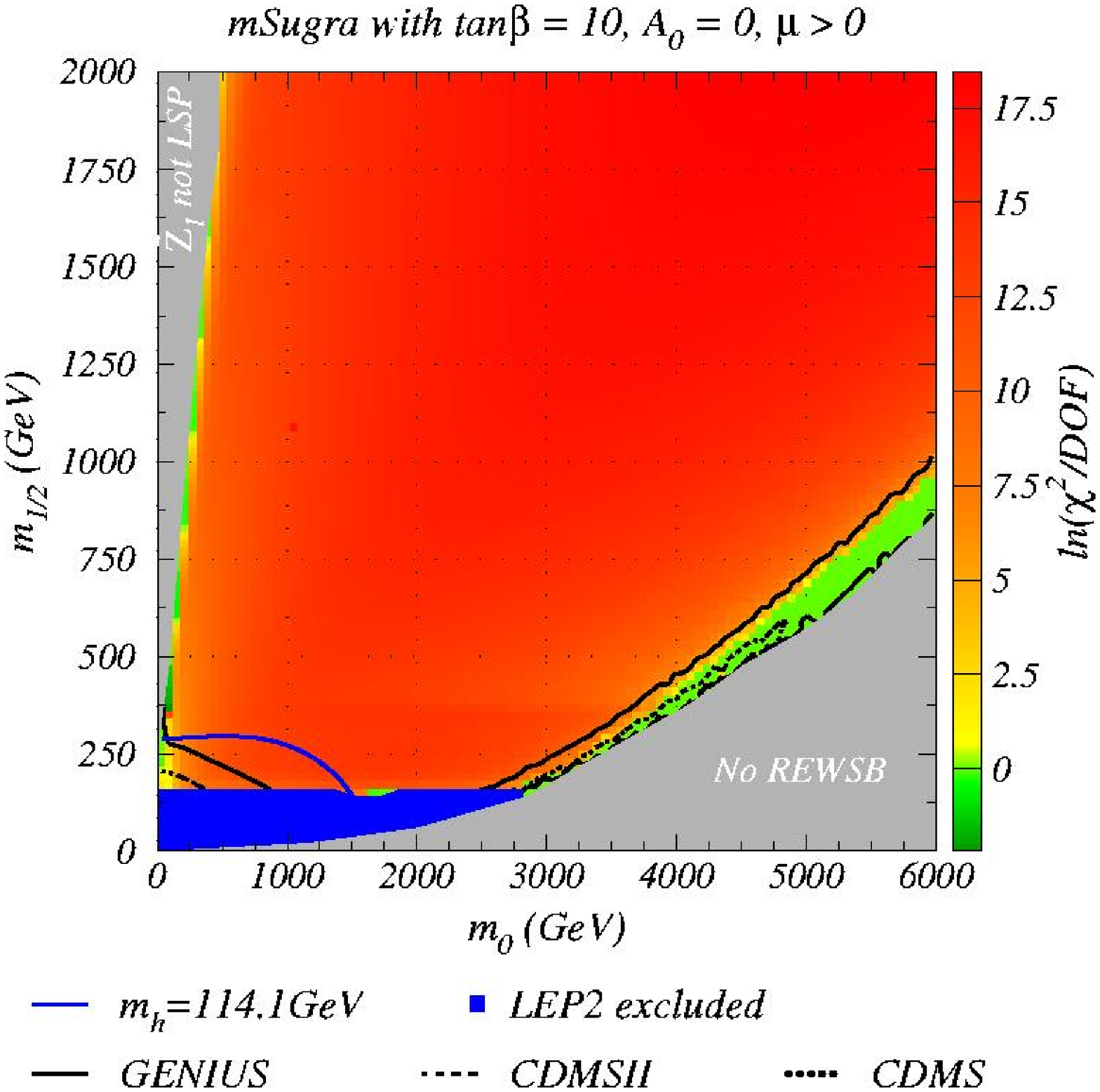,angle=0,width=0.4\linewidth}
\hfill \epsfig{file=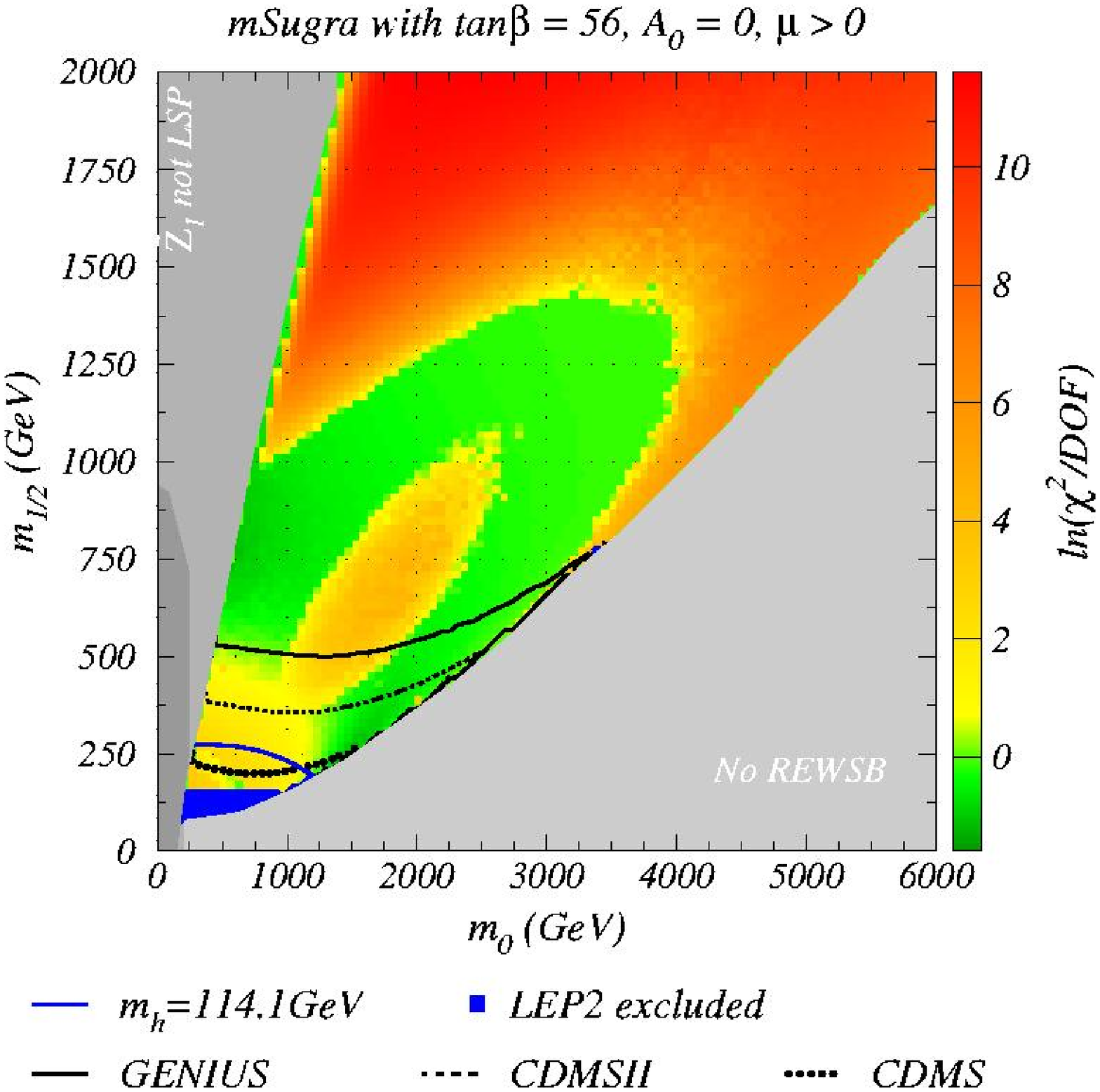,angle=0,width=0.4\linewidth}
\caption{ WMAP data seem to favor ($\frac{\chi^2}{dof}<4/3$)
(green) the HB/focus point
region (moderate to large values of $\mu$, large $m_0$ scalar masses)
for almost all tan$\beta$  (Left), as well as s - channel
Higgs resonance annihilation (Right) for $\mu>0$ and large
tan$\beta$.
}
\label{focus}
\end{figure}

\begin{figure}[htb]
\centering
\epsfig{file=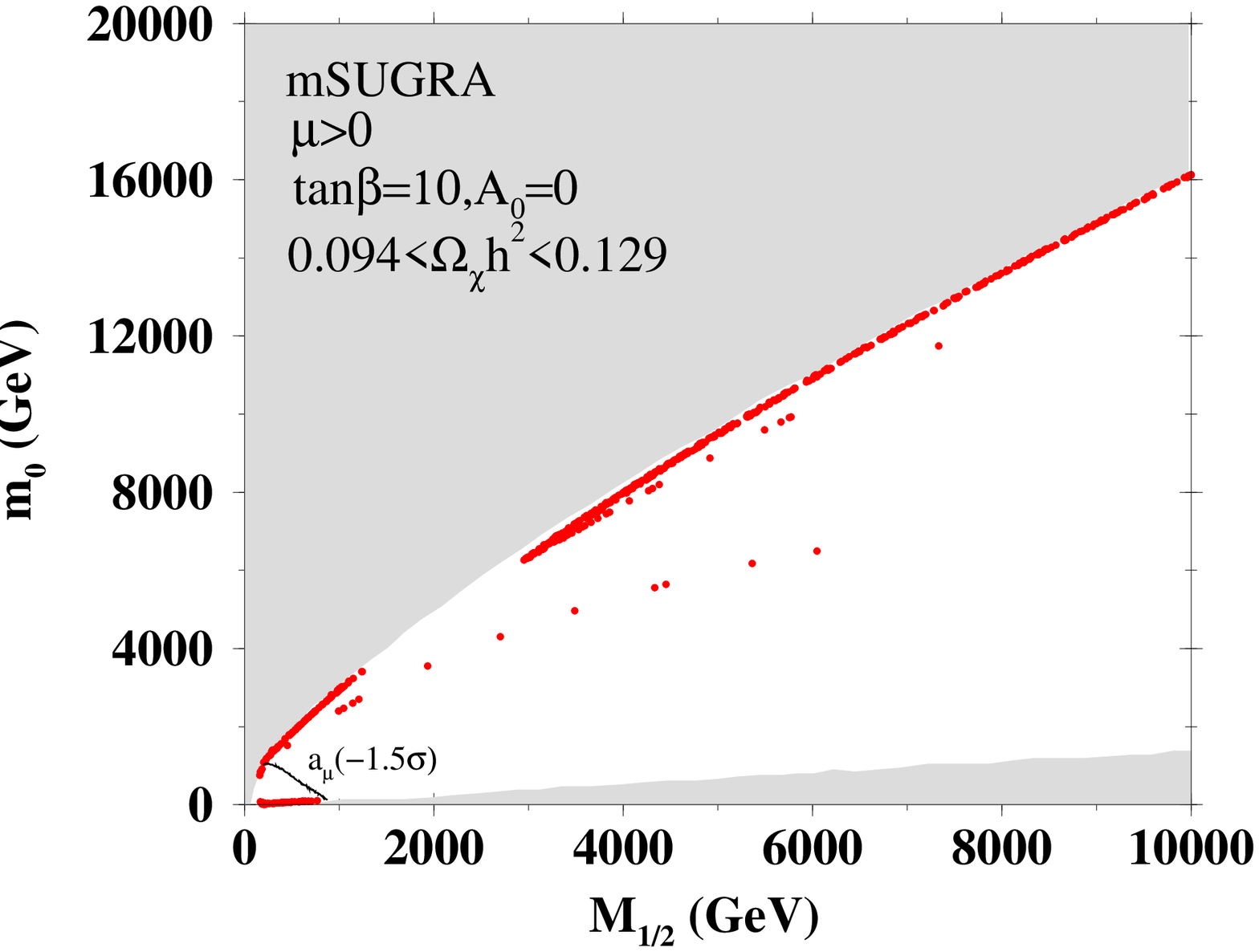,angle=0,width=0.41\linewidth,clip=}
\centering
\epsfig{file=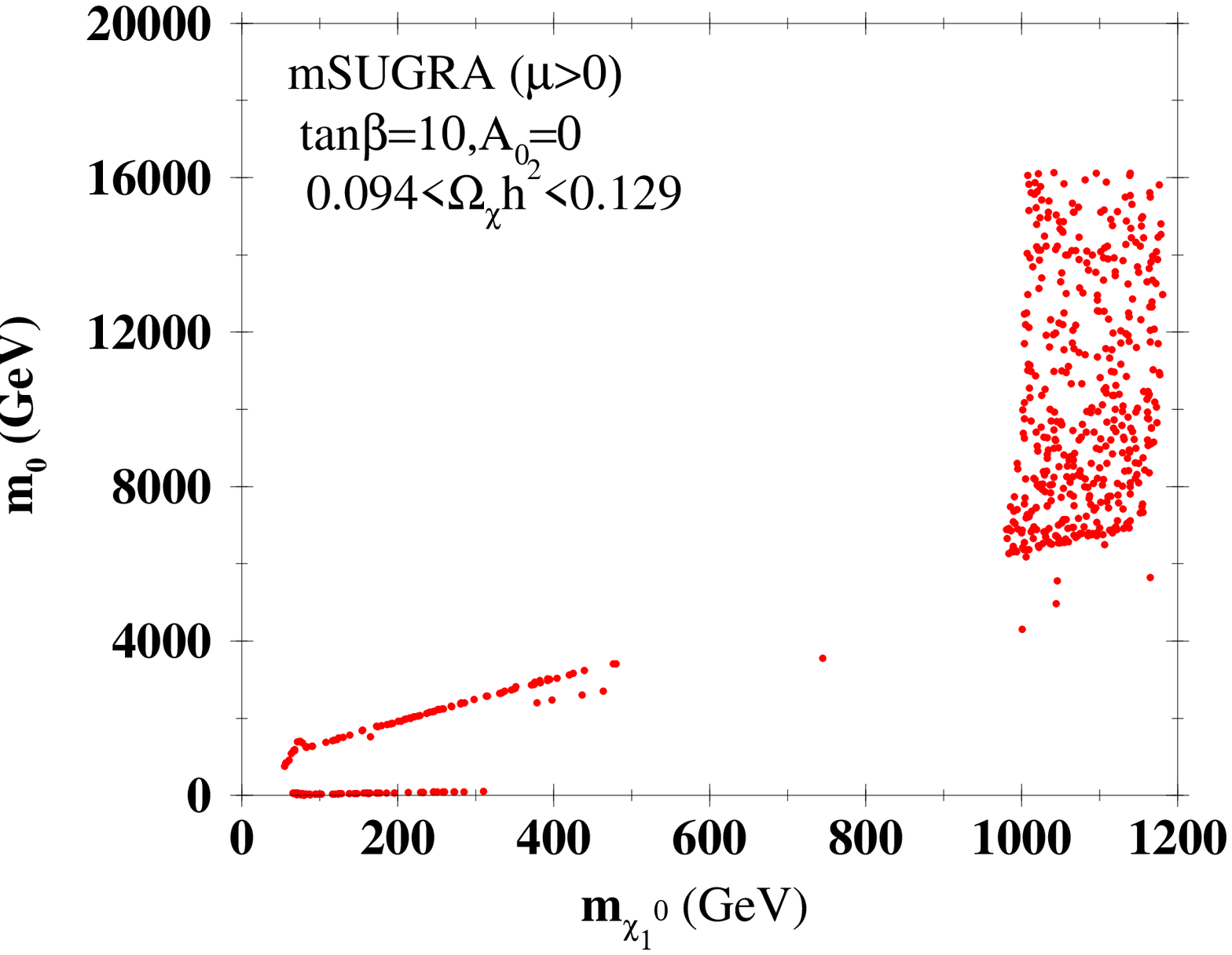,angle=0,width=0.41\linewidth,clip=}
\hfill\epsfig{file=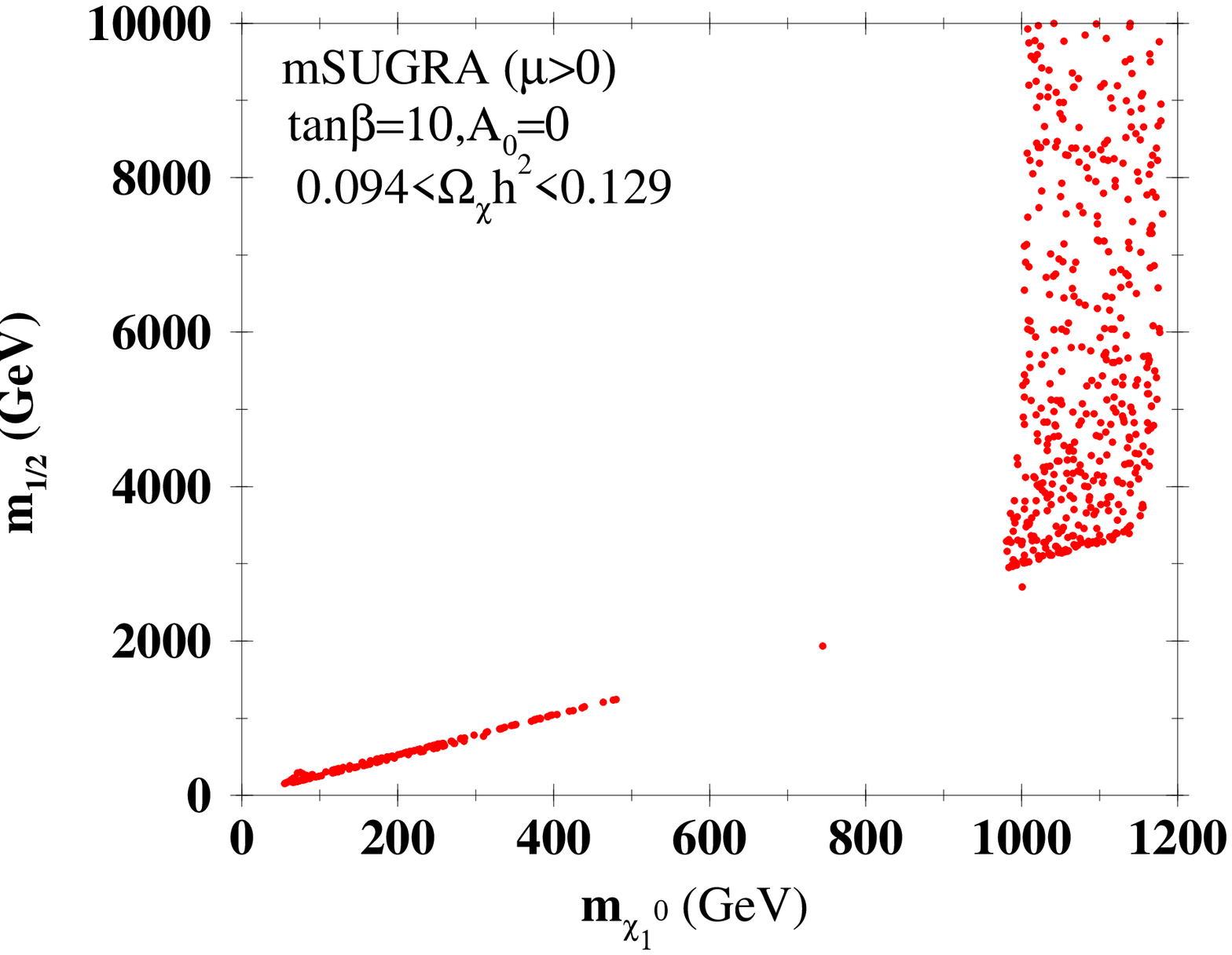,angle=0,width=0.41\linewidth,clip=}
\caption{{\it $m_0-m_{1/2}$ graph, and $m_0$ and $m_{1/2}$ vs.
$m_\chi$ graphs, including the HB of mSUGRA.
Such regions are favored by the WMAP data.}}
\label{chattofigure}
\end{figure}

Despite the above-mentioned good prospects of
discovering minimal SUSY models at future colliders, if the EB is realized,
however, things may not be that simple in Nature.
$\chi^2$ studies~\cite{baer} of mSUGRA in light of the recent
WMAP data has indicated (c.f. figure~\ref{focus}) that the HB/focus point region
of the model's
parameter space seems to be favored along with the neutralino resonance
annihilation region for $\mu>0$ and large tan$\beta$ values.
The favored focus point region corresponds to
moderate to large values of the Higgs parameter $\mu^2$, and large
scalar masses $m_0$ in the several TeV range.
\noindent
The situation in case the HB is included in the analysis
is depicted in figure \ref{chattofigure}~\cite{chatto},
where we plot the $m_0-m_{1/2}$ graphs,
as well as graphs of $m_0$, $m_{1/2}$ vs the neutralino LSP mass.
The neutralino density is that of the WMAP data.

We stress again that, in case the high zone (inversion) region
of the HB is realized,
then the detection prospects of SUSY at LHC are diminished significantly,
in view of the fact that in such regions slepton masses may lie
in the several TeV range (see figure \ref{chattofigure}). Fortunately, as already mentioned, last years's $g_\mu-2$ data~\cite{eidelman} (\ref{gm2}) seem to exclude
this possibility.

\subsubsection{Expected Reach of LHC and Tevatron}

\begin{figure}[htb]
\centering
\epsfig{file=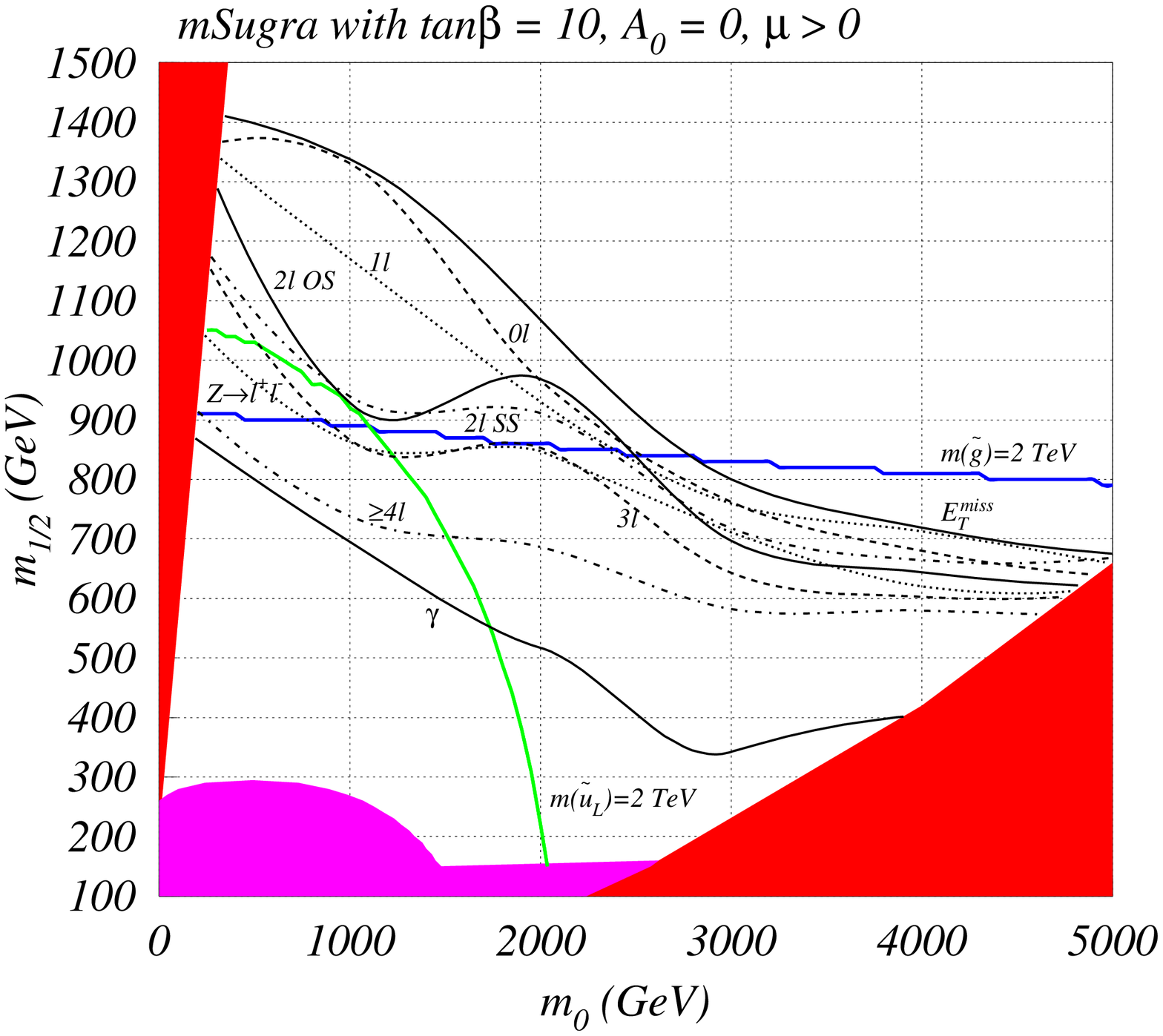,angle=0,width=0.35\linewidth}
\hfill \epsfig{file=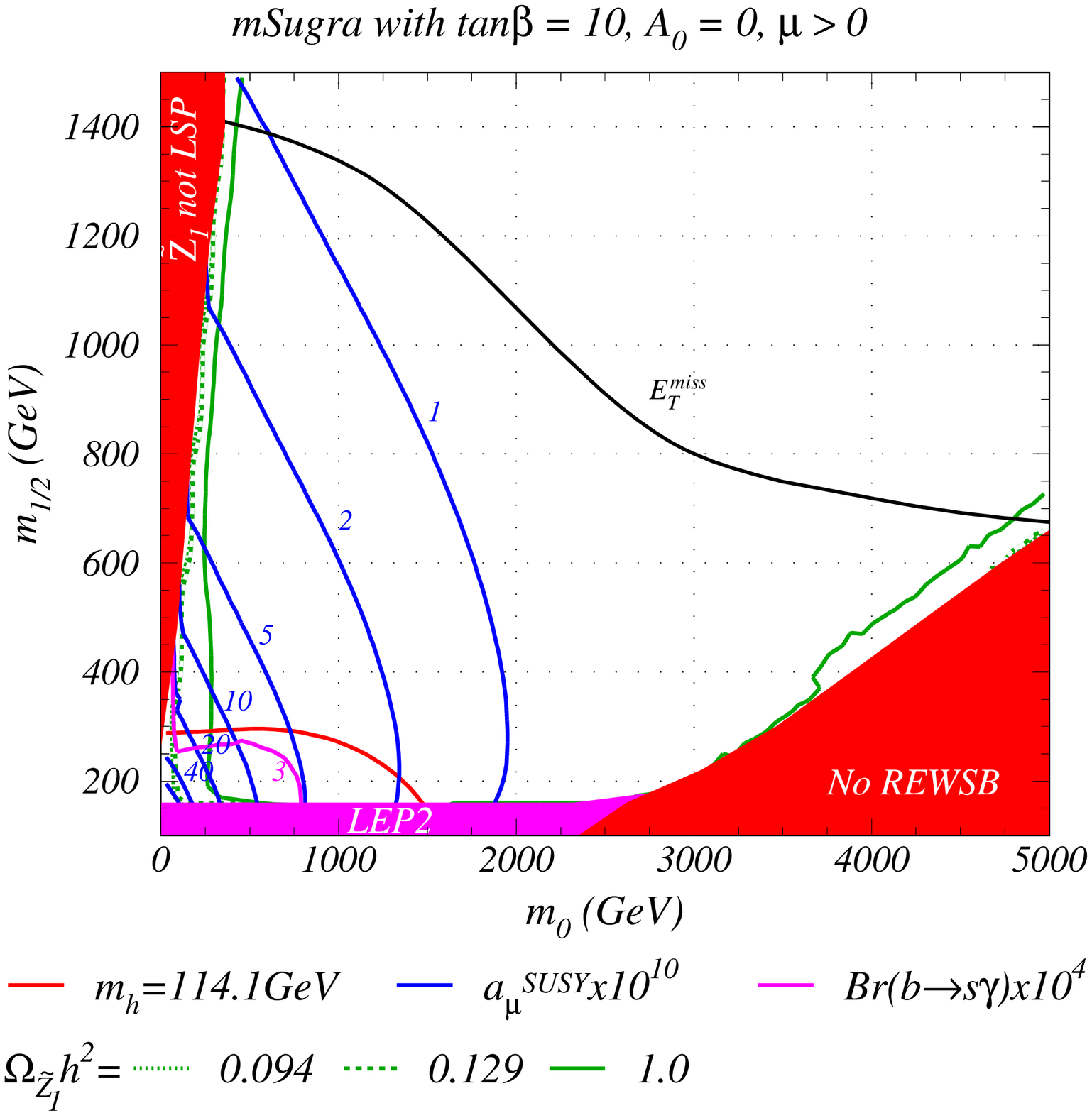,angle=0,width=0.35\linewidth}
\caption{{\it \underline{Left}:
The updated Reach in $(m_0, m_{1/2})$ parameter plane of mSUGRA
assuming 100 $fb^{-1}$ integrated luminosity. Red (magenta) regions
are excluded by theoretical (experimental) constraints.
\underline{Right}: Contours
(in view of the uncertainties) of several
low energy observables : CDM relic density (green),
contour of
$m_h=114.1$ GeV (red), contours of $a_\mu 10^{10}$ (blue)
and contours of $b \to s \gamma $ BF ($\times 10^{4}$)(magenta).}}
\label{reach}
\end{figure}

In view of the above results,
an updated reach of LHC in view of the recent WMAP
and other constraints discussed above
(see figure \ref{reach})
has been performed in \cite{reachlhc}, showing that
a major part of the HB, but certainly not its high zone (which though seems to be excluded by means of the recent $g_\mu -2$ data (\ref{gm2})),
can be accessible at LHC.
\noindent
The conclusion from this study~\cite{reachlhc} is that for an
integrated luminosity
of $100~fb^{-1}$ values of $m_{1/2} \sim 1400$ GeV can be probed
for small scalar masses $m_0$, corresponding to gluino masses
$m_{\tilde g} \sim 3$ TeV. For large $m_0$, in the hyperbolic branch/focus point region, $m_{1/2} \sim 700$ GeV can be probed, corresponding to
$m_{\tilde g} \sim 1800$ GeV. It is also concluded that
the LHC (CERN) can probe the entire stau co-annihilation
region and most of the heavy Higgs annihilation funnel
allowed by WMAP data, except for some range of $m_0,m_{1/2}$ in the case
tan$\beta \gsim 50$.
\begin{figure}[htb]
\epsfig{file=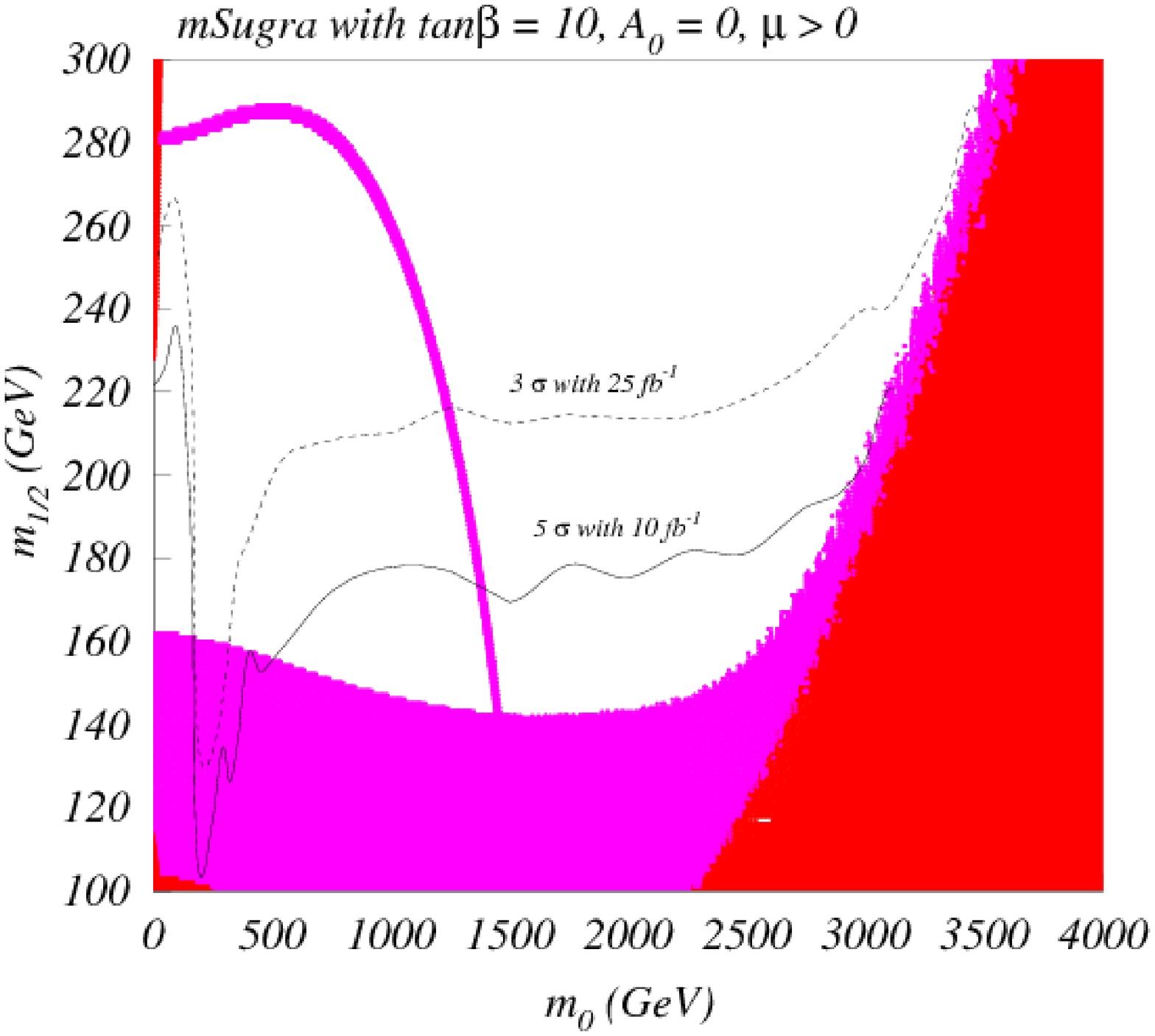,width=5cm} \hfill \hfill
\epsfig{file=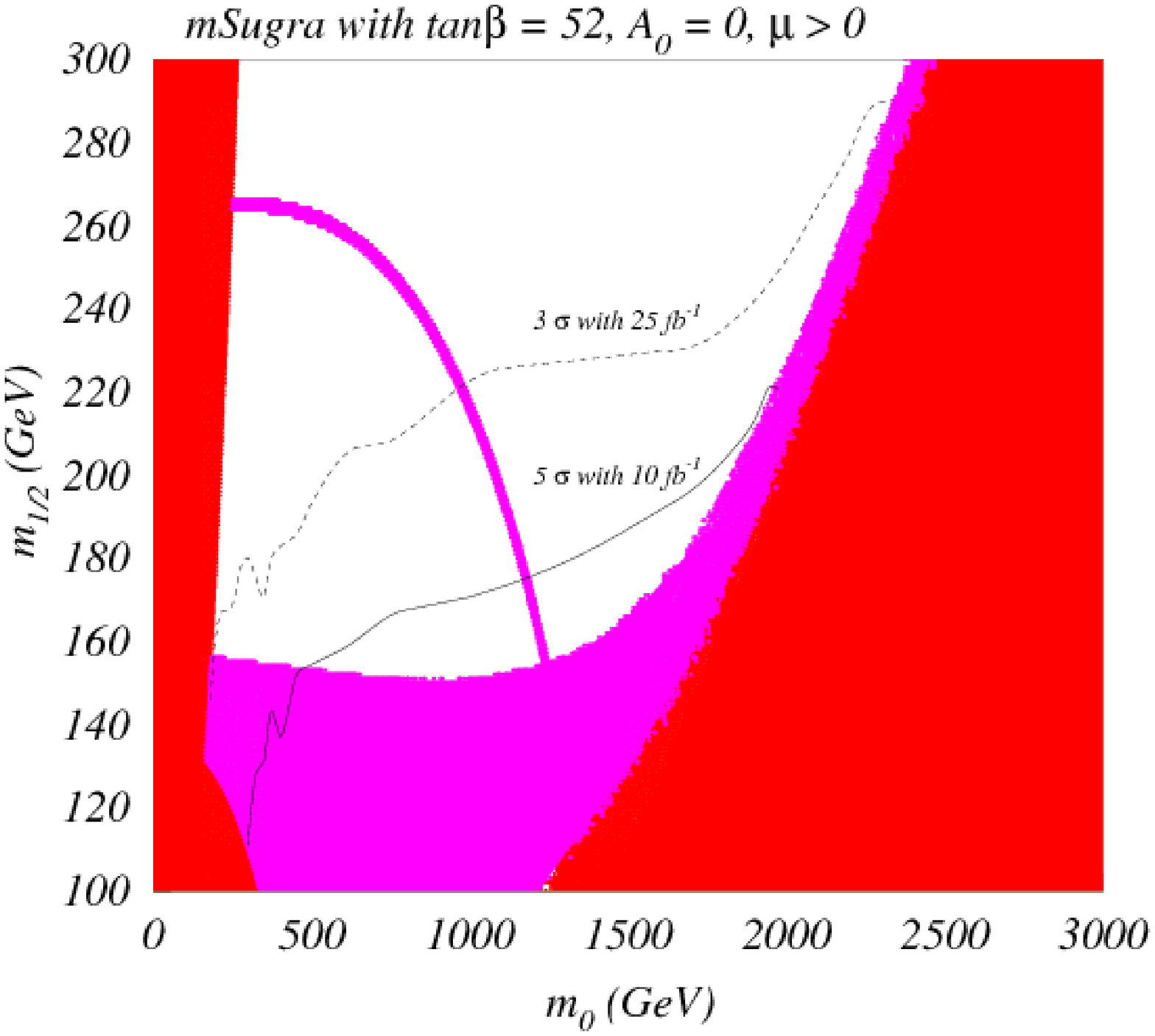,width=5cm}
\caption{\underline{Left}: The reach of Fermilab Tevatron in the
$m_0\ vs.\ m_{1/2}$ parameter plane
of the mSUGRA model, with
$\tan\beta =10$, $A_0=0$ and $\mu >0$,
assuming a $5\sigma$ signal at 10 fb$^{-1}$ (solid) and a $3\sigma$
signal with 25 fb$^{-1}$ of integrated luminosity (dashed). The red
(magenta) region is
excluded by theoretical (experimental) constraints. The region below
the magenta contour has $m_h <114.1$ GeV, in violation of Higgs mass
limits from LEP2. \underline{Right}: The reach of Fermilab Tevatron in the
$m_0\ vs.\ m_{1/2}$ parameter plane
of the mSUGRA model, with
$\tan\beta =52$, $A_0=0$ and $\mu >0$.
The red (magenta) region is
excluded by theoretical (experimental) constraints. The region below
the magenta contour has $m_h <114.1$ GeV, in violation of Higgs mass
limits from LEP2.
}
\label{fig:tevreach}
\end{figure}
A similar updated reach study in light of the new WMAP data
has also been done for the
Tevatron~\cite{reachtev}, extending previous analyses to large $m_0$ masses
up to 3.5 TeV, in order to probe the HB/focus region favored by the
WMAP data~\cite{baer}. Such studies
(c.f. figure~\ref{fig:tevreach}) indicate that for a $5\sigma$ (3$\sigma$) signal
with 10 (25) $fb^{-1}$ of integrated luminosity, the Tevatron reach in the
trilepton channel extends up to $m_{1/2} \sim 190$ $(270)$ GeV independent of
tan$\beta$, which corresponds to a reach in terms of gluino
mass of $m_{\rm g} \sim 575 (750)$ GeV.

\subsubsection{Astrophysical and Collider Dark Matter}

Above we have analyzed constraints placed on supersymmetric particle physics models, in particular MSSM, by WMAP/CMB  astrophysical data. The analysis made the assumption that neutralinos constitute exclusively the astrophysical
DM. It would be desirable to inverse the logic and ask the
question~\cite{arnowitt}: ``are  neutralinos
 produced at the LHC the
particles making up the  astronomically observed dark matter?''

\begin{figure}[htb]
\centering
\epsfig{file=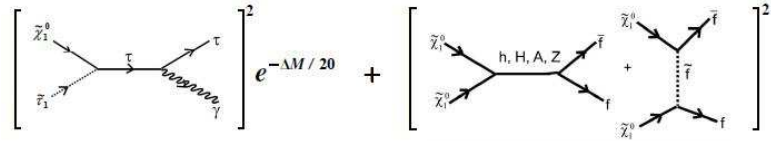,width=0.9\textwidth,clip=}
\caption{The Feynman diagrams for annihilation of neutralino dark matter in the
early universe. The Boltzmann factor $e^{-\Delta M/20}$ in the stau-co-annihilation graph is explicitly indicated.}
\label{neutralinofeynman}
\end{figure}

To answer this question,
let us first recall the relevant neutralino interactions (within the mSUGRA framework) that could take place in the Early universe (fig.~\ref{neutralinofeynman}).
As we have discussed previously, the
WMAP3 constraint  (\ref{relicdensity}) limits the parameter space to three main regions
arising from the above diagrams  (there is also a small ``bulk'' region):  (1) The stau-neutralino
($\tilde\tau_1-\tilde\chi^0_1$) co-annihilation region. Here $m_0$ is
small and $m_{1/2}\leq 1.5$ TeV. (2) The focus region where the
neutralino has a large Higgsino component. Here $m_{1/2}$ is small
and $m_0\geq 1$ TeV.  (3) The funnel region where annihilation
proceeds through heavy Higgs bosons which have become relatively
light. Here both $m_0$ and $m_{1/2}$ are large.
A key element
in the co-annihilation region is the  Boltzmann factor  from the
annihilation in the early universe at $kT\sim$ 20 GeV:  exp${[-\Delta M/20]}$, $\Delta M=M_{\tilde\tau_1}-M_{\tilde\chi^0_1}$ implying that
significant co-annihilation occurs provided $\Delta M\leq$ 20 GeV.

\begin{figure}[htb]\centering
 \epsfig{file=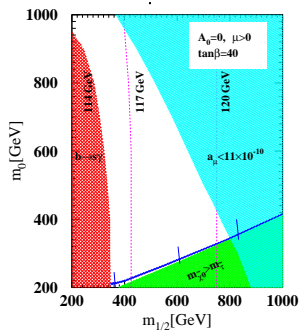,width=0.35\textwidth}
\caption{Allowed parameter space in mSUGRA. Dashed
vertical lines are possible Higgs masses (from [42]).} \label{wmapallreg}
\end{figure}

The accelerator constraints further restrict the parameter space and
if the muon  g$_\mu$-2  anomaly maintains~\cite{eidelman}, (c.f. (\ref{gm2})),  then $\mu>0$ is preferred and there
remains mainly the  co-annihilation region (c.f. figure~\ref{wmapallreg}).
Note the cosmologically allowed narrow co-annihilation band,
due to the Boltzmann factor for $\Delta M=5-15$ GeV, corresponding to
the allowed WMAP range for $\Omega_{\tilde\chi^0_1}h^2$.

 One may ask, then, whether: (i) such a small stau-neutralino mass
difference (5-15 GeV) arise in mSUGRA, since one would naturally
expect these SUSY particles to be hundreds of GeV apart and (ii)
such a small mass difference be measured at the LHC.
 If the answer to both these questions is  in the affirmative,  then the observation of such a
 small mass difference  would be a strong indication~\cite{arnowitt} that the
 neutralino is the astronomical DM particle,
since it is the
cosmological constraint on the amount of DM that forces the near
mass degeneracy with the stau, and it is the accelerator constraints
that suggest that the co-annihilation region is the allowed region.

As far as question (i) is concerned, one observes the following:
In the mSUGRA models, at GUT scale we expect no degeneracies, the $\Delta M$ is large, since  $m_{1/2}$ governs  the gaugino masses, while $m_0$
the slepton masses.  However, at
the electroweak scale (EWS), the Renormalization Group Equation can modify this: e.g.
the lightest selectron
$\tilde e^c$ at EWS has mass
$m^2_{\tilde e^c}=m_0^2+0.15 m_{1/2}^2+(37 \rm GeV)^2 
 \qquad{\rm while~the}\quad \tilde\chi^0_1 \quad{\rm has~mass}\qquad
m^2_{\tilde \chi^0_1}=0.16 m_{1/2}^2$
 The numerical accident that coefficients of $m_{1/2}^2$ is
nearly the same for both cases allows a near degeneracy: for
$m_0=0$, $\tilde e^c$ and $\tilde\chi^0_1$ become degenerate at
$m_{1/2}$=(370-400) GeV. For larger $m_{1/2}$, near degeneracy is
maintained by increasing $m_0$ to get the narrow
corridor in $m_0$-$m_{1/2}$ plane.
Actually the
case of the stau $\tilde\tau_1$ is more complicated~\cite{arnowitt}: large
t-quark mass causes left-right mixing in the stau mass matrix and
this results in the $\tilde\tau_1$ being the lightest slepton and not the
selectron. However, a result similar to the above occurs, with a
$\tilde\tau_1-\tilde\chi^0_1$ co-annihilation corridor appearing.

 We note that the above results depend only on the U(1) gauge
group and so co-annihilation can occur even if there were
non-universal scalar mass soft-breaking or non-universal gaugino
mass soft breaking at $M_G$.
Thus, co-annihilation can occur in a
wide class of SUGRA models, not just in mSUGRA.
Hence, in such models one has naturally near degenerate neutralino-staus, and hence the answer to question (i) above is affirmative.

Now we come to the second important question (ii), namely, whether LHC measurements have the capability of asserting  that the neutralino (if discovered) is the astrophysical DM.
\begin{figure}[htb]\centering
 \epsfig{file=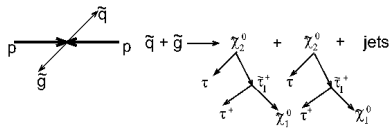,width=0.45\textwidth}
\caption{SUSY production of neutralinos and decay channels}
\label{neutrprod}
\end{figure}
To this end we note that,
in LHC, the major SUSY production processes of neutralinos are interactions of gluinos ($\tilde
g$) and squarks ($\tilde q$) (c.f. figure~\ref{neutrprod}),
e.g., $p+p\rightarrow \tilde g+\tilde
q$. These then decay into lighter SUSY particles. The final states involve two neutralinos $\tilde\chi^0_1$
giving rise to  missing transverse energy $E^T_{\rm miss}$)  and four
$\tau$'s, two from the $\tilde g$ and two from the $\tilde q$ decay chain for the
example of fig.~\ref{neutrprod}.

 In the
co-annihilation region, two of the taus have a high energy (``hard" taus)
coming from the $\tilde\chi^0_2\rightarrow\tau\tilde\tau_1$ decay
(since $M_{\tilde\chi^0_2}\simeq 2 M_{\tilde\tau_1}$), while the other two are
low energy particles (``soft" taus) coming from the
$\tilde\tau_1\rightarrow\tau+\tilde\chi^0_1$ decay, since $\Delta M$ is
small.

The signal is thus $E_T^{\rm miss}+$ jets +$\tau$'s, which
should be observable at the LHC detectors~\cite{arnowitt}.
\begin{figure}[htb]\centering
 \epsfig{file=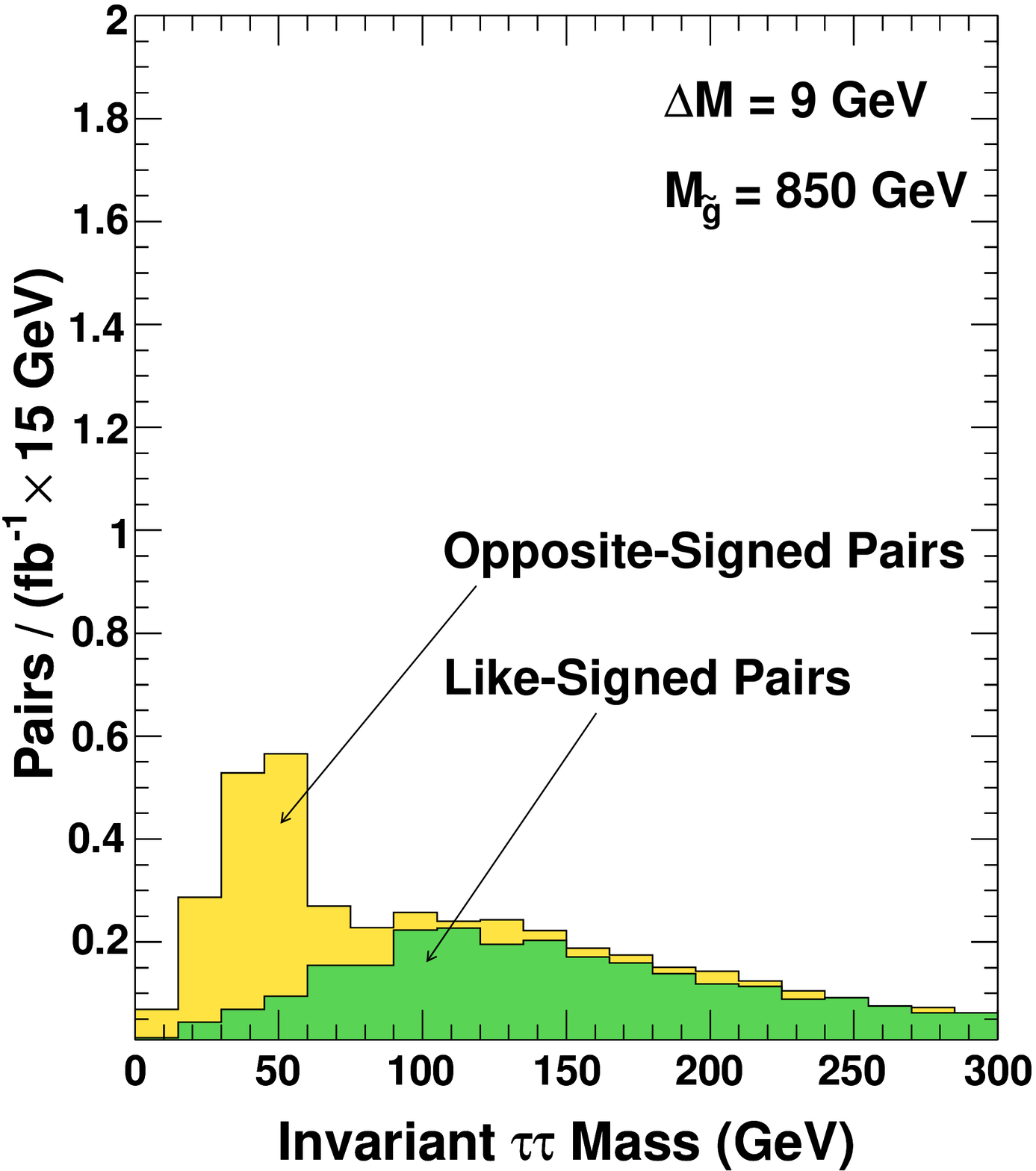,width=0.35\textwidth}\hfill
 \epsfig{file=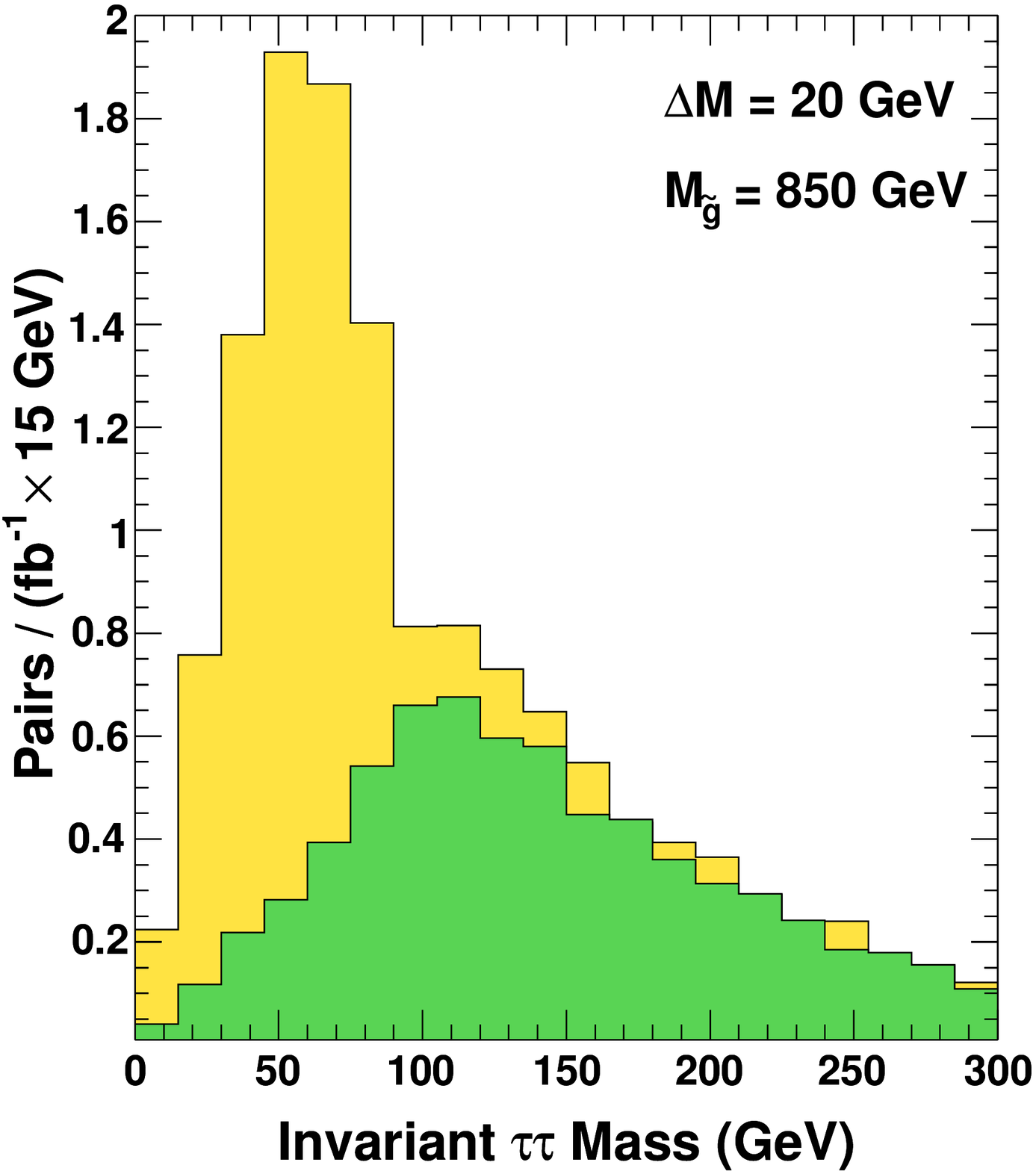,width=0.35\textwidth}
\caption{Number of tau pairs as a function of invariant $\tau\tau$
mass. The difference $N_{OS}$-$N_{LS}$ cancels for mass $\geq$ 100
GeV eliminating background events (from [42]).}
\label{tauspair}
\end{figure}
 As seen above, we expect two pairs of taus, each pair containing one
soft and one hard tau from each $\tilde\chi^0_2$ decay. Since
$\tilde\chi^0_2$ is neutral, each pair should be of  opposite
sign. This distinguishes them from  SM- and SUSY-backgrounds jets-faking taus,  which will have equal
number of like--sign as opposite--sign events~\cite{arnowitt}. Thus, one can suppress backgrounds statistically by considering the number of  opposite--sign
events $N_{OS}$ minus the like--sign events $N_{LS}$ (figure~\ref{tauspair}).

\begin{figure}[htb]
\begin{center}
 \epsfig{file=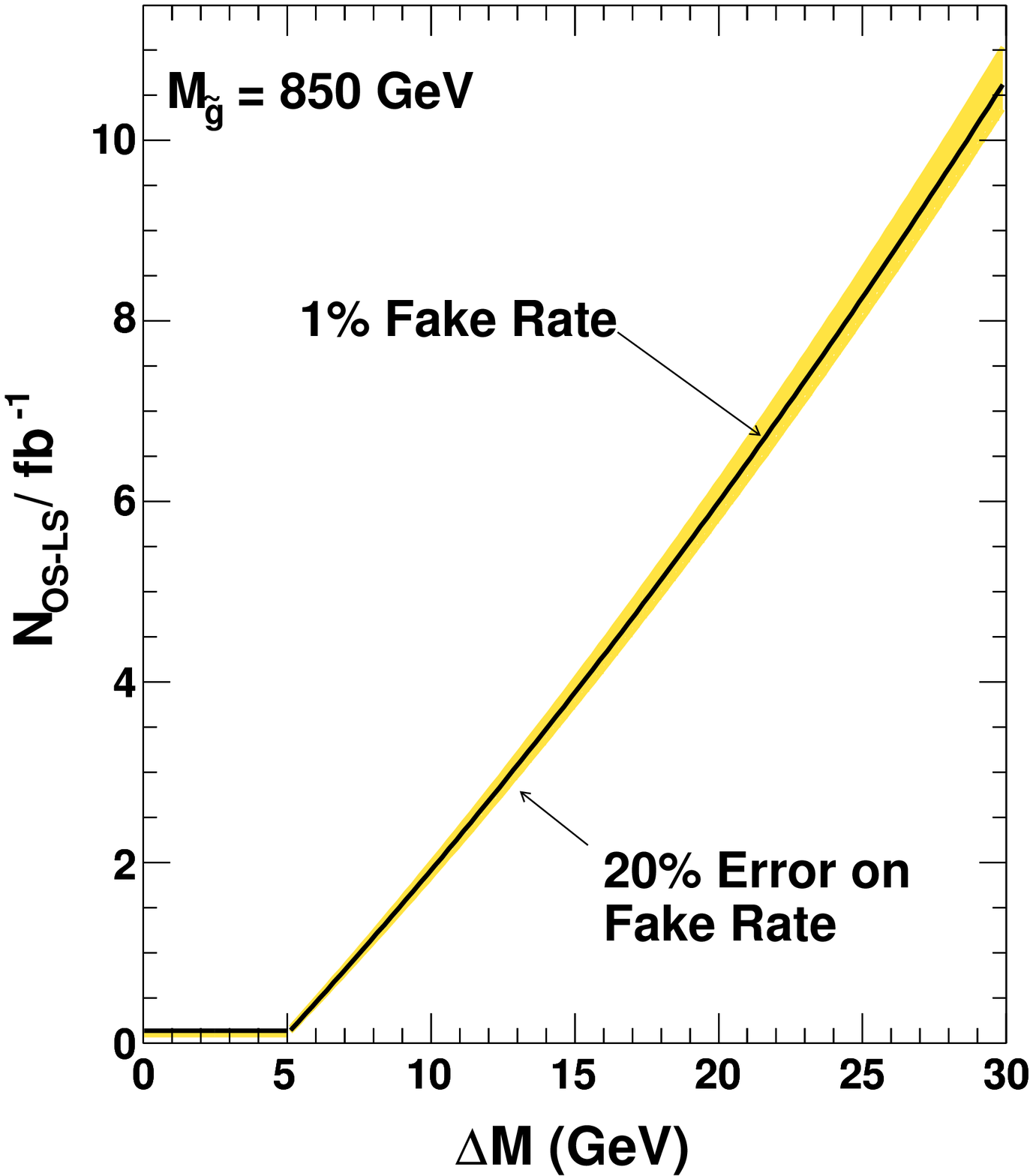,width=0.36\textwidth}\hfill
 \epsfig{file=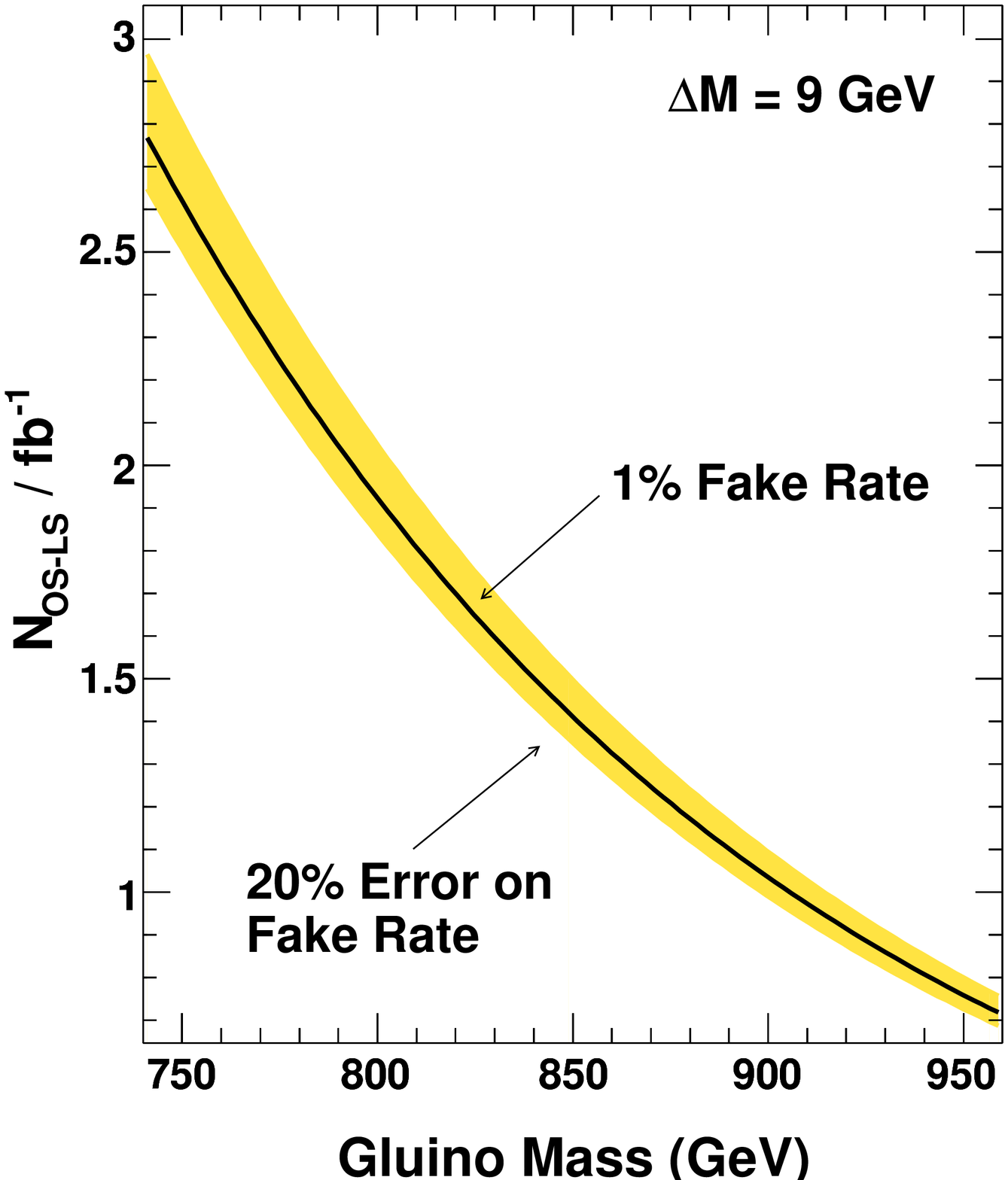,width=0.36\textwidth}
\end{center}
\caption{$N_{OS-LS}$ as function of $\Delta M$ (left graph) and as a
function of $M_{\tilde g}$ (right graph). The central black line
assumes a 1\% fake rate, the shaded area representing the 20\% error
in the fake rate  (from [42]). }
\label{taupair2}
\end{figure}

\begin{figure}[htb]
\begin{center}
\epsfig{file=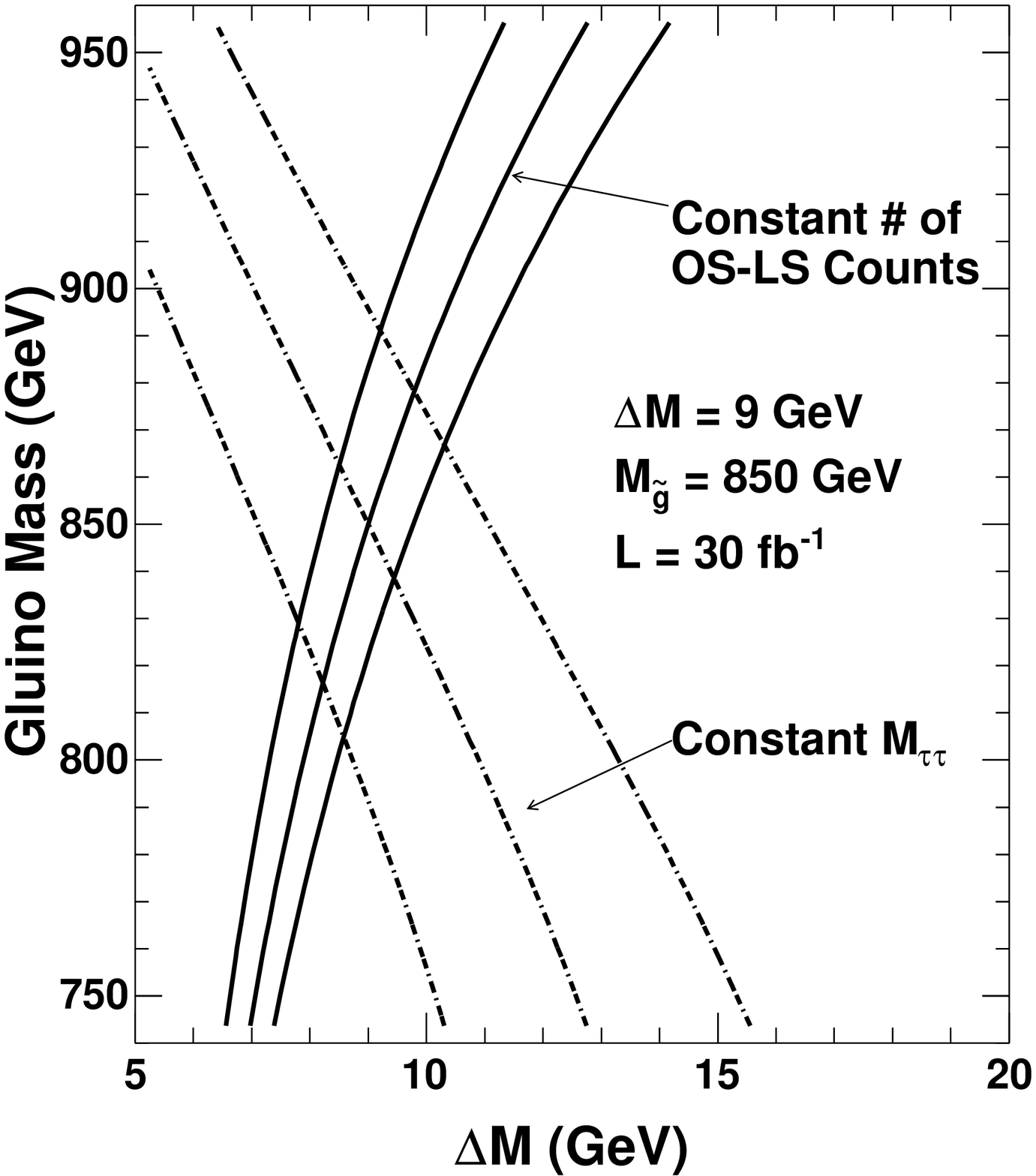,width=0.35\textwidth} \hfill
\epsfig{file=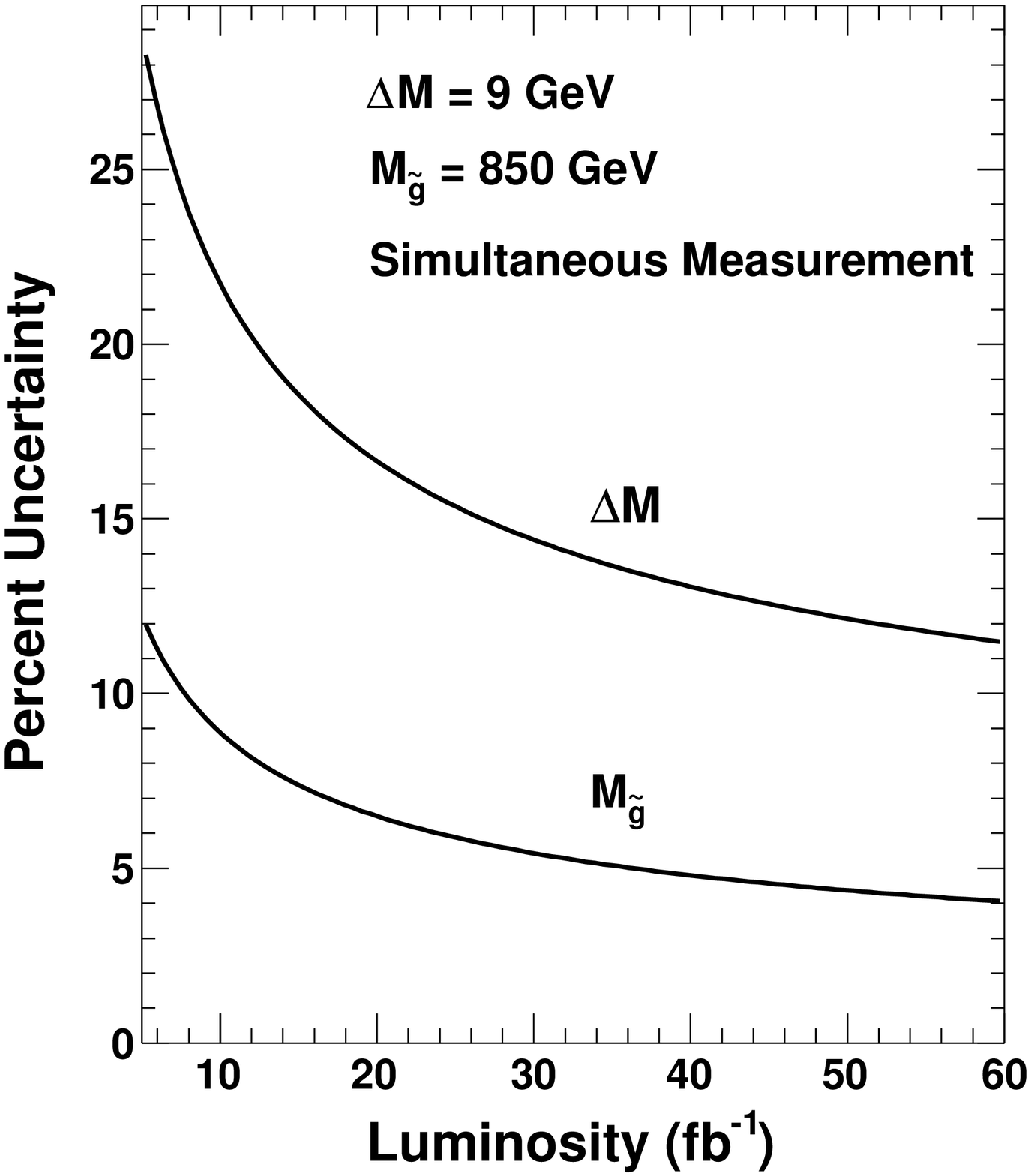,width=0.35\textwidth}
\end{center}
\caption{\underline{Left}: Simultaneous determination of $\Delta M$ and $M_{\tilde
g}$. The three lines plot constant $N_{OS-LS}$ and $M_{\tau\tau}^{\rm
peak}$ (central value and 1$\sigma$ deviation) in the $M_{\tilde
g}$-$\Delta M$ plane for the benchmark point of $\Delta M$=9 GeV and
$M_{\tilde g}$=850 GeV assuming 30 fb$^{-1}$ luminosity.
\underline{Right}: Uncertainty in the determination of $\Delta M$ and
$M_{\tilde g}$ as a function of luminosity (from [42]).}
\label{fig4}
\end{figure}

 The four $\tau$
final state has the smallest background but the acceptance and
efficiency for reconstructing all four taus is low. Thus to
implement the above ideas we consider here the three $\tau$ final
state of which two are hard  and one is soft.
There are two important features:  First, $N_{OS-LS}$
increases with $\Delta M$(since the $\tau$ acceptance increases) and
$N_{OS-LS}$ decreases with $M_{\tilde g}$(since the production cross
section of gluinos and squarks decrease with $M_{\tilde g}$).
Second, one sees that $N_{OS-LS}$ forms a peaked
distribution. The di-tau peak position $M_{\tau\tau}^{\rm
peak}$ increases with both $\Delta M$ and $M_{\tilde g}$. This
allows us to use the two observables $N_{OS-LS}$ and
$M_{\tau\tau}^{\rm peak}$ to determine both $\Delta M$ and
$M_{\tilde g}$ (c.f. figure~\ref{taupair2}).
As becomes evident from the analysis~\cite{arnowitt} (c.f. fig. \ref{fig4}) it is possible to simultaneously determine
$\Delta M$ and the gluino mass $M_{\tilde
g}$. Moreover, one sees that at LHC even with 10 fb$^{-1}$ (which should be
available at the LHC after about two years running) one could
determine $\Delta M$ to within 22\%, which should be sufficient to
know whether one is in the SUGRA co-annihilation region.
The above analysis was within the mSUGRA model, however
similar analyses for other SUGRA models can be made, provided the
 production of neutralinos is not suppressed.
In fact, the
determination of $M_{\tilde g}$ depends on mSUGRA
universality of gaugino masses at GUT scale, $M_G$, to relate
$M_{\tilde\chi^0_2}$ to $M_{\tilde g}$ thus  a model independent
method of determining $M_{\tilde g}$ would allow one to to  test the
question of gaugino universality.  However, it may not be easy to
directly measure $M_{\tilde g}$ at the LHC for high $\tan\beta$ in
the co-annihilation region due to the large number of low energy
taus, and the ILC would require a very high energy option to see the
gluino.

One can also measure~\cite{arnowitt}
$\Delta M$ using the signal
 $E_T^{\rm miss}$+ 2 jets+2$\tau$.
This signal has higher
acceptance but larger backgrounds.
 With 10 fb$^{-1}$
one can measure $\Delta M$ with 18\% error  at the
benchmark point assuming a separate measurement of $M_{\tilde g}$
with 5\% error has been made. While the benchmark
point has been fixed in \cite{arnowitt} at $M_{\tilde g}=850$ GeV(i.e. $m_{1/2}=$360 GeV), higher
gluino mass would require more luminosity to see the signal. One finds that with 100 fb$^{-1}$ one can
probe $m_{1/2}$ at the LHC up to $\sim 700$ GeV (i.e., $M_{\tilde
g}$ up to $\simeq 1.6$ TeV).
Finally it should be mentioned that measurements of
$\Delta M$ at the ILC could be made if a very forward calorimeter
is implemented to
reduce the two $\gamma$ background. In such a case, $\Delta M$ can be determined
with 10\% error at the benchmark point, thereby implying that~\cite{arnowitt}
in the
co-annihilation region, the determination of $\Delta M$ at the LHC is
 not significantly worse  than at the ILC.

\begin{figure}[htb]
\begin{center}
  \epsfig{file=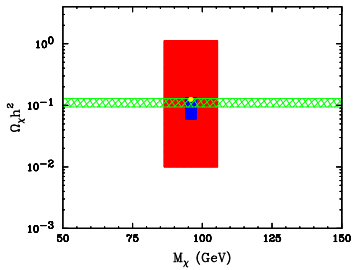,width=0.4\textwidth}
  \end{center}
\caption{Accuracy of WMAP (horizontal green shaded region), LHC (outer red rectangle)
and ILC (inner blue rectangle) in determining $M_\chi$, the mass of the
lightest neutralino, and its relic density $\Omega_\chi h^2$.
The yellow dot denotes the actual values of $M_\chi$ and
$\Omega_\chi h^2$ for a sample point in parameter space of mSUGRA:
$m_0 = 57$ GeV, $m_{1/2} = 250$ GeV, $A_0 = 0$, $\tan \beta = 10$ and ${\rm sign}(\mu) = +1$  (from A.~Birkedal {\it et al.},
  arXiv:hep-ph/0507214)}
  \label{summary}
\end{figure}
The results on the accuracy of determining DM mass in astrophysics and colliders within the mSUGRA framework is given in figure~\ref{summary}. We see that the cosmological measurement are at present the most accurate one, however, the reader should bear in mind the model-dependence of all these results. We now come to demonstrate this point by repeating the analysis for some class of stringy models.

\subsection{Stringy Models and Particle Physics Constraints}

String theory  (at least as we know it at present)
 lives in higher than four space-time dimensions (supersymmetric strings D=10). Low-energy field theory includes gravitational string mutliplet fields  (graviton,  dilaton $S$(scalar),  and  supersymmetric partners  in Supergravity  theories). There is an obvious need for compactification  to four dimensions. This happens dynamically through the
 Moduli scalar fields, $T^i$, of the string multiplets, which depend on the extra dimensions.

Originally, it was thought that the requirement of the absence of instabilities in the vacuum
necessitates  target-space supersymmetric strings (superstrings) in general.
However, tachyonic instabilities may be welcome in cosmological
scenarios, hence non supersymmetric target-space-time strings may be at play.
 Target-space supersymmetry needs breaking and must be  phenomenologically consistent, i.e. partners must have masses above a few TeV.
 Consistent breaking of SUSY in string-inspired SUGRA via gaugino is possible, and rigorous, and phenomenologically realistic models, in this respect, do exist. It is the purpose of this subsection to discuss particle physics constraints in the framework of one class of such models.
There are Modified Constraints on such string-inspired SUSY models from accelerator physics and Dark Matter which we shall discuss below.

The model we shall concentrate upon is a
Heterotic string with
Orbifold compactification from ten to four-dimensions  with standard model gauge group $SU(3)\times SU(2) \times U(1)$,  three generations,
and consistent SUSY breaking via gaugino condensate~\cite{heteropheno}.
Below we shall briefly review its most important features.

There is dominance of  one-loop soft-SUSY-breaking {\it non-universal} terms,  as a result of  superconformal anomalies (non-zero $\beta$-function).  This modifies predictions from mSUGRA, and implies and interpolation between Anomaly-Mediated SUSY Breaking models and mSUGRA.
One distinguishes two regimes/scenaria for the SUSY-breaking terms:

\noindent{\it \underline{Moduli-dominated:}}  SUSY breaking is driven by the compactification moduli fields $T^i$, whose
vacuum expectation value (v.e.v.)   $<T^i> \ne 0$ determines  the  size  of
the  compact manifold . In this regime,
there are light scalars and relatively heavy gauginos, whose nature
 depends completely on the value of
 the group-independent coefficient
of the  universal Green-Schwarz counterterm,
 $\delta_{\mathrm{GS}}$~\cite{heteropheno}.

 \noindent{\it \underline{Dilaton-dominated:}} The dilaton $S$ acquires a v.e.v.  $<S> \ne 0$, which in turn determines the value of  string coupling  $g_s$  at the string scale, and transmits, via the (SUSY) auxiliary fields,
 SUSY Breaking. There are  non-perturbative corrections
to K\"ahler potential which stabilize the dilaton
in the presence of  gaugino condensation .
The associated
phenomenology is completely different from the moduli-dominated case.  We are in a domain of  heavy squarks and sleptons (of order
of the gravitino scale)  and  {\it light gaugino masses},  driven by the dilaton auxiliary field v.e.v.'s

\begin{figure}[ht]
\centering
       \epsfig{file=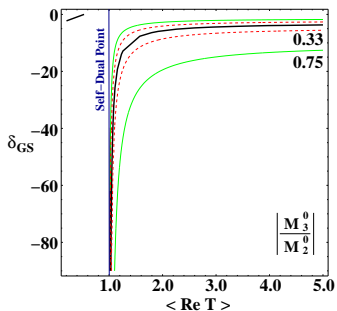,width=0.4\textwidth}
          \caption{Contours of relative running gaugino masses
          $M_3/M_2$ in the $(\lang \re \; t \rang ,\,\; \delta_{\GS})$ plane. These soft
          masses are at the initial (GUT) scale. The heavy (dark) contour is the limit
          of vanishing gluino mass (there is another such contour in the upper left
          corner on the other side of the self-dual point). For $\lang \re \; t \rang\ >1$
          we also give contours of $|M_3/M_2| = 0.33$ (dashed) and 0.75 (solid).}
        \label{fig:gluinomass}
\end{figure}

 There is a diverse origin of SUSY Breaking terms in this class of models:
{\bf (i)} Some come from the  superconformal anomalies , and hence
are non--universal  (proportional to
 the  $\beta$-- function  of the  $SU(3) \times SU(2) \times U(1)$  groups);
{\bf (ii)}  some are
  independent  of the gauge group considered  (Green--Schwarz counterterm, v.e.v.
 of the condensate). This interplay between universality and non--universality implies a {\it rich phenomenology}, new trends in the
  search of supersymmetric particles in accelerator and astro-particle physics.

\begin{figure}[ht]
\centering
       \epsfig{file=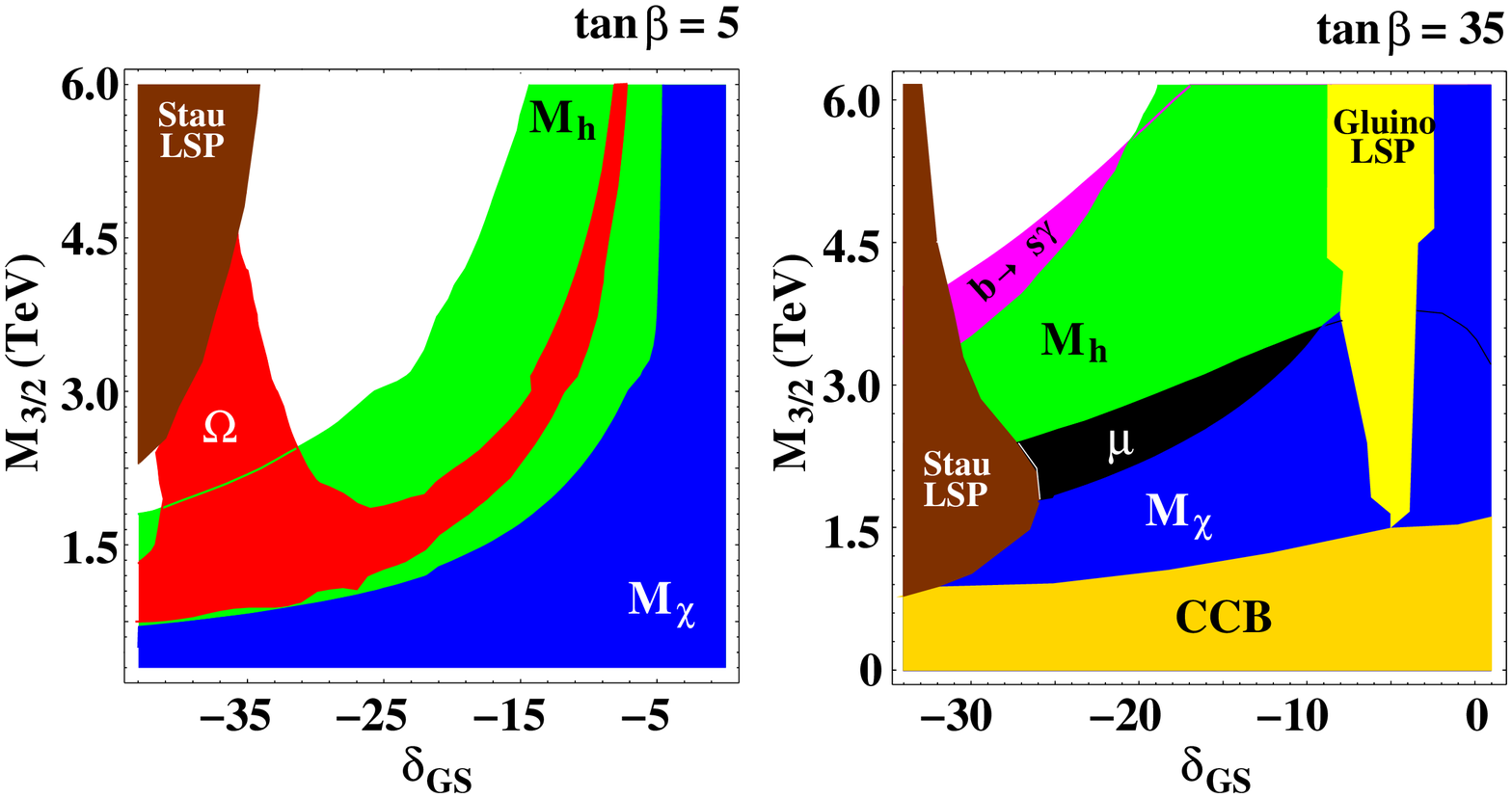,width=0.6\textwidth}\hfill\hfill
\epsfig{file=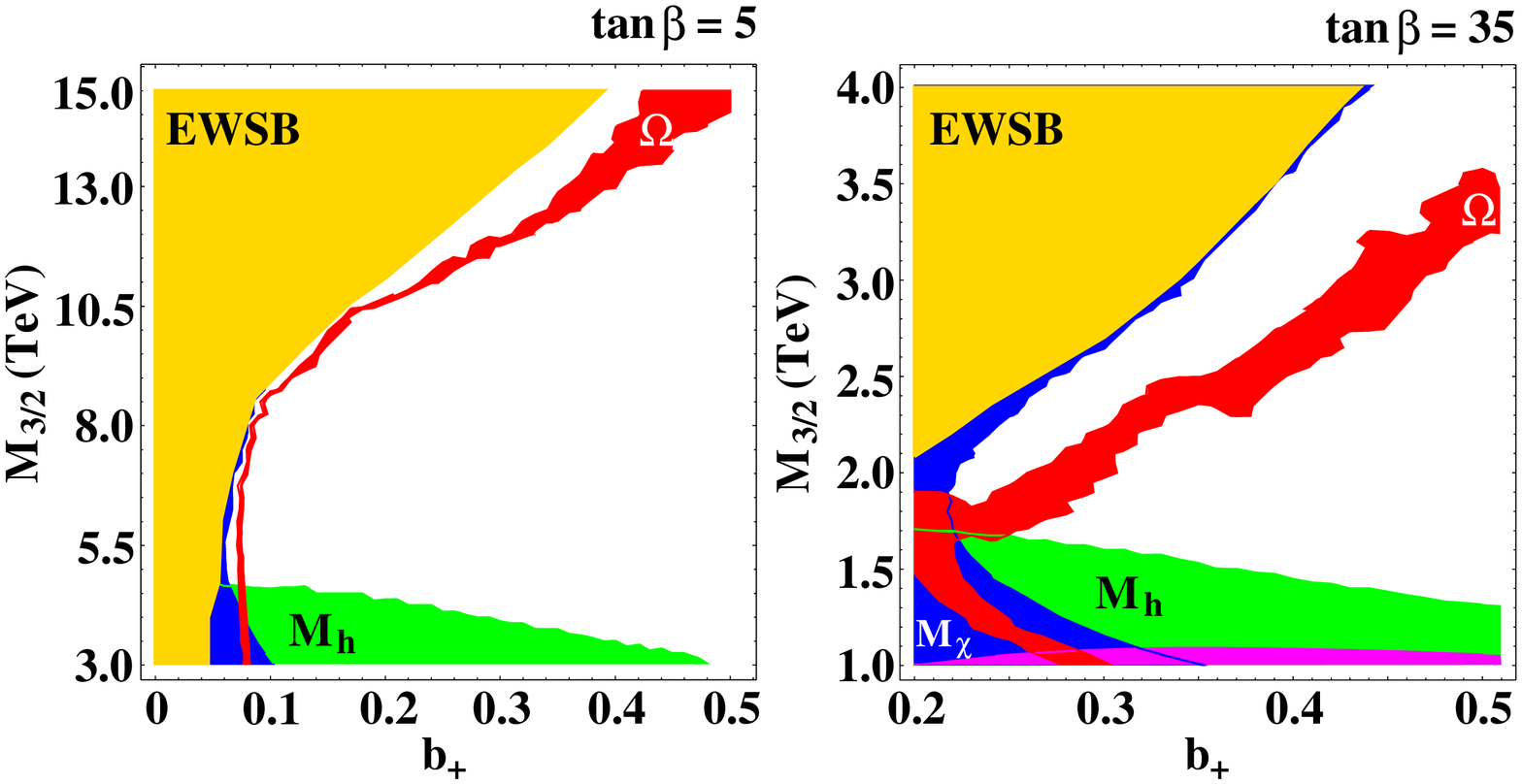, width=0.6\textwidth}
          \caption{\underline{Upper}: Constraints on the moduli-dominated
          parameter space for $\tan\beta=5$ (left) and $\tan\beta=35$ (right) with
          $p = 0$ and $\lang \re \; t \rang = 2.0$. Constraints on the
          ($M_{3/2},\; \delta_{\GS}$) plane are given for $\mu > 0$. The dark shaded
          regions on the left have a stau LSP. The $\tan\beta=35$ plot also has a
          region with a gluino LSP. For $\tan\beta=5$ the region labeled
          ``$\Omega$'' has the cosmologically preferred relic density of neutralinos.
          No such region exists for the higher $\tan\beta$ plot. In that case the
          exclusion contours are due to (from bottom right to upper left) Charge Conjugation Breaking  vacua,
          the chargino mass, too large SUSY contributions to $(g_{\mu}-2)$, the Higgs
          mass limit and too large a $b \to s \gamma$ rate.
          \underline{Lower:}Constraints
on the dilaton-dominated parameter space for
 $\tan\beta=5$ (left) and $\tan\beta=35$ (right). Constraints on the
($M_{3/2},\; b_+$)
 plane are given for $\mu > 0$ [with $b_+$ the largest $\beta$-function coefficient among condensing gauge groups of the hidden sector].}
        \label{fig:modp0t2}
\end{figure}

The heterotic string models we shall analyze here make use of the so-called Pauli-Villars -Anomaly Mediated Supersymmetry Breaking (PV-AMSB) scenario~\cite{heteropheno}.
There is an important feature of the AMSB-string models which turns out to be important for DM searches, namely the presence of
{\it non-thermal} LSP relic densities.
Indeed, in AMSB studies it was found that the
 neutralino thermal
relic density  is  generically too small
 to explain the amount of
dark matter: the wino content of the LSP is quite high.
Additionally, co-annihilation between the LSP and the lightest
chargino is also very efficient.  Both of these effects combine to
make the thermal relic density of LSP negligible.
Thus the
 anomaly-mediated character of the gaugino sector
 in this model
necessitates a  non-thermal production mechanism for neutralino
LSPs,  or another candidate for the cold dark matter must be
postulated.

We next remark that, in the moduli-domination scenario of the heterotic string models, the various K\"ahler
moduli are not stabilized, i.e.
$\lang \re\;t
\rang$ is not fixed, and
one must use the
soft terms. The value of the
Green-Schwarz coefficient $\delta_{\GS}$, then,  becomes relevant for the
determination of gaugino masses~\cite{heteropheno}, which are ``running''
with it (c.f. figure~\ref{fig:gluinomass}).

The relevant astro-particle physics constraints for one indicative
example of heterotic orbifold models~\cite{heteropheno}, in some range of the parameter space, are summarized in figure \ref{fig:modp0t2}.  The reader's attention is drawn to the strong suppression (or disappearance) of thermal neutralino relic densities in some regions of the parameter space.

\subsection{Non-critical (non-equilibrium) Stringy Q-Cosmology}

The results described on the previous sub section were based on a critical string theory, where the dilaton field is stabilized. However, one may encounter situations in some non-equilibrium stringy cosmologies, described by non-critical strings~\cite{qcosmol} (Q-cosmologies),
in which the dilaton is not stabilized.
Such cosmologies might arise, for instance, in colliding brane world scenarios, where the cosmically catastrophic early universe event of collision suffices to induce departure of the associated string theory, describing various  excitations on our brane world, from conformal invariance (on a world sheet) and hence non-criticality.
In such models, the dilatons are time dependent and are not stabilized. At late eras of the Universe, such a time dependent dilaton may behave as quintessence-like field leading to acceleration of the Universe~\cite{diamandis,qcosmol,veneziano}. The departure of criticality has also other consequences for the low-energy field theory, namely the existence of off-shell terms~\cite{qcosmol}, i.e. the variations of the relevant effective action with respect to the fields in the gravitational string multiplet (gravitons, dilatons) may be non zero. A detailed analysis~\cite{lmnbol} shows that
in such models, where the dilaton is not stabilized, it is possible to have thermal relic abundances of DM particles, such as neutralinos in supersymmetric cases,
whose density, however, obeys a {\it modified} Boltzmann equation by the dilaton and off-shell source terms:
\begin{equation} {\frac{d n}{dt} +
3\left(\frac{\dot a}{a}\right)n =  \dot \Phi n  -\frac{1}{2}
\left(e^{-\Phi}g^{\mu\nu}{\tilde \beta}_{\mu\nu}^{\rm Grav}
+ 2 e^{\Phi} {\tilde
\beta}^\Phi
\right)
n + \int \frac{d^3p}{E}C[f]} \label{mbol}
\end{equation}
with  $n= \int d^3p~f$ the species number density.
The $\beta$-terms denote the
off-shell terms in the gravitational (graviton, dilaton) multiplet of the string: ${\tilde \beta}_{\mu\nu}^{\rm Grav}
\propto \frac{\delta S}{\delta g^{\mu\nu}},~ {\tilde \beta}^\Phi
\propto \frac{\delta S}{\delta \Phi} \ne 0$, where $S$ is the low-energy string-inspired target-space  effective action.

If we set, for brevity, $\Gamma (t) \equiv \dot
\Phi - \frac{1}{2} \left(e^{-\Phi}g^{\mu\nu}{\tilde
\beta}_{\mu\nu}^{\rm Grav} + 2 e^{\Phi} {\tilde
\beta}^\Phi \right),$ acting as a source term on the right-hand-side of the Boltzmann equation (\ref{mbol}),
then we can solve this equation to obtain for the thermal DM relic abundance (assuming a single species, e.g. neutralino $\tilde \chi$)~\cite{lmnbol}:
\begin{eqnarray}
\Omega_{\tilde{\chi}} h_0^2 \;=\; \left( \Omega_{\tilde{\chi}} h_0^2 \right)_{no-source}
\times
{\left(  \frac{{\tilde g}_{*}}{g_{*}}   \right)}^{1/2} \;
{\rm exp}\left(\int_{x_{0}}^{x_f}  \frac{\Gamma H^{-1}}{x} dx \right)
\end{eqnarray}
 with
 $\left( \Omega_{\tilde{\chi}} h_0^2 \right)_{no-source}$
the standard (equilibrium) cosmology result (\ref{standardbol});
the star notation denotes quantities at the freeze-out point,
${\tilde g} = g  +
\frac{30}{\pi^2} T^{-4} \Delta \rho $, with $T$ the temperature, $g$ is the effective number of (thermal) degrees of freedom in this non-critical string Universe and $\Delta \rho$ denotes collectively the dilaton and off-shell (due to the $\beta$-functions) terms in the effective modified Friedman equation of the Q-cosmology~\cite{qcosmol,lmnbol}: $H^2=\frac{8 \pi G_N}{3}\; (\rho + \Delta \rho)$.

\begin{figure}[htb]
\begin{center}
\epsfig{file=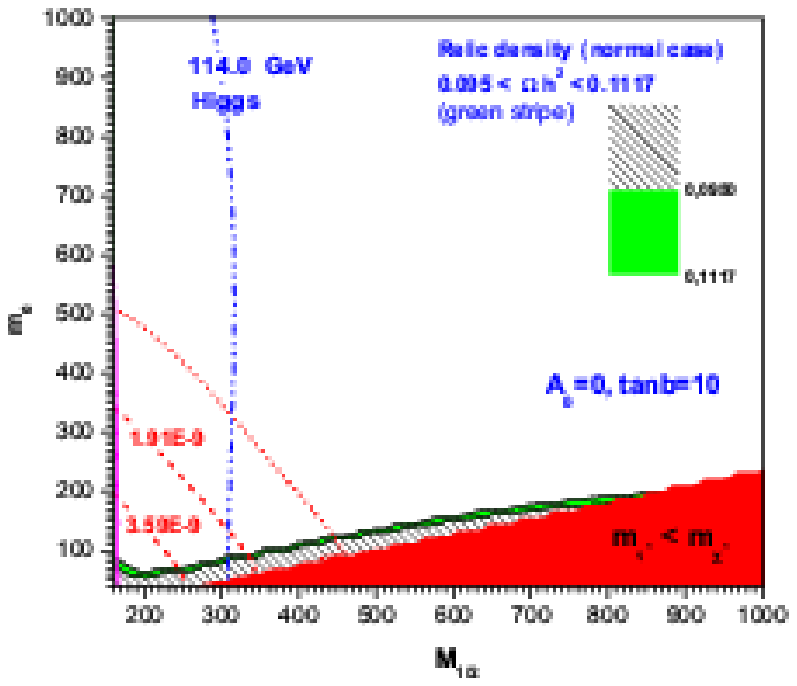, width=0.45\textwidth} \hspace{0.5cm}
\epsfig{file=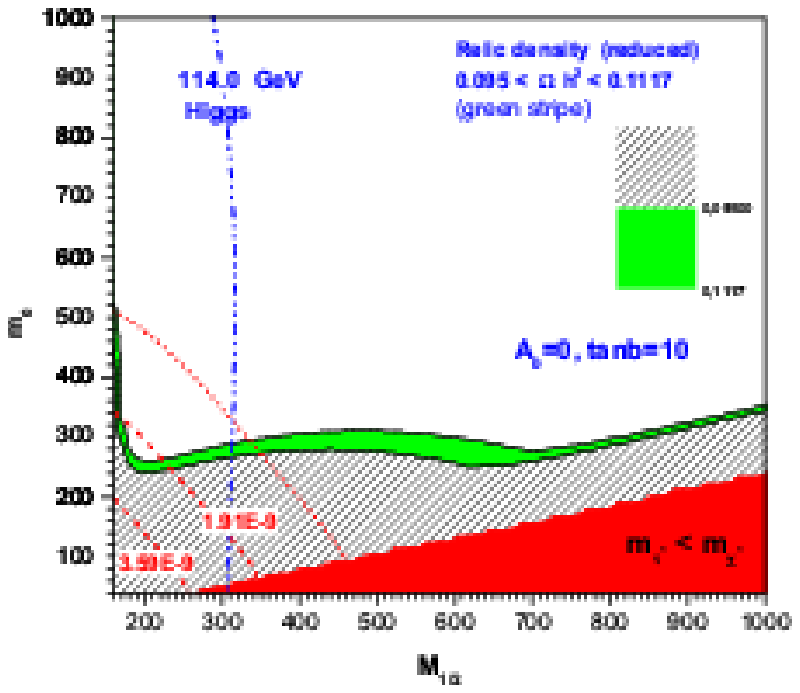, width=0.45\textwidth}
\end{center}
\vspace{-0.2cm}
\caption{{\underline{Left:}  In the thin green (grey) stripe the neutralino  relic density is within the  WMAP3 limits
 for values $A_0=0$ and $tan \beta=10$, according to the  source-free $\Gamma$=0 {\it conventional Cosmology}. The dashed lines (in red) are the $1\sigma$ boundaries for the allowed region by the $g-2$ muon's data as shown in the figure. The dotted lines (in red) delineate the same boundaries at the $2 \sigma$'s level. In the hatched region $0.0950 > \Omega_{CDM}h^2$, while in the dark (red) region at the bottom the LSP is a stau.   \underline{Right}: The same as in left panel, but according to the  non-critical-string  calculation, in which the relic density is reduced in the presence of dilaton sources $\Gamma = \dot \phi \ne 0$.}}
\label{lmn:fig1}
\end{figure}

Depending on their signature,
the Source Terms $\Gamma$ have different influence on the relic abundance, with profound consequences on the prospects for detecting supersymmetry in such models at colliders. For instance, for a given model~\cite{lmnbol} and within a certain region of the parameters, there is a reduction of neutralino relic abundance, as compared with conventional
cosmologies, by a factor of about 1/10 (c.f. figure~\ref{lmn:fig1}), leading to a relaxation  on some of the constraints regarding SUSY detection prospects at LHC imposed by mSUGRA models.

\section{Conclusions and Looking Ahead}

In this set of lectures I have reviewed various astrophysical methods for constraining the Universe energy budget, and used such results to constrain
interesting particle physics models of cosmological relevance.
I have discussed the issue of calculating thermal-relic DM abundances in those models, with the intention of placing constraints on interesting particle physics models, such as supersymmetry, exploiting astrophysical data. I mainly concentrated on DM searches at LHC.
I came to the conclusion that LHC could shed some light on the issue as to whether the
astronomically observed DM is the neutralino,  but this is a highly model dependent statement.
I have analyzed briefly the
phenomenology of various
SUGRA models (mSUGRA, and some string inspired ones, including some relaxation non-equilibrium dilaton-quintessence models). The associated  phenomenologies are  very different,  depending crucially on the details of the underlying dynamics, such as the type of the SUGRA model, the  way of SUSY breaking {\it etc}. There are model independent methods for testing DM at
colliders, but they pertain to subdominant processes at colliders.

For future directions it would be
desirable to
 explore in more detail SUSY models with  {\it CP violation}, which recently
 started attracting attention~\cite{belanger},
since, due to bounds on Higgs $m_H > 114$ GeV,
we now know that the amount of CP Violation in the Standard Model is  not sufficient  to generate the observed  baryon asymmetry of the Universe~\cite{PilaftsisWagner}, and hence SUSY CP violation might play an important r\^ole in this respect. At this point I mention that parameters
in SUGRA models
that can have CP phases are the gaugino and higgsino masses and trilinear sfermion-Higgs couplings.
 CP phases affect co-annihilation scenaria, and hence the associated particle physics dark matter searches at colliders~\cite{belanger}.
Another direction is to constrain
 SUSY GUTs models (e.g. flipped SU(5)) using astrophysical data~\cite{lmnreview}, after
taking, however, proper account of the observed dark energy in the Universe.
Personally, I believe that this dark energy is
due to some quintessence (relaxing to zero (non-equilibrium) field).
 WMAP data point towards an equation of state
of quintessence type, $w=p/\rho \to -1 $ (close to that of a cosmological constant, but not quite $-1$).
Such  features may be shared by dilaton quintessence, as discussed briefly above in the context of string theory. The issue is, however, still wide open and constitutes one of the pressing future directions for theoretical research in this field.

On the experimental side,
LHC and future (linear) collider, but also direct~\cite{zacek}, dark matter searches could shed light on the outstanding issue of the nature of the  Cosmological Dark Sector  (especially Dark Matter),  but one has to bear in mind that such searches are highly theoretical-model dependent.
To such ideas one should also add the models invoking Lorentz violation as alternative to dark matter. Clearly, particle physics can play an important r\^ole in constraining such alternative models in the future, especially in view of the upcoming high-precision terrestrial and extraterrestrial experiments, such as Auger, Planck mission, high-energy neutrino astrophysics {\it etc}.

Nothing is certain, of course,
and careful interpretations of possible results are essential.
Nevertheless, the
future looks promising, and certainly particle physics and astrophysics will proceed together and provide a fruitful and complementary experience to each other and exchange interesting sets of ideas for the years to come.

\section*{Acknowledgments}

I would like to thank the organizers of the Lake Louise Winter Institute 2007 (Lake Louise (Canada), February 19-24 2007) for the invitation and support and for providing such a successful and thought-stimulating meeting. This work is
partially
supported by the European Union through the Marie Curie Research and Training Network \emph{UniverseNet} MRTN-CT-2006-035863.

\end{document}